\documentclass[final,authoryear,5p,times,twocolumn]{elsarticle}
\usepackage{hyperref}
\makeatletter
\journal{Elsevier}
\def\ps@pprintTitle{%
	\let\@oddhead\@empty
	\let\@evenhead\@empty
	\def\@oddfoot{}%
	\let\@evenfoot\@oddfoot}
\makeatother
	
\usepackage{numcompress}\biboptions{authoryear}

\usepackage{graphicx} 
\usepackage{amsfonts}
\usepackage{amsmath} 
\usepackage{bm}
\usepackage{bbm}
\usepackage{amssymb} 
\usepackage{hyperref} 
\usepackage{siunitx} 
\usepackage{booktabs} 
\usepackage{cleveref}
\usepackage{float}
\usepackage[caption = false]{subfig} 
\usepackage{multirow}
\usepackage{booktabs}
\usepackage{tabu}
\usepackage{color}
\usepackage{todonotes}
\usepackage{filecontents,pdfpages}
\usepackage{float}
\restylefloat{table}

\begin{document}
	
	\begin{frontmatter}
		
		\title{
		%Forecasting innovation dynamics using technological interdependencies\\
		%Predicting innovation dynamics in the technological ecosystem\\
		Technological interdependencies predict innovation dynamics}
		
		\author[inet,math,csh]{Anton Pichler} \ead{anton.pichler@maths.ox.ac.uk}
		\author[inet,math]{Fran\c{c}ois Lafond}
		\author[inet,math,sfi]{J. Doyne Farmer} 
		
		\cortext[cor]{Corresponding author}

		\address[inet]{Institute for New Economic Thinking at the Oxford Martin School, University of Oxford, Manor Road, OX1 3UQ, UK}
		\address[math]{Mathematical Institute, University of Oxford, Woodstock Road, Oxford OX2 6GG, UK}																																																														
		\address[csh]{Complexity Science Hub Vienna, Josefst\"adter Stra\ss e 39, A-1080, Austria}
		%\address[oms]{Oxford Martin School, University of Oxford, 34 Broad Street, Oxford OX1 3BD, UK}
		\address[sfi]{Santa Fe Institute, 1399 Hyde Park Road, Santa Fe, NM 87501, USA}

\begin{abstract}
We propose a simple model where the innovation rate of a technological domain depends on the innovation rate of the technological domains it relies on. Using data on US patents from 1836 to 2017, we make out-of-sample predictions and find that the predictability of innovation rates can be boosted substantially when network effects are taken into account. In the case where a technology's neighborhood future innovation rates are known, the average predictability gain is 28\% compared to simpler time series model which do not incorporate network effects. Even when nothing is known about the future, we find positive average predictability gains of 20\%. The results have important policy implications, suggesting that the effective support of a given technology must take into account the technological ecosystem surrounding the targeted technology.
\end{abstract}


\end{frontmatter}
%\linenumbers

\section{Introduction}

Technological evolution is often described as a recursive process whereby the recombination of existing components leads to new or improved technological components \citep{schumpeter1939business,usher1954,kauffman1993origins,fleming2001recombinant, arthur2009nature,mcnerney2011role,tria2014dynamics,youn2015invention,fink2019much}. A simple hypothesis, therefore, is that technological domains that tend to recombine elements from fast-growing technological domains should themselves grow faster. In other words, a technology will tend to progress faster if the technologies it relies on are themselves making fast progress. 
%From a system perspective, these technological interdependencies form a technological network or ecosystem where a burst of activity in ``upstream'' technologies should lead to more activity in ``downstream'' technologies. 
While these ideas are well established, very little has been done to establish empirically that technological interdependencies help predict future innovation dynamics. Being able to demonstrate this relationship would be very helpful, as it would allow us to support key technologies and overall technological progress by designing and supporting technological ecosystems.

In this paper, we establish that knowing the technological ecosystem helps predict the dynamics of future innovation. We use a simple model where the innovation rates of a technological domain depends on efforts within the domain, but also on the stock of knowledge in the domain and in supporting domains. We test the model on the record of United States Patent Office (USPTO) patents from 1836 to 2017, which includes more than 10M patents, 40M classifications in $\sim 650$ technological domains, and 90M citations. As predicted by the model, we find that that the growth rates of the number of patents in a technological domain depends strongly on the growth rates of its knowledge sources. Given the volatile nature of growth rates, this strong relationship is remarkable. We use this insight for making out-of-sample predictions of patenting activities and find that integrating network effects can improve predictions substantially compared to independent time series models. The results have important policy implications, suggesting that research policy targeted at fostering innovation in a technological domain has to take its surrounding technological ecosystem into account.

Several studies have put forward the idea of network-dependent innovation dynamics. For instance, \cite{cowan2003dynamics} and \cite{konig2011recombinant} have proposed models of innovation arising from the recombination of knowledge in R\&D partnership networks.

\cite{acemoglu2016innovation} find that upstream patenting levels are highly correlated with downstream patenting levels. \cite{taalbi2018evolution} finds broadly similar results using innovation counts and a network constructed to reflect which industry uses innovation from another industry. Patenting levels are persistent, and so are automatically very predictable. In contrast, we focus on changes in patenting levels, which are not persistent and are much more difficult to predict. 

\iffalse
Empirically, \cite{acemoglu2016innovation} find that upstream patenting levels are highly correlated with downstream patenting levels. Similar results are found by \cite{taalbi2018evolution} who studies patent citation data from Sweden. Due to the persistence of patenting levels, this finding is not entirely unexpected.
In contrast to these studies, we quantify network effects on changes in patenting levels, which are not very persistent and are much more difficult to predict.
\fi

Our work on innovation dynamics is also related to evolutionary models such as \cite{farmer1986autocatalytic}, \cite{bagley1992evolution} and \cite{jain2001model}. In these models, the exponential growth trajectories of nodes arise due to the existence of autocatalytic sets, i.e. a subset of the network where nodes have at least one positive incoming link from a node of the same subgraph, leading to sustainable self-reinforcing dynamics. 
Evidence for the presence of autocatalytic sets in technology systems was recently provided by \cite{napolitano2018technology} who find that the autocatalytic structure of the patent network has grown in time and that patent classes belonging to the autocatalytic set show higher levels of innovation activity.

Our contribution to the literature can be summarized as follows. First, we show strong empirical evidence of coupled innovation growth in the presence of network linkages and demonstrate that the observed effects are far from what would be expected by chance (Section \ref{sec:empirical}). Second, we propose a simple theoretical model of network-dependent knowledge production which is able to explain the observed empirical pattern. We then show how the model can be estimated from empirical data, and quantify how network effects in innovation dynamics have become more important over the last 70 years (Section \ref{sec:model}). Finally, we validate the model by predicting future patenting levels (Section \ref{sec:predict}). We find that independent time series models can be substantially improved in case network information is available.

\section{Empirical evidence} \label{sec:empirical}

\begin{figure*}[!tbp]
	\centering
	\includegraphics[width=0.9\textwidth]{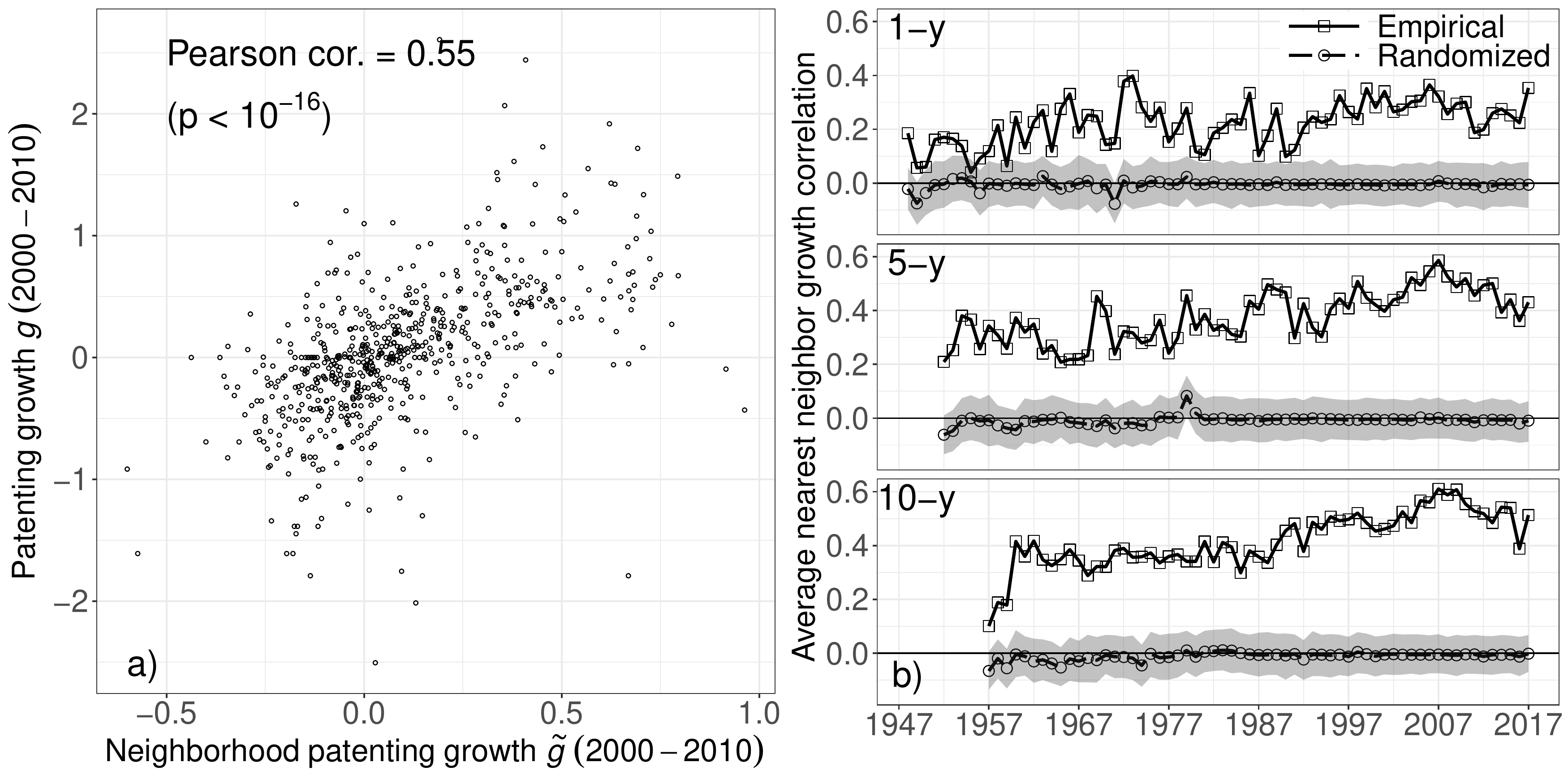} 
	\caption{
		a) Technologies' 10-year growth rates $g$ for the horizon 2000 to 2010 plotted against average nearest neighbor growth rates $\tilde{g}$ (ANNG). The neighbors' growth rates are calculated based on the technology network in 2000. The plot thus suggests that technologies experienced a similar growth rate over the next ten year as the technologies they were linked to in 2000.
		b) Network assortativity over time with respect to 1-year (upper panel), 5-year (center panel) and 10-year (bottom panel) growth rates. There is pronounced correlation between growth rates of nodes and their neighbors in the technology network, while the randomized assortativity measure hovers around zero. The shaded area around the randomized network predictions depict the 0.05-0.95 quantile range of the 1,000 sample realizations. The correlations tend to get stronger in time in the empirical network.} 
	\label{fig:avnng}
\end{figure*}

\subsection{Data and network construction}
We use data on the whole universe of granted United States patents starting in 1836 up to 2017, where the year refers to the publishing date of the patent. The dataset covers 9.83 million patents and over 91 million citations.
Each patent is categorized by the patent office into one of several Cooperative Patent Classification (CPC) codes. We regard each 4-digit CPC code as a distinct technological domain\footnote{
To check robustness of our results with respect to alternative classification schemes and aggregation levels, we redo the entire analysis presented in the main text on the level of CPC 3-digits, International Patent Classification (IPC) 3- and 4-digits and report the results in the supplementary information.
}.
We construct a directed network $C_t$ where an element $C_{ij,t}$ is the sum of all citations from technology $j$ at time $t$ to technology $i$. In \ref{app:tempnet} we show how this matrix can be derived from patent citations by straightforward algebra\footnote{
The analysis shown here does not rule out alternative technological distance measures. Instead of using technology citation networks, other networks such as co-classification and co-citation networks could be used.
}. In our analysis, we use row-normalized matrices $\{W_t: t=1947,...,2017\}$ 
where each element $W_{ij,t}$ is defined as 
\begin{equation} \label{eq:tech_nw_ext}
{W}_{ij,t} := \frac{C_{ij,t}}{\sum_{j =1 } C_{ij,t}}.
\end{equation}
Thus, the row $\{W_{ij,t}: j=1\dots N\}$ can be interpreted as technology $i$'s dependence in its patenting activity at time $t$ on all other technologies: each entry is the share of citations from technology $i$ at time $t$ to another technological domain $j$. This is a temporal network with yearly snapshots \citep{masuda2016guide}. We do not construct networks for years before 1947, because earlier citations made by patents are not well documented. But citations \emph{to} earlier patents are well reported. For example, a citation from a patent granted after 1947 to a patent of the 19$^{\text{th}}$ century is included in the network\footnote{
\cite{alstott2017mapping} argue that links of technology networks can be driven by domain sizes and patent ages and suggest a network normalization procedure to control for these impinging factors. We have applied this method as explained in detail in \ref{app:significant}, but our results are not sensitive with respect to the network normalization.
}.

\subsection{Preliminary evidence}
We now investigate whether technological domains connected to fast-growth technological domains also grow faster. In other words, we study the assortativity of the temporal network with respect to patenting growth rates. For each technology, we compute its neighborhood patenting growth rate as the average nearest neighbor growth rate (ANNG),
\begin{equation}
\tilde{g}_{i,t} = \sum_{j=1}^N \frac{C_{ij,t} (1-\delta_{ij})}{\sum_{k =1 } C_{ik,t} (1-\delta_{ik}	)} g_{j,t},
\end{equation}
where $\delta_{ij}$ denotes the Kronecker delta (the matrix diagonal is excluded in the summation so that the ANNG measures neighbor growth rates only). We then test if there is a positive correlation between growth rates of patenting activity in technological domains and their ANNG. 

Fig. \ref{fig:avnng}a) plots the 10-year growth rates against the 10-year ANNG for the technology network in 2000, revealing a highly significant positive relationship in growth rates from 2000 to 2010 between technologies and their neighborhood. Since the technology network is based on the year 2000, the figure suggests that if a technology draws a significant share of knowledge from another technology in 2000, we would expect them to exhibit similar growth rates over the next ten years. This pronounced positive relationship is striking given that it is based on growth rates, which tend to be much noisier than patenting levels. While one could expect the network to be assortative with respect to degrees and patenting levels, assortativity with respect to growth rates is far less trivial.

Although interesting, we should test if this relationship holds over time and if we really can be sure that the observed relationship is significantly different from what a null model would produce. To benchmark our results, we calculate the correlations between growth rates and ANNG rates for a randomized version of the given technology network. For each year $t$ and a fixed row $i$ in the matrix ${W}_t$, we resample all off-diagonal entries without replacements. In the randomized control network the nodes still have the same weighted outgoing links, but now randomly pointing to other nodes (excluding self-loops). We repeat the randomization process 1,000 times and report the average correlation between growth rates and ANNG as the corresponding null model. Fig. \ref{fig:avnng}b) plots the Pearson correlation over time between growth rates and ANNG, once computed in the given technology network and once in the randomized control. We report the results for three different growth rates, 1-year (upper panel), 5-year (center panel) and 10-year growth rates (bottom panel). The positive correlation structure between knowledge sources growth and own growth rates over time is far from what one would expect from a network where nodes distribute their out-links randomly over the whole set of potential knowledge sources.
Intuitively, one would expect that network effects materialize over the long run instead of showing immediate effects.  The figure confirms this hypothesis, but even for 1-year growth rates we find relatively strong and significant positive correlations.
Another interesting aspect is that for all three growth rate types, the positive relationship tends to get stronger over time.\footnote{
It should be noted that technology citation networks are by construction correlated with co-classification networks (see \ref{app:tempnet}), and thus, their marginal effects are difficult to disentangle. To test whether there is signal in the citations themselves, we computed the same ANNG measures for USPC main classes. When considering only USPC main classes, there is no co-classification, allowing us to separate these effects. Although slightly noisier, we also find strongly positive ANNG values for the less granular USPC main classes.}
Further evidence on the impact of a technology's neighborhood on its patenting dynamics is discussed in the supplementary information.

\section{Network-dependent knowledge creation} \label{sec:model}

\subsection{The model}

To explain the empirical observations, we now introduce a simple model of network-dependent knowledge creation and discuss its implications for the long-term evolution of technologies. Let us consider $N$ distinct technological domains. The creation of new knowledge in a technological domain requires active research effort, and depends positively on the existing stock of knowledge, from the same domain or from specific other domains.

More precisely, we assume that the creation of new knowledge follows the dynamical system
\begin{equation} \label{eq:know_prod}
\dot{K}_{i}(t) = \theta_i R_{i}(t)^{\alpha} \prod_{j =1}^N K_{j}(t)^{\beta W_{ij}} ,
\end{equation}
where $\theta_i$ is a technology-specific productivity parameter, 
$R_{i}(t)$ the research effort in domain $i$ at time $t$ and
${K}_{i}(t)$ is the stock of knowledge in $i$ at $t$.
The technological ecosystem is represented by the weighted adjacency matrix $W$ which is normalized to be row-stochastic and for mathematical convenience assumed to be fixed in time. A technological domain is then a node in the network whose innovation rates depend on its location in the technological ecosystem.
If a technology $i$ draws knowledge from the knowledge stock of node $j$ in the innovation process, the two nodes are connected through a directed edge from $i$ to $j$. The directed link has the weight $W_{ij}$ denoting the share of technology $j$'s knowledge in the creation of new knowledge in technology $i$.
$\theta_i$ captures the fact that intrinsic characteristics of technologies affect how easy innovation rates can be influenced.
$\alpha\ge 0$ denotes the elasticity of knowledge output with respect to research efforts. The elasticity of knowledge output in $i$ ($\dot{K_i}$) with respect to the knowledge stock in $j$ ($K_j$) is $\beta W_{ij}$, thus $\beta$ denotes the sum over all domains $j$ of these elasticities (since $\sum_j W_{ij}=1$).

Eq. (\ref{eq:know_prod}) also relates to the knowledge production functions used in classical endogenous growth models (\cite{romer1990endogenous}, \cite{aghion1990model}, \cite{grossman1991quality} and \cite{jones1995r} ), but incorporates network effects. 
It simplifies to the standard Cobb-Douglas knowledge production function in case each technology uses only its own knowledge stock ($W = \mathbb{I}$). Similar equations have been estimated empirically within the ``R\&D spillovers'' literature which estimates the impact of R\& D in one entity on outcomes in another entity, such as countries, regions, firms, and sectors \citep{ertur2007growth, hall2010measuring, ho2018international}.

Research effort is considered to be an exogenous policy variable which we assume to grow at a constant rate $R_{i}(t) = R_{i,0}e^{\lambda_i t}$.
Dividing Eq. (\ref{eq:know_prod}) by $K_i(t)$, taking logs and the derivative with respect to time, we obtain after rearranging the nonlinear autonomous system
\begin{equation} \label{eq:know_prod_Kgrow}
\dot{g}_{i}(t) =  \alpha \lambda_{i} {g}_{i}(t) + (\beta W_{ii} - 1) g_{i}(t)^2 + 
\beta g_{i}(t)  \sum_{j=1}^N W_{ij} g_{j}(t) (1-\delta_{ij}),
\end{equation}
where $g_{i}(t) := \dot{K}_i(t)/K_i(t)$ is the growth rate of the knowledge stock.
Eq. (\ref{eq:know_prod_Kgrow}) has been extensively studied without network effects in traditional endogenous growth models and in this case its dynamics are well-understood (\cite{romer2012advanced}, ch.3). When solving the model without network effects, the steady state growth rate (``balanced growth path'') is $g_{i}^* = \alpha \lambda_{i} / (1 - \beta)$ which is globally stable under the standard assumption of $\beta < 1$. Here, the long run-growth rate of knowledge in technology $i$ can only be increased by increasing the long-run growth rates of research efforts in the particular technology.

When network effects are included, a technology's long-term growth path depends also on the growth rates of other technologies. By setting $\dot{g}_{i} =0$ for all $i$, we find the steady state of the form
\begin{equation}\label{eq:steady_state_nw}
g_{i}^* = \frac{ \alpha \lambda_{i} }{1- \beta W_{ii}} + \frac{\beta }{1- \beta W_{ii}} \sum_{j =1}^N W_{ij} g_{j}^*(1-\delta_{ij}).
\end{equation}
The steady state growth path for technology $i$ can be understood as a sum of two components. The first part is the idiosyncratic term, which equals the endogenous growth result without network effects. The second component suggests that the long-run growth of a technology depends positively on its neighbors' growth rates.

\begin{figure}[t]
	\centering
	\includegraphics[width=.23\textwidth]{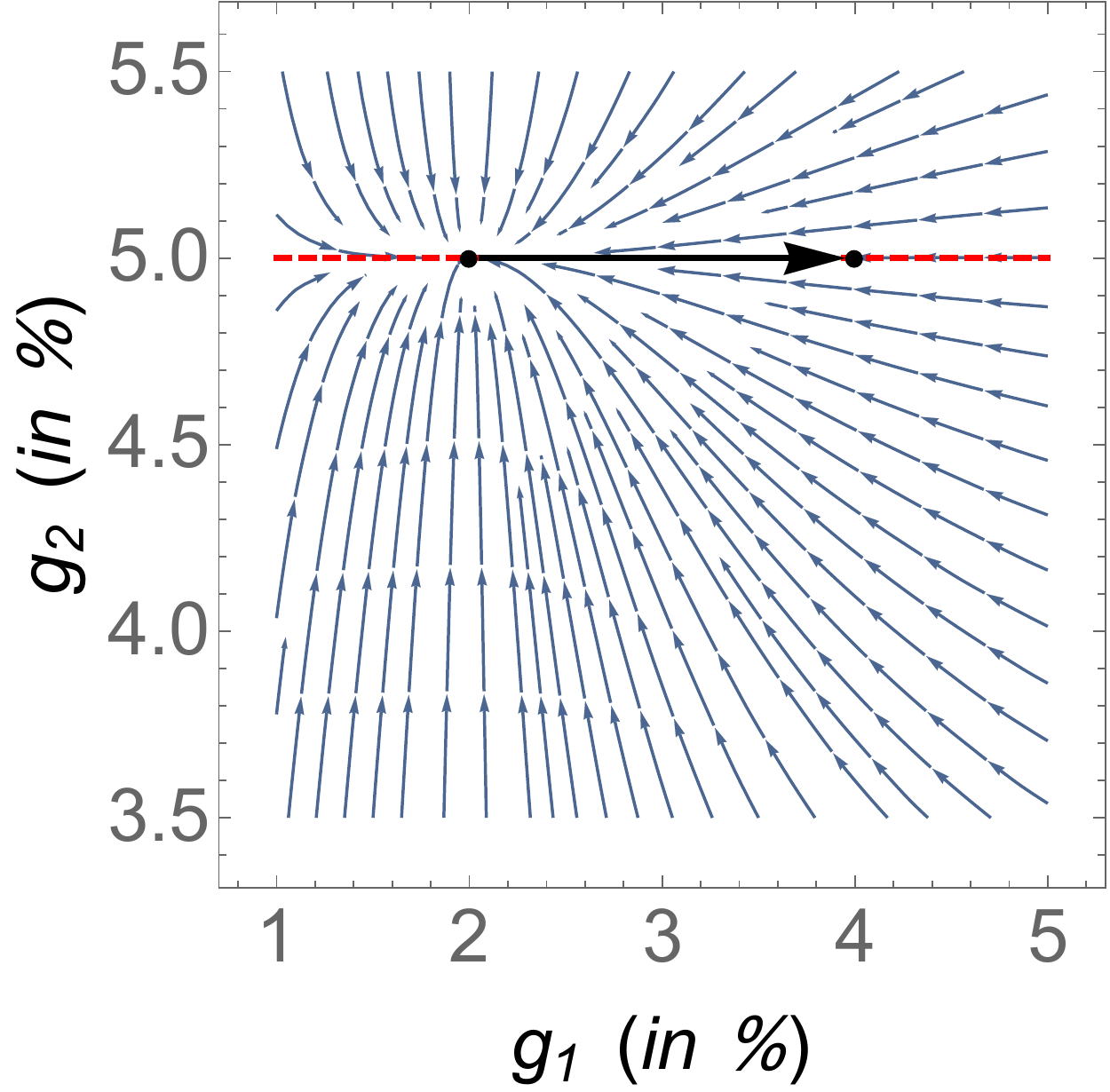} 
	\includegraphics[width=.23\textwidth]{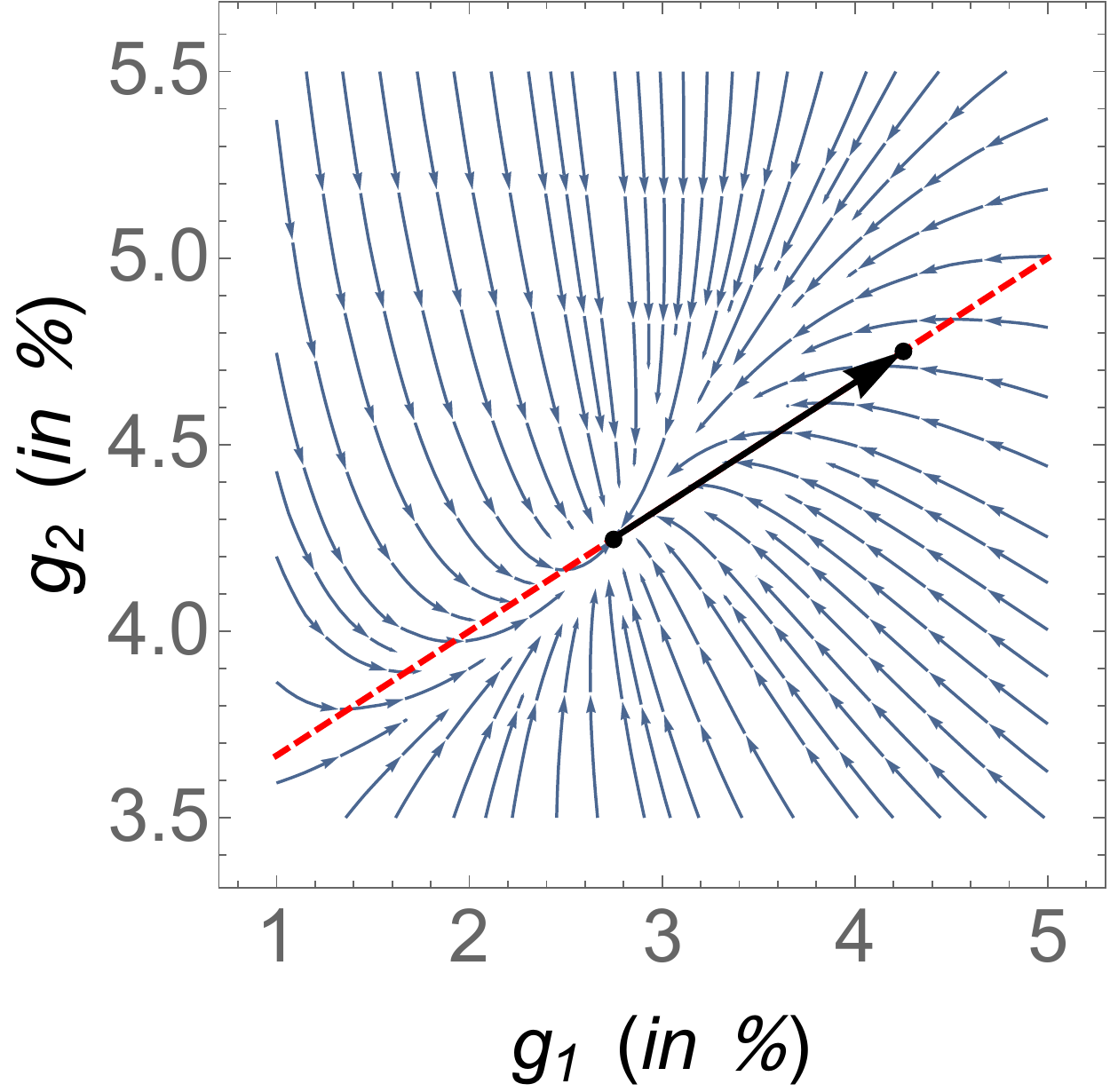}
	\caption{Phase portraits of growth trajectories for a simple example of only two technologies. 
		 	The horizontal and vertical axes denote the growth rates of technology 1 and technology 2, respectively.
			In the left panel there are no network effects, $W=\mathbb{I}$. In the right panel a network is present with elements $W_{ij}=0.5$ for all $i,j$.
			The phase diagrams are shown for initial research growth rates $\lambda_1 = 0.02$ and $\lambda_2 = 0.05$. The remaining parameters are set as follows: $\alpha = \beta = 0.5$.
			With this parametrization all trajectories converge to the steady state $(g_1^*, g_2^*) = (0.02, 0.05)$ in the no-network case and to the steady state $(g_1^*, g_2^*) = (0.0275, 0.0425)$ if network effects are included. In the absence of network effects, the steady state growth rates are simply equal the research effort growth rates. If network effects are included, technology 1 is growing faster in steady state since it draws knowledge from technology 2 which experiences higher research effort growth. In contrast, for technology 2 the steady state growth rate is lower due to the network effects.
			The black arrows indicate how the steady state of the system moves if research effort growth $\lambda_1$ is doubled to $\lambda_1' = 0.04$, but leaving $\lambda_2 = 0.05$ unchanged. In the left panel, this increases only the steady state growth rate for technology 2, but leaves technology 2 unaffected. In the right panel where network effects are included both technologies benefit from the increase in research effort in technology 1. The red dashed line shows the trajectory of the steady state as a function of $\lambda_1$. 
	} 	
	\label{fig:phase}
\end{figure}

To see the difference between the network-dependent model and a simple endogenous growth model version without network effects, let us assume that we could increase the constant growth rate of research effort in a technology $i$ from $\lambda_{i}$ to $\lambda_{i}'$. Without network effects, this would simply increase the growth rate of $i$ by $\alpha  (\lambda_{i}' - \lambda_{i}) /(1-\beta)$ with no impact on other technologies. Yet if network effects are included, this initial increase of $i$'s growth rate will also impact neighboring technologies which draw upon the knowledge stock of $i$, which in turn, will again affect their downstream neighbors and so on. If node $i$ points to any of the affected technologies, it will again experience a change in its growth rate and trigger the process again. We see that the convergence (if any) to the steady state after a shock in the research effort variable is more involved if the innovation rate of a technology depends on other technologies.
The phase portraits depicted in Fig. \ref{fig:phase} visualize how a change in research efforts impact innovation dynamics in a simple system of only two technologies, one with network effects (right panel) and one without (left panel).

\subsection{Calibration}

While the dynamical system can have a non-trivial transient, for simplicity we calibrate the model based on the steady state results, using maximum likelihood. To do so, we reformulate Eq. (\ref{eq:steady_state_nw}) as the econometric model

\begin{equation}\label{eq:econometric_model}
g_{i,t} = \frac{ a_{i} }{1- \beta W_{ii,t}} + \frac{\beta }{1- \beta W_{ii,t}} \sum_{j =1}^N W_{ij,t} g_{j,t}(1-\delta_{ij}) + \epsilon_{i,t},
\end{equation}
where $\epsilon_{i,t} \sim N(0,\sigma^2)$ and $a_i$ is capturing the composite variable $\alpha \lambda_{i}$. 

The specification is related to spatial econometric models using panel data \citep{elhorst2014spatial, wang2015estimation}, but there are also significant differences. To estimate the parameters, we therefore derive an estimator which is outlined in more detail in \ref{app:regress}.
We allow the technology network to be time-varying, since we observe changing citation networks over time. In the supplementary information we present results of extensive robustness checks, covering alternative model specifications such as using time-fixed networks and spatial autoregressive models. There we also control for further possible explanatory variables, such as technological domain sizes.

The model is derived in terms of knowledge stocks $K_{i,t}=\sum_{\tau=0}^t P_{i,\tau}$, but knowledge stocks grow very smoothly, leaving little variance to exploit for estimating the parameters of interest. In the steady state, however, by definition growth rates are constant over time within each domain. In that case, the (deterministic) growth rates of stocks and flows are the same. Thus, for empirical convenience, we let $g_{i,t} := \ln(P_{i,t}/P_{i,t-\tau})$ be the $\tau$-year patenting growth rate of the technology class $i$ at a time $t$.

We fit the model to the whole time series up to 2017 and report the estimated parameters in Table \ref{t:regress_res}. Since the results of Section \ref{sec:empirical} suggest varying magnitudes of network effects for different time lags, we estimate the parameters for 1-year, 5-year and 10-year growth rates (columns one to three, respectively). Unsurprisingly, we find research growth parameters $a_i = \alpha \lambda_i$ which are positive on average and higher for larger time lags.
The highly significant network parameter $\beta$ is large, ranging from 0.85 to 0.94, depending on the growth rate lag. This value is large, because if $\beta > 1$, we would expect exploding dynamics where an increasing knowledge stock leads to ever larger growth rates. $\beta$ is smaller, but close to one, exemplifying the importance of the existing technology network for future innovation dynamics.

\begin{table}[t]
	\centering
	\resizebox{.35\textwidth}{!}{		
		\begin{tabular}{lccc}
			\\[-1.8ex]\hline 
			\hline \\[-1.8ex] 
			\multicolumn{4}{r}{\emph{Dependent variable: patenting growth}} \\ 
			\cline{2-4} 
			\\[-1.8ex] & (1-y) & (5-y) & (10-y)\\ 
			\hline \\[-1.8ex]
			average $\hat{a}_i$ & 0.01 & 0.03 & 0.04 \\[1ex] 
%			& (XXX) & (XXX) & (XXX) \\ 
			$\hat{\beta}$ & 0.85$^{***}$ & 0.93$^{***}$ & 0.94$^{***}$\\ 
			& (0.0059) & (0.0060) & (0.0076) \\[1ex] 
			$\hat{\sigma}^2$ & 0.10 & 0.19 & 0.29\\ 
			%& 0.0007 &  0.0030 & 0.0062 \\[1ex]
			 \hline \\[-1.8ex]
			Log-likelihood & -13,383 &-5,464  &-3,612  \\ 
			Observations & 43,651 & 8,734 & 4,348 \\
			\hline 
			\hline \\[-1.8ex] 
			\emph{Note:}  & \multicolumn{3}{r}{$^{***}$p$<10^{-16}$}
		\end{tabular}
			\caption{Results from estimating the econometric network model (Eq. \ref{eq:econometric_model}). The results are shown for 1-, 5- and 10-year growth rates. 
			}
	\label{t:regress_res}
	}
	
\end{table}

To explore the network impact on innovation growth more systematically, we rewrite the derived result in Eq. (\ref{eq:steady_state_nw}) into matrix notation,
\begin{equation} \label{eq:steady_state_nw_matrix}
g^* = \alpha L \lambda,
\end{equation}
where $L := \left[\mathbb{I} - \beta W \right]^{-1}$.
The matrix representation is useful as it shows how the long-run trajectory of innovation in a given technological domain depends on the research efforts in the entire technological ecosystem. In particular, if research effort is subject to policy, we can use Eq. (\ref{eq:steady_state_nw_matrix}) to study two interesting policy experiments. 

First, Eq. (\ref{eq:steady_state_nw_matrix}) allows us to identify the key supporting technologies for each technology. The na\"ive way to foster innovation in a technology is to increase research efforts in the particular technological domain. But Eq. (\ref{eq:steady_state_nw_matrix}) shows that knowledge growth depends also on research in other technologies and that there could even be cases where knowledge spillovers from research in other technologies will be more beneficial than simply devoting additional research efforts in the focal technology's domain\footnote{
When estimating the model based on 1-year growth rates, we find five technologies with larger off-diagonal entries than diagonal elements. Although these technologies are very different, ranging from \emph{multipurpose hand tools} (B25F) to \emph{emergency protective circuit arrangements} (H02H), they all rely heavily on the general purpose technology class A61B, \emph{diagnosis, surgery and identification}, which exhibits the largest downstream spillovers, $\sum_j L_{ij} (1-\delta_{ij})$, of all technologies.
}. 
%\red{proof?? are there lines of L where Lii is not the largest?}. 
Knowledge spillovers from technology $j$ to technology $i$ are made explicit when looking at the Jacobian matrix
\begin{equation} \label{eq:spillover}
\frac{\partial g^*_{i}}{\partial \lambda_{j}} = \alpha L_{ij}.
\end{equation}
The matrix element $L_{ij}$ therefore informs us on how much we would expect the long-term innovation growth rate of technology $i$ to change as a consequence of a marginal change in technology $j$. A large row sum of off-diagonal elements $\sum_{j =1} L_{ij}(1-\delta_{ij})$ indicates a large dependence of technology $i$'s growth on external research activities. Ordering technologies based on their size in a given column yields a ranking from the most to least important supporting technologies for the focal technology.

As a second policy experiment, we can ask how a sustained change in R\&D in a particular sector affects the long-run growth rate of the entire technological ecosystem. This can be relevant for a policy-maker who wants to devote her research investments in an efficient manner. For example, if a choice has to be made in what technology to invest for fostering overall innovation activities, the decision can be supported with Eq. (\ref{eq:steady_state_nw_matrix}). 
If an economy's total knowledge is the sum of sectoral knowledge stocks $K_{tot}(t) = \sum_{i=1}^N K_i(t)$, the growth rate of the total knowledge stock can be expressed as $g_{tot}(t) = \sum_{i=1}^N g_{i}(t) \frac{K_{i}(t)}{K_{tot}(t)}$. The impact on total knowledge growth as a consequence of a sustained change in research effort growth in a single domain $i$ is then given by
\begin{equation} \label{eq:multiplier}
\frac{\partial g^*_{tot}}{\partial \lambda_{j}} = 
\alpha \sum_{i=1}^N L_{ij} \frac{K_i}{K} +
\alpha \lambda_{j} \sum_{i=1}^N \frac{\partial (K_i/K)}{\partial \lambda_{j}} L_{ij}.	
\end{equation}
The impact on overall innovation growth is again a sum of two components. The first part in the sum takes the form of a output multiplier, analogous to the output multiplier in input-output economics. In contrast to traditional input-output economics \citep{leontief1936, miller2009input}, however, the output multiplier has to be weighted with the relative domain sizes since the model is in growth rates instead of levels. Increasing research efforts in a domain with a high weighted output multiplier entails large positive effects on overall innovation rates. The second part in the sum accounts for the fact that the weights themselves change due to changes in research effort.

\section{Predicting innovation rates} \label{sec:predict}
The empirical evidence for network effects on innovation dynamics is strong. In this section we take advantage of this finding to test the predictive power of the estimated network model. We use the model to make out-of-sample forecasts of patenting levels and evaluate their performance by comparing them with predictions based on time series models which do not incorporate network effects explicitly. 
We test the network model in two separate prediction exercises. In the first prediction exercise we ask whether knowledge on innovation dynamics around a focal technology can help predicting its growth rates. Or to put it differently, conditional on growth rates of $i'$s neighbors, what is $i'$s growth rate? We call these the \emph{conditional} forecasts. To make this more precise, we define the conditional predictor as 
\begin{align} \label{eq:predex1}
	\hat{g}_{i,t+\tau}^{cond.} :=  \frac{\hat{a}_i}{1- \hat{\beta} W_{ii,t}} + \frac{ \hat{ \beta}}{1- \hat{\beta} W_{ii,t}} \sum_{j=1}^N W_{ij,t} g_{j,t+ \tau} (1-\delta_{ij}).
\end{align} 
We also make forecasts where we do not assume knowledge on the neighbors' growth rates, and use only information available up to $t$. We call these the \emph{unconditional} forecasts. As explained in detail in \ref{app:predict}, predictions in this case are based on the estimator
\begin{equation} \label{eq:predex2}
\hat{g}_{t+\tau}^{uncond.} := 
\left(\sum_{k=0}^{k'} \hat{\beta}^k W_t^k \right) \hat{a} \approx
L(\hat{ \beta})\hat{a},
\end{equation}
where the approximation is exact in case $k' \to \infty$.
The predictions are essentially agnostic with respect to the used technology network. While our focus lies on knowledge spillovers, the model is easily extendable to alternative technological distance measures, such as co-classification or co-citations networks.

The forecasts of both prediction exercises are compared to the forecasts obtained from standard ARIMA($p$,1,$q$) time series models
\begin{equation} \label{eq:arima}
g_{i,t} = \nu_i + \sum_{s=0}^{p} \phi_{i,s} g_{i,t-s} + \sum_{s=0}^{q} \psi_{i,s} u_{i,t-s} + u_{i,t},
\end{equation}
where $\phi_{i,0} = \psi_{i,0} = 0$. Note that an ARIMA simply reduces to a geometric random walk if $p=q=0$.
%\red{is this formally correct? that means the index s in the first sum, for instance, runs from 1 to 0}. 
As discussed in more depth in the \ref{app:predict}, we choose $p$, $q$ and $k'$ such that the models yield the best forecasting performance in a validation set. The chosen parameters are shown in detail in the supplementary information.
Note that predictions from the ARIMA model are the same for both prediction exercises, since here forecasts for a specific domain $i$ are completely independent of other domains.

We estimate all models based on the data between 1947 and 2002 and use the result to predict yearly patenting from 2003 to 2017. 
%\red{are results robust to change of date? I'd like to know even if we don't write it down}. 
To assess the predictive performance of the network models, we calculate the predictability gain for each year and technology as
%\begin{align}
	%PG_{i,t} &=  \frac{ | {P}_{i,t} - \hat{P}_{i,t}^{ARIMA} | - | {P}_{i,t} - \hat{P}_{i,t}^{network} | }{ |{P}_{i,t} | }, \label{eq:pg}
%\end{align} 
\begin{align}
	PG_{i,t} &=  \frac{ | {P}_{i,t} - \hat{P}_{i,t}^{ARIMA} |}{ {P}_{i,t} } - \frac{| {P}_{i,t} - \hat{P}_{i,t}^{network} | }{ {P}_{i,t} }, \label{eq:pg}
\end{align} 
where $\hat{P}_{i,t}^{ARIMA}$ denotes the predicted number of patents from the ARIMA model and $\hat{P}_{i,t}^{network}$ the predictions from the network models. Eq. (\ref{eq:pg}) is simply the difference between absolute percentage errors from the ARIMA and the network model predictions.

When taking time averages of Eq. (\ref{eq:pg}) for each time series, we find for 86\% of technologies positive mean predictability gains in the conditional forecasts. In the unconditional forecasts, we get positive predictability gains in 63\%. The result looks similar when taking time medians instead of averages where we find positive predictability gains for 84\% of all conditional forecasts and for 62\% of unconditional forecasts. For each technology, we also conduct a one-sided t-test to check if the predictability gains are significantly greater than zero. We find that predictability gains are significantly larger than zero (on the 5\% level: note that we have only 15 observations per series) in 69\% of all cases for the conditional forecasts and in 41\% for unconditional forecasts. On the other hand, only 6\% of all time series are found to have significantly negative predictability gains in the conditional forecasting setup and 18\% in the unconditional one.

\begin{figure}[!tbp]
	\centering
	\includegraphics[width=.49\textwidth]{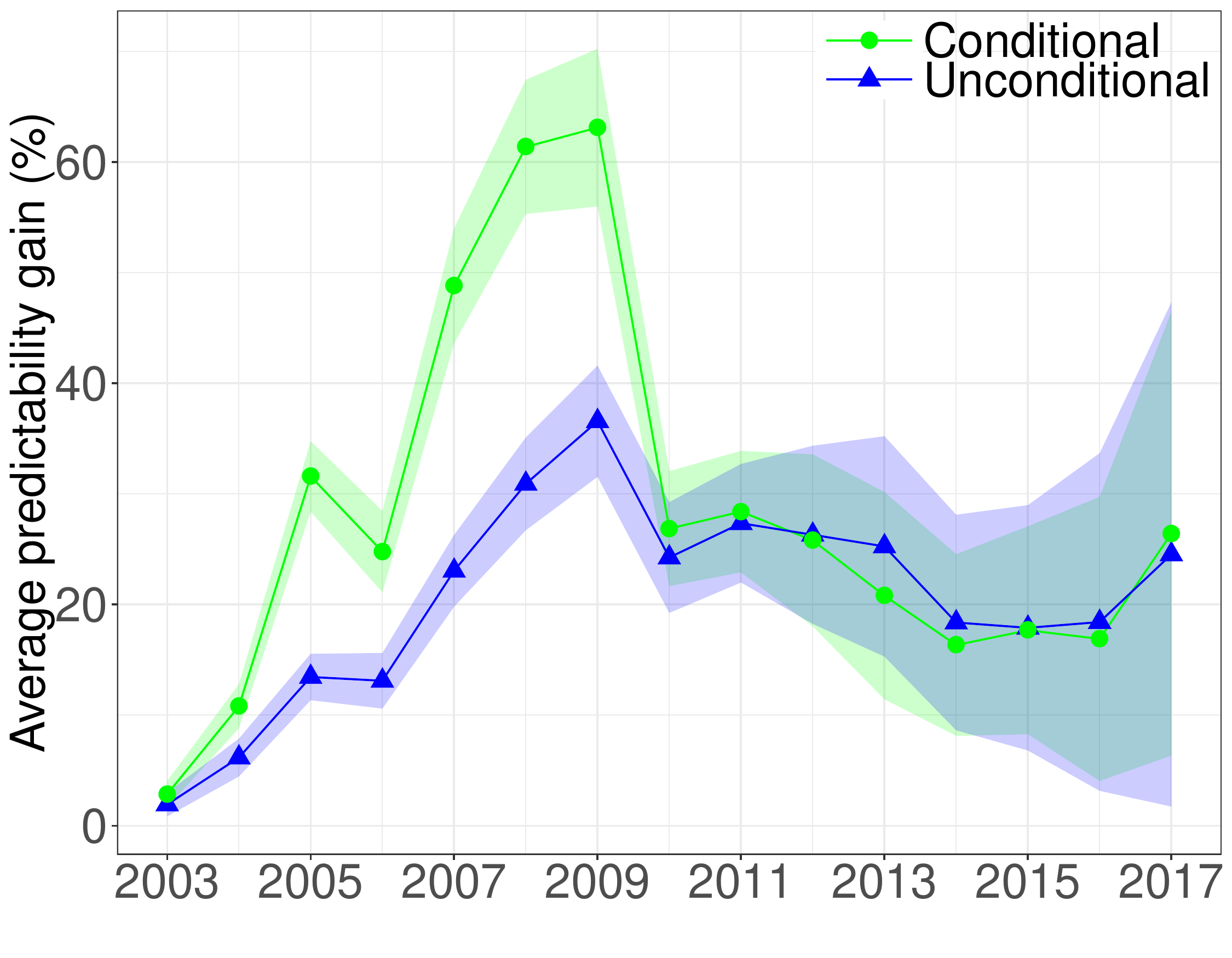}
	\caption{Average predictability gain in \%, $\sum_i PG_{it}/N$, over ARIMA time series models for the conditional (Eq. \ref{eq:predex1}) and unconditional forecasts (Eq. \ref{eq:predex2}). Results are based on forecasts made in 2002 for each year in the future. Note that conditional forecasts use future information on neighborhood growth rates. Shaded areas indicate twice the standard errors.
	} 	
	\label{fig:pg}
\end{figure}

The average predictability gains for each year are shown in Fig. \ref{fig:pg}, demonstrating the substantial predictive power of network model. In the conditional forecasts, the predictability gain is relatively small for one year growth rates ($\approx 3\%$), but reaches a maximum of roughly 63\% in 2009. Note that if a technology's evolution is not influenced by its surrounding technological ecosystem, we would not expect any systematic predictability gain at all.
Obviously, it is harder to beat the ARIMA models in the unconditional forecasting scenario where no information on future innovation dynamics of neighboring technologies is available. Nevertheless, the network model performs significantly better every single year, with an average predictability gain of around 20\%.
In the supplementary information we investigate further aspects of the predictability gain distribution and present results for alternative forecasting benchmarks.

The results of the prediction exercises add further support to the analysis of the previous sections. Innovation rates are network-dependent, and having knowledge on the technology network helps improving forecasts of patenting dynamics. 

\section{Discussion} \label{sec:conclusio}
We have provided evidence that technological domains co-evolve if they are linked through the patent citation network. This empirical observation can be explained with a simple model of network-dependent knowledge creation. Motivated by the recursive nature of the innovation process, technologies take advantage of distinct knowledge sources in the technological ecosystem. We also validated the model by making out-of-sample prediction of growth rates, conditional and unconditional of neighboring growth rates.
We have shown that the network model improves forecasts of patenting growth rates substantially when compared to standard time series models. Thus, the network of technological interdependencies is informative for inferring future patenting dynamics.

It is also important to point out the limitations and caveats of our analysis. 
%We have drawn on the notion of ecosystems when motivating the evolution of technologies. But in many ecosystems, interactions between organisms are not exclusively mutually beneficial \-- they can also be negative (e.g. predator-prey or resource competition dynamics). The same holds true for the technological ecosystem. While many technologies depend positively on each other, there can be technologies which are in direct competition. For instance, advancements in a given technology might extend its functional scope and therefore take over functionalities of other technologies, making them substitutable or redundant. Such negative feedback loops and their impact on technological evolution are much less understood, giving rise to ample research opportunities in the future. \red{I would just remove this. I have already downplayed the ecosystem motivation, precisely because it is not that good.}
By keeping the number of nodes fixed in the technological network, we ignored the emergence of breakthrough technologies. In our approach, technological domains are taken as given and well-defined by the latest technology classification system. But technological classifications themselves are subject to evolutionary forces emerging from the changing technological base \citep{lafond2019long}. New technology classes are created, old ones merge or are abandoned and innovations can be reclassified -- mechanisms which were not made explicit in this work.

In our empirical framework, we allowed the technology network to vary in time, but we did not model this explicitly. An interesting avenue for future research is to gain a better understanding of the mechanisms driving these rewiring processes. Furthermore, it would be important to understand how patenting activity and technology network metrics such as centrality translate into cost reductions of technologies \citep{farmer2016predictable, triulzi2018estimating}. Finally, another interesting direction of research is to further investigate the interaction of network-dependent innovation dynamics with the real economy by coupling innovation network with input-output networks. This could illuminate new aspects of how structural change happens in the economy and improve economic forecasts.

We conclude by discussing the policy implications of our analysis. Since innovation dynamics in technological domains reveal strong network dependence, the allocation of research resources has to consider the ecosystem in which the focal domain is embedded in. Innovation in a domain is not an isolated process, but results from complex mechanisms involving research efforts, institutional settings, and technological interdependencies. Facing the enormous challenge of transitioning the current economic system to a low-carbon economy, we will need substantial improvements in key low-carbon technologies such as photovoltaics and wind energy. Better understanding their technological interdependencies emerging from the technological ecosystem and making it fruitful for research policy will be a crucial step in this direction.

\section*{Acknowledgments}
We are grateful to Christian Diem, Torsten Heinrich, Giorgio Triulzi and the INET Complexity Economics research group for helpful discussions.
This work was supported by Baillie Gifford, 
the Oxford Martin Programme on the Post-Carbon Transition, 
European Union’s Horizon 2020 research and innovation programme under grant agreement No. 730427 (COP21 RIPPLES)
and Partners for a New Economy. 
This research is based upon work supported in part by the Office of the Director of National Intelligence (ODNI), Intelligence Advanced Research Projects Activity (IARPA), via contract no. 2019-1902010003. The views and conclusions contained herein are those of the authors and should not be interpreted as necessarily representing the official policies, either expressed or implied, of ODNI, IARPA, or the U.S. Government. The U.S. Government is authorized to reproduce and distribute reprints for governmental purposes notwithstanding any copyright annotation therein.

\section*{References}
\bibliography{tech_ref}

%\clearpage
\appendix 

\section{Temporal technology network} \label{app:tempnet}
The technology citation matrix $C_t$ which quantifies knowledge spillovers between technologies at each point in time is not directly observable, but has to be derived from the patent citation matrix $H_t$. 
We first show for the static case how simple matrix algebra can be used to construct technology networks from patent data.
Let us define the patent citation network $H$, where $H_{pq}$ is one if patent $p$ cites patent $q$ and zero otherwise. To obtain the number of citations between technologies, the patent citation matrix, $H$, has to be projected onto the technology space.
The patents-technology class relationship can be represented as a binary bipartite network where the link $\tilde{B}_{pi}$ means that patent $p$ belongs to technology class $i$. 
This setup is useful for deriving technology networks. Technological distances can be quantified in different ways, for example, with co-classification networks, co-citation networks or technology citation networks.

%Let $H_{pq,t}$ be one if patent $p$ cites patent $q$ in year $t$ and zero otherwise. The technology citation matrix $C_t$ is different from the patent citation matrix $H_t,$ since multiple technology domains can be assigned to a single patent. We therefore need to project the citation matrix from the patent space onto the technology space.
%The patents-technology class relationship can be represented as a binary bipartite network where the link $\tilde{B}_{pi,t}$ means that patent $p$ at time $t$ is associated with the technology class $i$. 

A simple one-mode projection of the bipartite network onto the set of CPC classes which does not take the citation matrix into account yields the co-classification network and is given by 
\begin{equation}
	C^{\text{co-classification}} = \tilde{B}^\top \tilde{B}.
\end{equation}
The diagonal of $C^{\text{co-classification}}$ is then simply the count of patents in a given domain.
As alternative measure of technological distance, the co-citation network can be derived by computing
\begin{equation}
C^{\text{co-citation}} = (\tilde{B} H)^\top \tilde{B} H.
\end{equation}
The co-citation network quantifies how often two technology codes are cited together.

Since we are interested in the knowledge flows between technologies, the focus of this work is the technology citation network which can be obtained as
\begin{equation}
C^{\text{tech-citation}} = \tilde{B}^\top H \tilde{B}.
\end{equation}
This projection yields a integer-valued citation matrix, but inflates the total number of citations in the technology-based network if multiple technology classes are assigned to a patent. 

\begin{figure}[!ht]
	\centering
	\includegraphics[trim={1cm .75cm 1.5cm .2cm},clip, width= .45\textwidth]{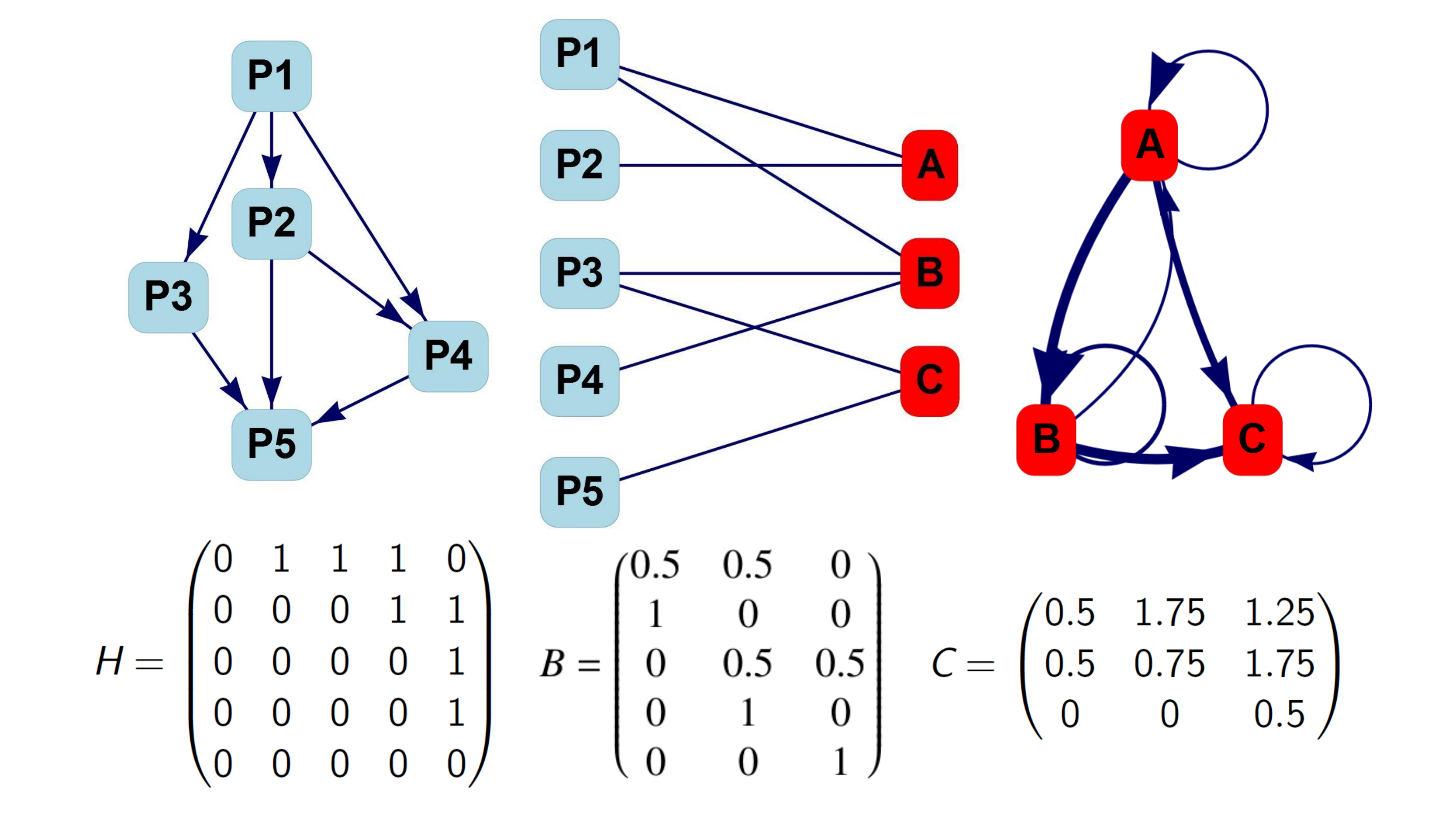} 
	\caption{A toy example of the one-mode projection in Eq. (\ref{eq:projection}). 
		The left network is a patent citation network $H$ where a directed edge from $i$ to $j$ indicates that patent $i$ cites $j$. The network is hierarchical as older patents cannot cite more recent patents. Below the network the corresponding adjacency matrix is shown.
		In the middle is the row-normalized bipartite patent-classification network $B$ with some patents (blue) assigned to multiple technology classes (red). The projection yields the weighted and directed technology citation network $C$ with self-loops shown on the right. 
	} 
	\label{fig:projection}
\end{figure}

To keep total number of citations constant ($\sum_{ij} H_{ij} =  \sum_{ij} C_{ij} $), we use the row normalized bipartite network in the projection. Note that the summing across rows $\sum_i \tilde{B}_{pi}$ yields the total number of CPC classes per patent $p$ (the degree of $p$ in the bipartite network) and the normalized entry $B_{pi} := \tilde{B}_{pi} / \sum_i \tilde{B}_{pi}$ is the ``share'' of technology $i$ in patent $p$. 
Given the row-normalized bipartite network $B$, we obtain the technology citation matrix considered in the analysis by
\begin{equation} \label{eq:projection}
C =  B^\top H B.
\end{equation}
Fig. \ref{fig:projection} illustrates the network projection in a schematic toy example.

To obtain a temporal representation of knowledge flows, we index the patent citation network by time, $H_t$, where the time index refers to the \emph{citing} patent. The time-indexed bipartite network $B_t$ includes all the patents which cite and have been cited at time point $t$, as well as the corresponding technology classes. The time dependent technology citation matrix $C_t$ is obtained by applying Eq. (\ref{eq:projection}) to these matrices. 
The temporal network $W_t$ row-normalizes the technology citation matrix as shown in the main text (Eq. \ref{eq:tech_nw_ext}). The $i^{\text{th}}$ row of $W_t$ can be interpreted as the technological reliance of technology $i$ at time $t$ as it contains all the links from patents in $i$ to all other technology classes at the particular time point. The structural stability of the network is investigated further in the supplementary information.

\section{Significance sampling} \label{app:significant}

Technology citation networks are prone to be biased due to impinging factors such as the number of patents in a class or the age of patents.
Size effects can drive the number of citations between technologies, because the expected number of citations between two technological classes increases with the number of patents in the classes. Another important source of bias is the age of a patent. A new patent might have few forward citations because other inventions simply did not have the time to cite the patent. To eliminate the size and age biases, \cite{alstott2017mapping} suggest an algorithm to sample random network realizations conditional on domain sizes and patent ages. The algorithm yields networks which preserve these impinging factors, but are random otherwise. The random realizations can then be used to derive a z-score for each link in the network. 
Links between technological domains which one would expect by chance, will have small z-scores and can be removed.

The randomized controls are generated as follows. First, identify all citations with a specific time lag which have been made in a particular year. An example would be to take all the citations made by patents in the year 2000 which cite patents three years earlier. Then, second, reshuffle all the \emph{cited} patents. In the example, patents issued in 2000 would point randomly to the set of patents issued three years earlier.
Larger classes will receive more citations on average. Note that the total number of citations for a given time lag are preserved by this procedure.
By repeating the algorithm many times, an expected number of citations between classes can be derived, as well as the standard deviation. The link expectations and standard deviations are then used to compute the z-scores.

Sampling a large number of random networks for every year between 1947 and 2017 is computationally costly. Therefore, we analytically derive z-scores for every potential inter-domain link as suggested by \cite{alstott2017mapping} who have shown that the analytical derivation approximates the numerical randomization process very well.
The randomization procedure can be stated as drawing from a hypergeometric distribution. Recall that a hypergeometric distribution is the probability 
of observing a certain number of successes in $n$ draws, without replacement, from a sample of size $s$ containing $k$ objects with the success feature.

In our context, the hypergeometric distribution describes the probability of observing a certain number of citations from class $i$ to class $j$ at $t$ with lag $l$. Let us denote the number of citations between two classes at $t$ and given lag $l$ with $C_{ij,t(t-l)}$.
The sample size $s$ is then simply the total number of inter-domain citations, $s_{t(t-l)} = \sum_{i,j} C_{ij,t(t-l)} (1-\delta_{ij})$.
The number of draws $n$ corresponds to the number of citations made by a given class $i$. Thus, $n_{i,t-l} = \sum_{j\neq i} C_{ij,t(t-l)}$.
The number of successes $k$ is the number of citations received by $j$ at $t-l$, i.e. $k_{j,t-l} = \sum_{i\neq j} C_{ij,t(t-l)}$. 

The expected number of patents from $i$ to $j$ at $t$ for all possible lags can then be found by
\begin{equation}
\mathop{\mathbb{E} [C_{ij,t} ] } = 
\sum_{l=0}^{t-1836} \frac{k_{j,t-l}}{s_{t-l} } n_{i,t-l},
\end{equation}
and, by assuming independent random variables, the standard deviation is given by
\begin{equation}\label{stdev}
\sigma_{ij,t} =
\left(
\sum_{l=0}^{t-1836} n_{i,t-l} \frac{k_{j,t-l}}{s_{t-l} } 
\left(1 - \frac{k_{j,t-l}}{s_{t-l} } \right)
\frac{s_{t-l} - n_{i,t-l}}{s_{t-l}-1}
\right)^{1/2}.
\end{equation}
For each entry in the citation matrix we calculate z-scores based on 
\begin{equation}
z_{ij,t} = \frac{C_{ij,t} - \mathop{\mathbb{E} [ C_{ij,t} ] }}{\sigma_{ij,t}}.
\end{equation}
We eliminate all citations in the technology citation matrix $C_{t}$ with z-score smaller 2, roughly corresponding to the 97.5\% quantile of the standard normal distribution. As we show in more detail in the supplementary information, the significance sampling yields substantially sparser networks.

\section{Estimation of network model} \label{app:regress}
The econometric network model, Eq. (\ref{eq:econometric_model}), is estimated using maximum likelihood since the OLS estimator is biased and inconsistent due to the reasons outlined in \cite{anselin2013spatial}. The model is related to spatial autoregressive (SAR) models for panel data. But in contrast to most other spatial econometric applications, our model uses a time-varying network term and includes diagonal elements which complicate the estimation. We restate the econometric model in matrix form which will prove useful in the derivation of the estimators.

The model and its corresponding data generating process can be written as
\begin{align}
g_t &= V_{t}a + \beta V_t \tilde{g}_t + \epsilon_t, \quad t \in \{ 1,2,...,T \}, \label{eq:econometric_model_app}\\
g_{t} &= \left[\mathbb{I} - \beta W_t \right]^{-1}  a  + \left[\mathbb{I} - \beta W_t \right]^{-1} \epsilon_t, \label{eq:dgp} \\
\epsilon_t &\sim N(0,  \sigma^2 \mathbb{I}) \nonumber ,
\end{align}
where the matrix $V_{t} = V_t(\beta) \in \mathbb{R}^{N \times N}$ is diagonal with elements $V_{ii,t} := (1- \beta W_{ii,t})^{-1}$. $g_t$, $a$, $\tilde{g}_t$ and $\epsilon_t$ are column vectors of length $N$. The $i^{\text{th}}$ element of $\tilde{g}_t$ is defined as $\tilde{g}_{i,t} := \sum_{j =1}^N W_{ij,t} g_{j,t}(1-\delta_{ij})$.

The likelihood function is given by
\begin{align} \label{eq:likelihood}
& \text{LogL}\left( (\sigma^2, a, \beta) |\{g_{t}\} \right) =  \\
& -\frac{NT}{2} \ln(2 \pi \sigma^2)  + \sum_{t=1}^{T} \ln |\mathbb{I} - \beta V_t {W}_t| - \frac{1}{2 \sigma^2}  \sum_{t=1}^{T} \epsilon_{t}^\top \epsilon_{t}. \nonumber
\end{align}
To find the network specific parameter $\beta$, we first optimize the log-likelihood function with respect to $\sigma^2$ and $a_i$. Following the usual procedure in spatial autoregressive models, we then numerically optimize the concentrated log-likelihood function with respect to $\beta$ \citep{ord1975estimation, elhorst2014spatial, wang2015estimation}.
Maximizing the log-likelihood function with respect to $\sigma^2$ and $a_i$ yields the following expressions:
\begin{align*}
\hat{\sigma}^2 &= \frac{\sum_{t} \epsilon_{t}^\top \epsilon_{t}}{NT}, \\
\hat{a} &=  \left(\sum_t V_t^2 \right)^{-1} \sum_t V_t g_{t} -  \beta \left(\sum_t V_t^2 \right)^{-1} \sum_t V_t^2 \tilde{g}_{t}.
\end{align*}

By plugging the found expressions into Eq. (\ref{eq:likelihood}), we find parameter $\beta$ through the following optimization problem
\begin{align}
\underset{\beta}{\text{argmax}} \quad
& constant  + \sum_{t=1}^T \ln |\mathbb{I} - \beta V_t {W}_t| 
-  \frac{NT}{2} \ln  \sum_{t=1}^T \tilde{\epsilon}_{t}^\top \tilde{\epsilon}_{t} ,
\end{align}
with $\tilde{\epsilon}_{t} := g_t - V_t \left(\sum_t V_t^2 \right)^{-1} \sum_t V_t g_t 
- \beta V_t \left(\sum_t V_t^2 \right)^{-1} \sum_t V_t^2 \tilde{g}_t$.
The (asymptotic) variance-covariance matrix is obtained as the Fisher information matrix, i.e. by inverting the numerically computed Hessian of the negative log-likelihood function.

\section{Details on predictions} \label{app:predict}
For both the ARIMA 
%\red{ why for the ARIMA?} 
and the unconditional network model forecasts we fine-tune the models in-sample to achieve good out-of-sample forecasts of patenting levels. 
As frequently done in statistical learning, we split the data set into three parts: a training set (1948-1987), a validation set (1988-2002) and a test set (2003-2017). 
We first calibrate the models to the training set and use different specifications, as outlined below, to predict patenting in every domain in the validation set. We then choose the models which yield the lowest median absolute percentage error in the validation set forecasts. We use median absolute percentage errors over mean absolute percentage errors, since the well-known downward bias of the mean absolute percentage error \cite[]{armstrong1985long} tends to favor predictors which are systematically underestimating.
After choosing the best model specifications, we re-estimate the models to the whole data set up to 2002. Using these parametrizations of the ``optimal'' models, we then predict patenting activities for every single year and technology from 2003 to 2017.

In case of the ARIMA predictions, a choice has to be made regarding how many MA and AR terms to include in the predictions. Following the proposed principles, we first estimate all possible ARIMA models with lags $\{(p,q) | \; p,q \in \{0,...,5\} \}$ for a given technology in the training set. Second, we use all estimated configurations to predict patenting levels in the validation set. The model specification which reduces the median absolute percentage error in the validation set is then chosen to predict patenting activities in the test set. This procedure is repeated for every single time series. 

For the unconditional network model forecasts we follow a similar approach to obtain the optimal order $k'$. To see why we optimize the network model with respect to $k'$, note that the data generating process in Eq. (\ref{eq:dgp}) for given parameters implies $\mathbb{E}[g_t] = \left(\mathbb{I} - \beta W_t \right)^{-1} a$ and $\mathbb{V}[g_{t}]
= \sigma^2 \left[( \mathbb{I} - \beta W_t^\top ) ( \mathbb{I} - \beta W_t )\right]^{-1}$, entailing an exploding variance for $\beta$ approaching one.
Since our estimations suggest that $\beta$ is actually close to one, forecast errors are expected to be large. To alleviate this issue, we take advantage of the power series expansion, $\left(\mathbb{I} - \beta W_t \right)^{-1} = \sum_{k=0}^\infty \beta^k W_t^k $, by defining the predictor as
$\hat{g}_{t}^{uncond.} = ( \sum_{k=0}^{k'} \hat{ \beta}^k W_t^k ) \hat{a}$. Increasing $k'$ reduces the bias, versus reducing $k'$ lowers the variance. To find the best choice of $k'$, we first estimate the model based on the training set and use the defined predictor to forecast patenting activities in the validation set. We then choose $k'$ such that the median absolute percentage error, across time and technologies, in the validation set is minimized. The selection procedure yields $k' =3$.
As discussed in the main text, we found different magnitudes of network dependence for different growth rate lags. We therefore estimated the network model for every growth rate lag (1-year, 2-years, ..., 15-years) separately.

This approach assumes that the optimal choice in the validation set also corresponds to the best choice in the test set which does not always have to be the case. The dependence between variables can change over time, an aspect which is not taken into account here. Alternative approaches could be considered in future research. For the conditional forecasts, no further model fine-tuning has to be done.

\clearpage
\IfFileExists{SI/SI.pdf}{\includepdf[pages=-]{SI/SI.pdf}}{}

\end{document}

% TTLEPAGE

\begin{center}
\huge{\textbf{Supplementary Information}}
\end{center}
\vspace{0.05cm}

\begin{center}
\LARGE{Technological interdependencies predict innovation dynamics}
\end{center}
\vspace{0.05cm}

\begin{center}
\large Anton Pichler, Fran\c{c}ois Lafond, J. Doyne Farmer
\end{center}

\tableofcontents
\newpage

%-----------------------------------------------------------------------------------------
\section{Data sources and preprocessing}

Our database contains all US utility patents granted between 1836 and 2018, except for rare exceptions. These patents are classified in two different classification systems: Cooperative Classification Scheme (CPC) and International Patent Classification (IPC) (up to 2015, they were also classified in the US Patent Classification System (USPC)).
%; we use this only in Section \ref{section:acemogluetal2016} where we discuss the results of \cite{acemoglu2016innovation}. Since we use their data we do not discuss this further here.) 

CPC classifications are retrieved from the Master Classification File available from the USPTO/Reed Tech (version 2019-04-01). IPC classifications and patent dates are retrieved from Google Patents Public Data, provided by IFI CLAIMS Patent Services, accessed June 2018. 

For the citation data, we merge the citation database of \cite{kogan2017technological} with the citation data collected by \cite{lafond2019long} and with the database provided by the USPTO (patentsview.org/download, accessed: 14/05/2019).

\section{Robustness to alternative classification systems}
We check the robustness of the results by applying our method to two different technology classification schemes, at two different levels in each case. In the main text, we discuss results based on the Cooperative Classification Scheme (CPC) 4-digits codes. Here, we also show results based on the CPC 3-digits, International Patent Classification (IPC) 3-digits and IPC 4-digits codes scheme. In the following, we describe the data in detail and how we have processed it for the presented analysis.

Fig. \ref{fig:patenting_classes}a) shows total patenting rates in the US from 1836 to 2018 for distinct classification codes of patents. While we find that CPC codes for almost every patent back to 1836 (the solid and the dotted line are almost indistinguishable in the plot), reclassification of IPC codes seems to be less rigorous for patents before 1920. For 2018 the data is not yet entirely updated, indicated by the sharp drop in patenting rates at the end of the time series. We therefore exclude the year 2018 in further analysis. Fig. \ref{fig:patenting_classes}b) shows the citations counts per year. The figure illustrates that reliable data on citations made is available only from 1947 on, but we have information on citations received by earlier patents.

\begin{figure*}[!ht]
	\centering
	\includegraphics[width=.65\textwidth]{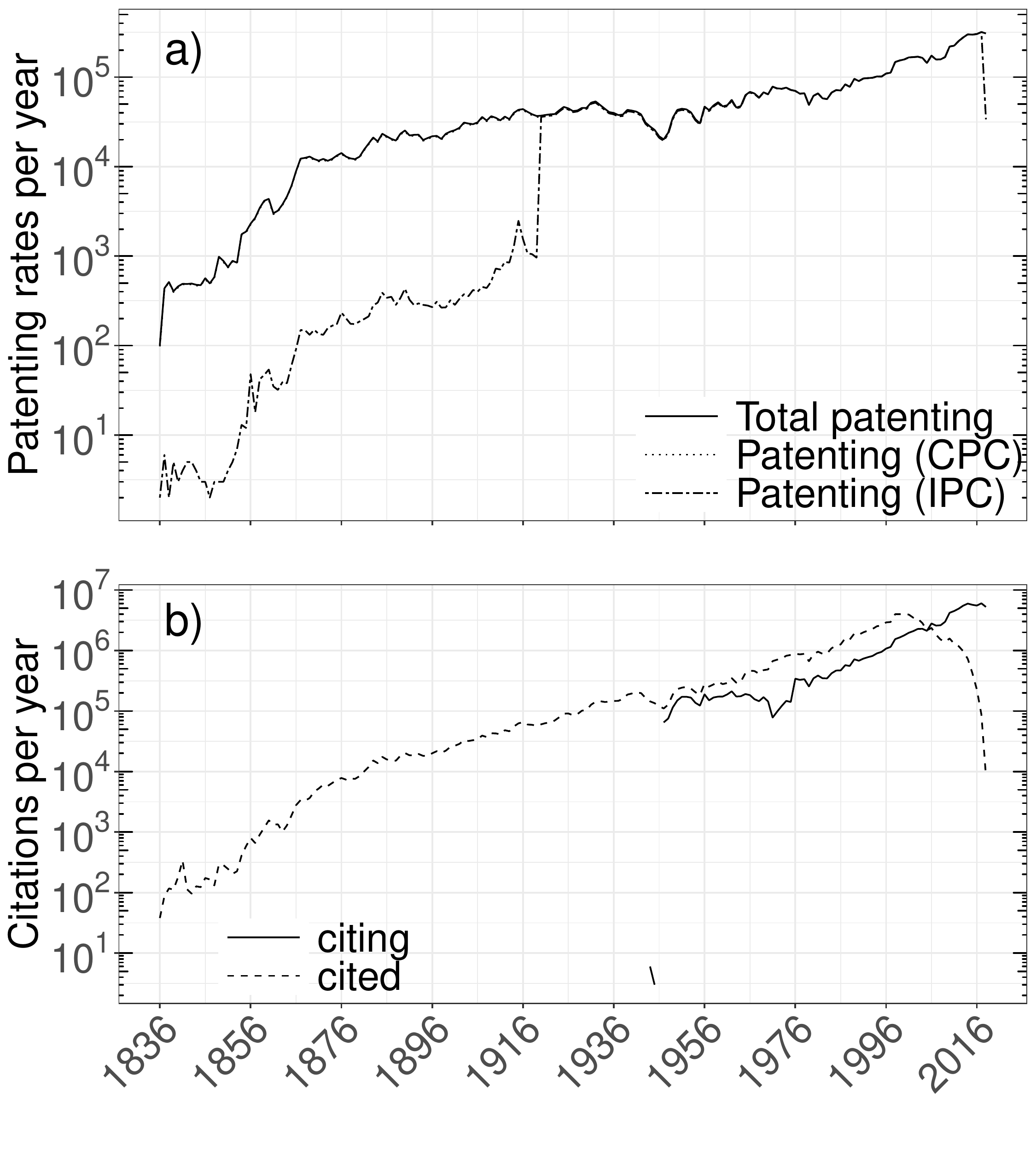} 
	\caption{a) Yearly time series of US patents granted between 1836 and 2019. The solid line is the number of total granted patents, the dotted line is the number of patents with CPC attributes and the dotdashed line the number of patents having IPC attributes. 
		b) Time series of citations per year in the data set. \textit{Citing} refers to the number of citations made in a given year and \textit{cited} refers to the number of citations received.
	}	
	\label{fig:patenting_classes}
\end{figure*}

We map patents onto their CPC and IPC codes and eliminate duplicates if the same patent was classified multiple times with the same 3- or 4-digits code. We only include technological classes which were at least ten times assigned to patents between 1836 and 2000. This removes no class at all in the CPC 3-digits case and only a single class in the CPC 4-digits case (A61P with only a single patent up to 2000). The cleaning is more effective in the IPC regimes where there are several classes which are very infrequently used. Introducing this threshold eliminates 143 IPC 3-digits and 389 IPC 4-digits classes. For the CPC classes, we also eliminate all Y-codes which are tags rather than separate technology classes. Table \ref{tab:classes} summarizes some basic statistics of the used data, after excluding the year 2018.

\begin{table}[!ht]
	\centering
	\begin{tabular}{lrrr}
		Classification & \# classes & \# observations & \# patents \\ 
		\hline
		CPC 3-digits & 125  & 12,912,549 & 9,774,895\\ 
		CPC 4-digits & 649  & 14,422,099 & 9,774,895\\ 
		IPC 3-digits & 121  & 12,000,519 & 8,449,988\\ 
		IPC 4-digits & 633  & 13,697,093 & 8,449,720\\ 
		\hline
	\end{tabular}
	\caption{ Descriptive statistics of used data after initial cleaning.}
	\label{tab:classes}
\end{table}

We obtain a temporal network for each of the four classification schemes. To avoid links between technological domains which one would expect by chance, we apply the significance sampling procedure \citep{alstott2017mapping} described in the main text. 
In Fig. \ref{fig:density_classes} we show the evolution of network density, i.e. the number of existing links divided by the number of all possible links ($N^2$), It becomes evident that the density is reduced substantially after applying the significance sampling procedure, but is also increasing in time. Unsurprisingly, the network density is higher when using the more aggregate 3-digits classification. 

\begin{figure*}[!ht]
	\centering
	\includegraphics[width=\textwidth]{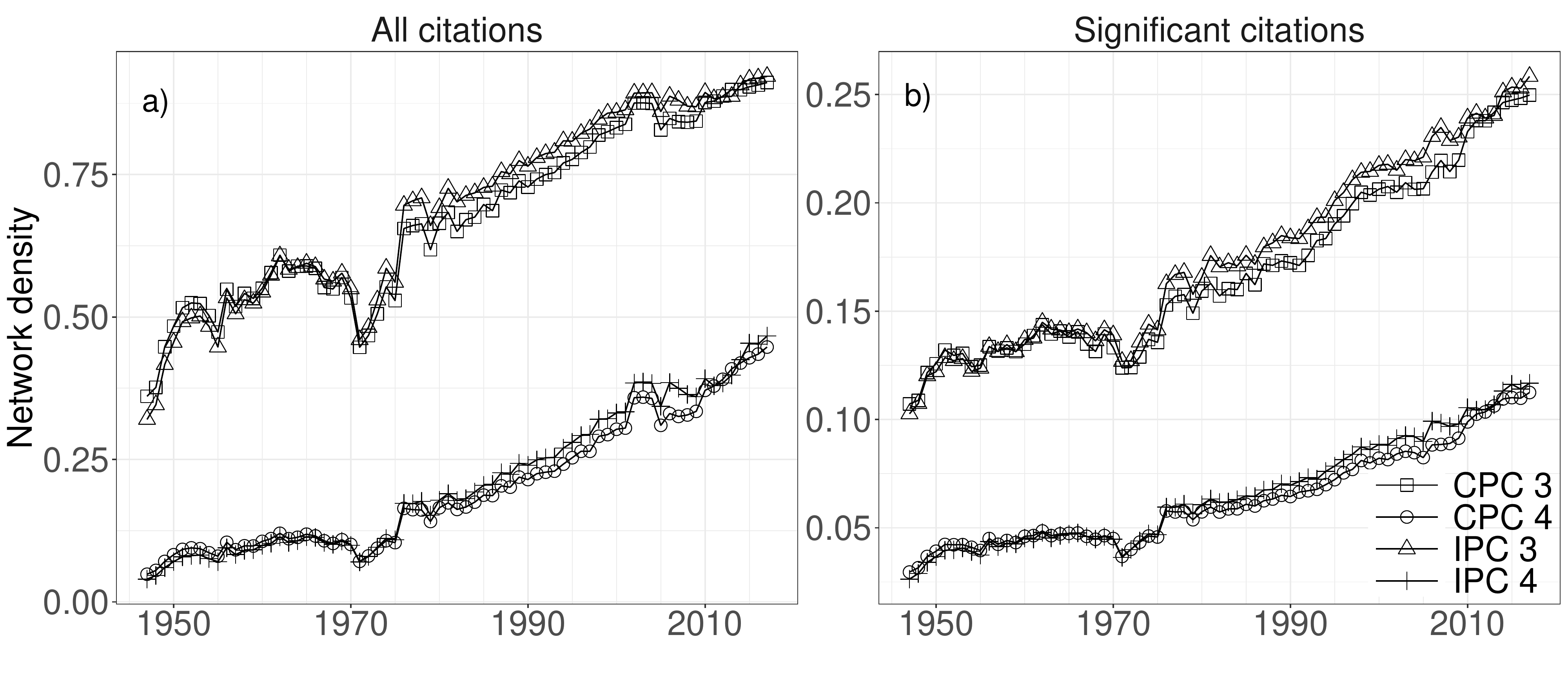} 
	\caption{a) The network densities, number of existing links divided by the number of all possible links, over time for different classification schemes when all citations are included. b) The network densities after removing non-significant links. }	
	\label{fig:density_classes}
\end{figure*}

\section{Temporal network stability}
As a measure of temporal stability, Fig. \ref{fig:nw_stability} shows the average cosine similarity between a row $i$ of the technology matrix at $t$ and $t-l$, or in more mathematical terms
\begin{equation*}
\text{similarity}_{t,t-l}^\text{av.} =
\frac{1}{N} 
\sum_{i=1}^N
\frac{\sum_j W_{ij,t} W_{ij,t-l}}{\sqrt{\sum_j W_{ij,t}^2} \sqrt{ \sum_j W_{ij,t-l}^2} }.
\end{equation*}
We observe that the network tends to be fairly stable in time, i.e. the future technological dependence of a given technology is likely to be similar to its current one. After the temporal stability of the network was increasing for most parts of the second half of the last century, it became weaker in the last 20 years. As one would expect, we find that the average similarity decreases for higher lags, although there is still a pronounced positive relationship between technological dependences over a 10-year horizon.
A qualitative difference between 3-digits and 4-digits classification schemes is apparent. The networks based on the more aggregated 3-digits classifications exhibit higher correlations in time for all shown lag choices. Moreover, the differences in temporal correlations between different time lags become less pronounced. This suggests that it is important to use more aggregated technology classification if the research is interested on the temporal evolution and structural change of the technological ecosystem.
\begin{figure*}[!ht]
	\centering
	\includegraphics[width=.44\textwidth]{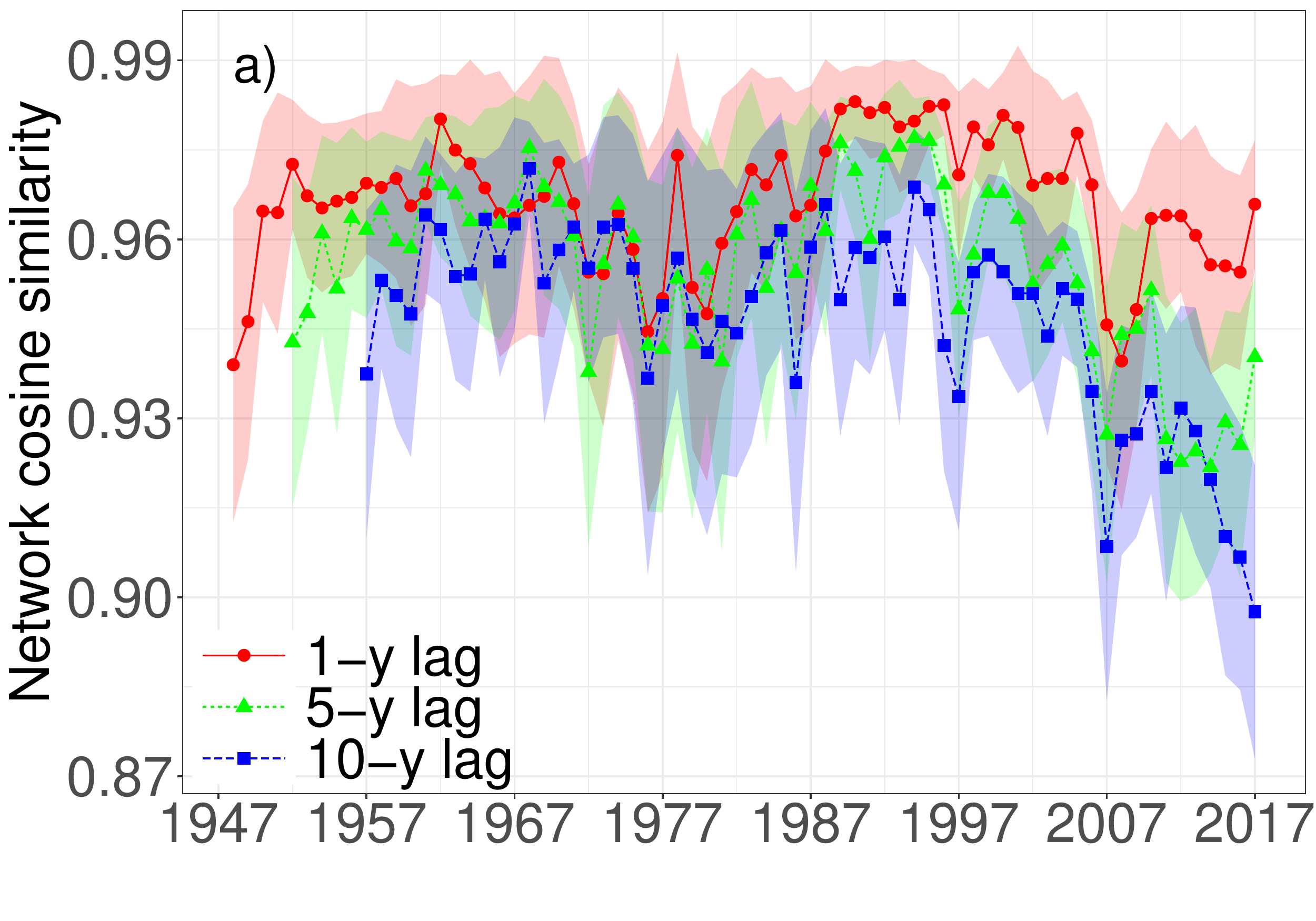}
	\includegraphics[width=.44\textwidth]{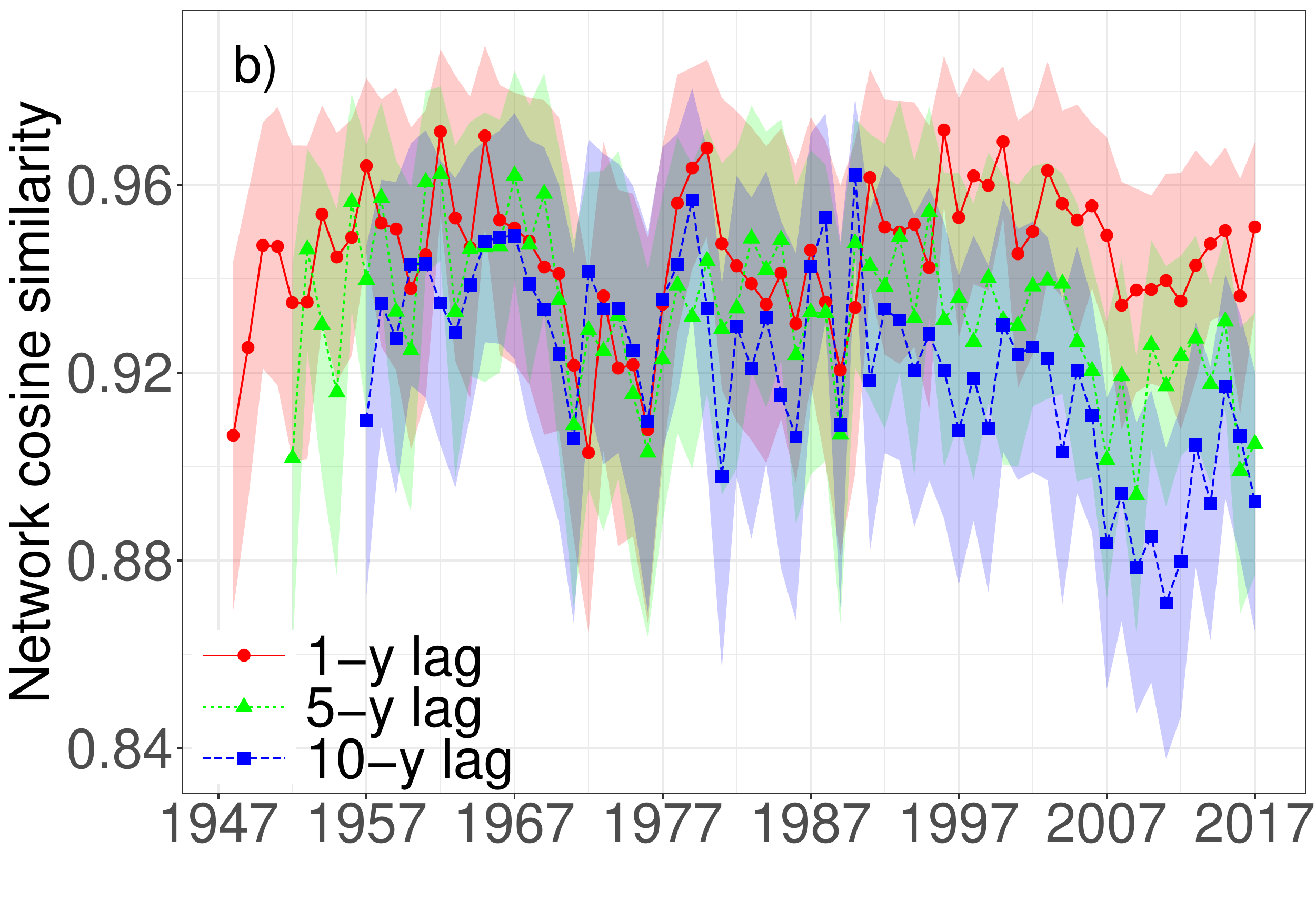} \\
	\includegraphics[width=.44\textwidth]{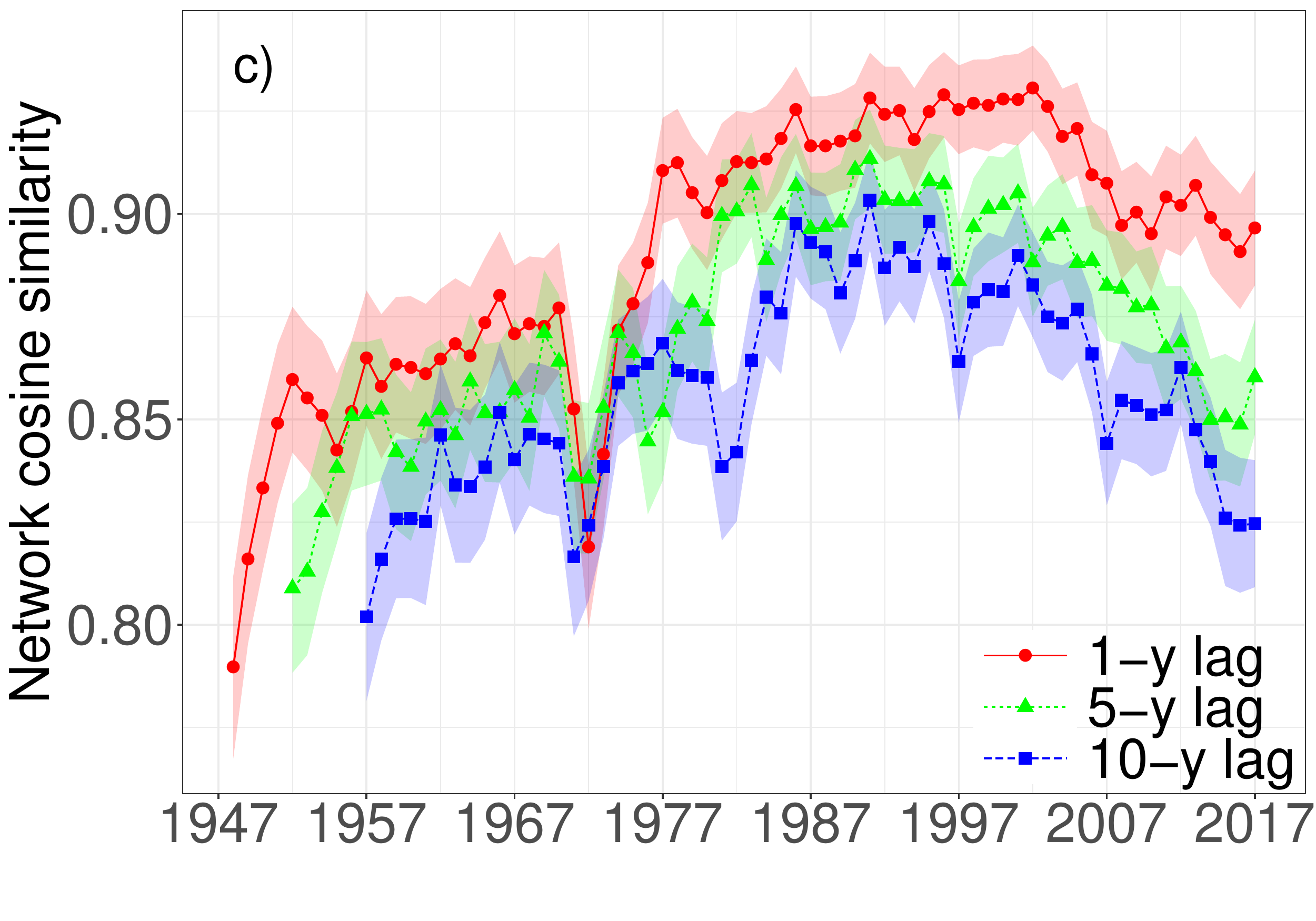}
	\includegraphics[width=.44\textwidth]{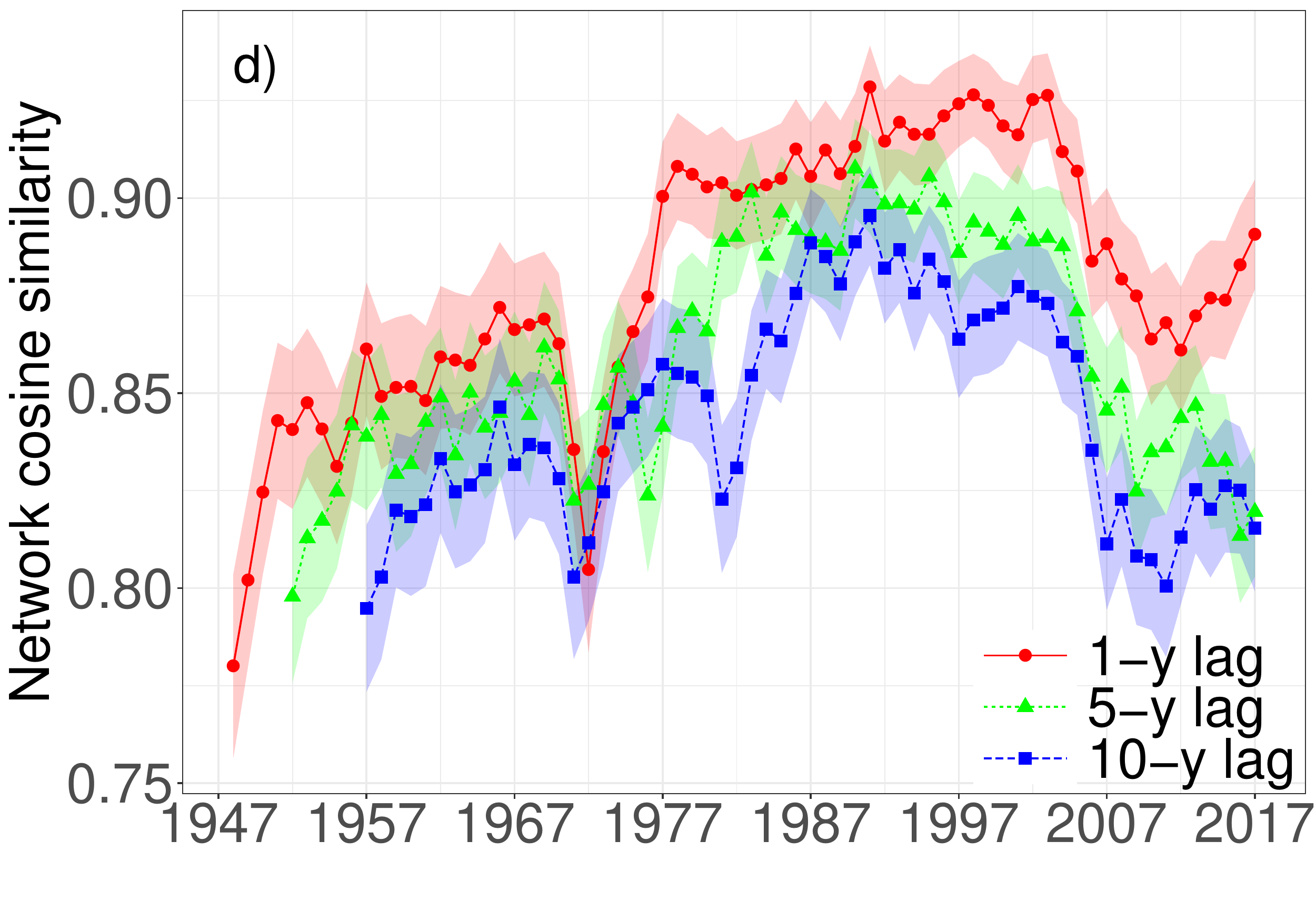}
	\caption{Average temporal cosine similarity of technology networks over time. The shaded area shows the range of the average plus/minus twice the standard errors. The results are shown for networks based on the a) 3-digits CPC codes, b) 3-digits IPC codes and c) 4-digits CPC codes and d) 4-digits IPC codes. }	
	\label{fig:nw_stability}
\end{figure*}

\section{Nearest neighbor growth correlations}
In the main text we have shown that the technology network is highly assortative with respect to growth rates, suggesting that patenting growth rates are similar if the technologies are connected. Fig. \ref{fig:assortativity} plots the average nearest neighbor growth rates correlation over time for the differently aggregated networks. The positive assortativity pattern is more noisy in the more aggregated networks, panel a) and b), but still pronounced for most of the cases. The average nearest neighbor growth rates correlation is relatively insensitive with respect to different growth rate lengths choices for the more aggregated networks, again pointing at the importance of using more granular networks for discerning temporal effects in the technological evolution.

Table \ref{tab:avnng} summarizes the nearest neighbor growth correlations for different classifications and and lags. We see that the correlation magnitude is similar for different classification schemes, but systematically less noisy when using the more granular 4-digits classification. Data aggregation often reduces noise by averaging out extrema. Thus, it is not obvious a priori that the more fine-grained network representations exhibit less variation in their average nearest neighbor growth rates correlations.

\begin{figure*}[!ht]
	\centering
	\includegraphics[width=.44\textwidth]{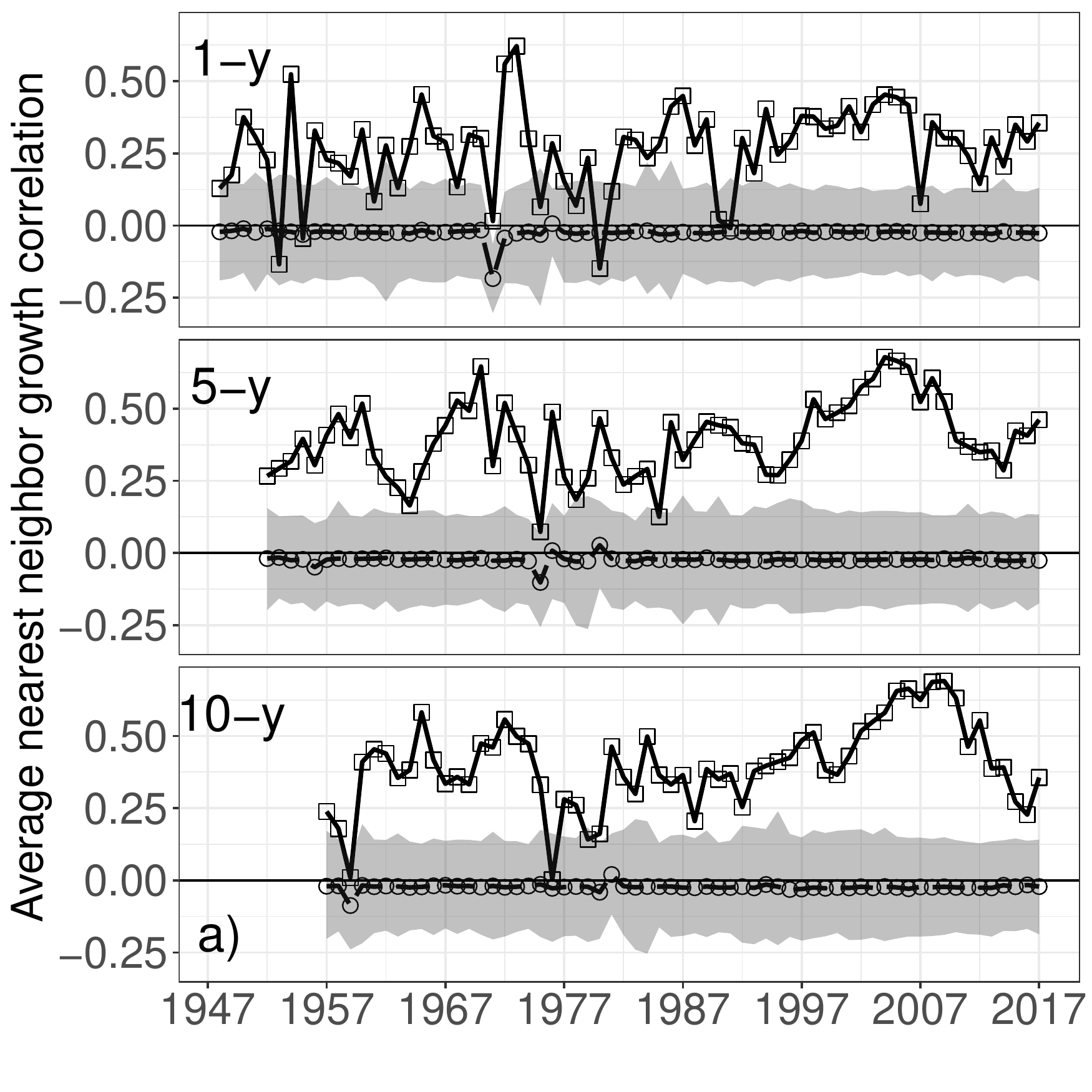}
	\includegraphics[width=.44\textwidth]{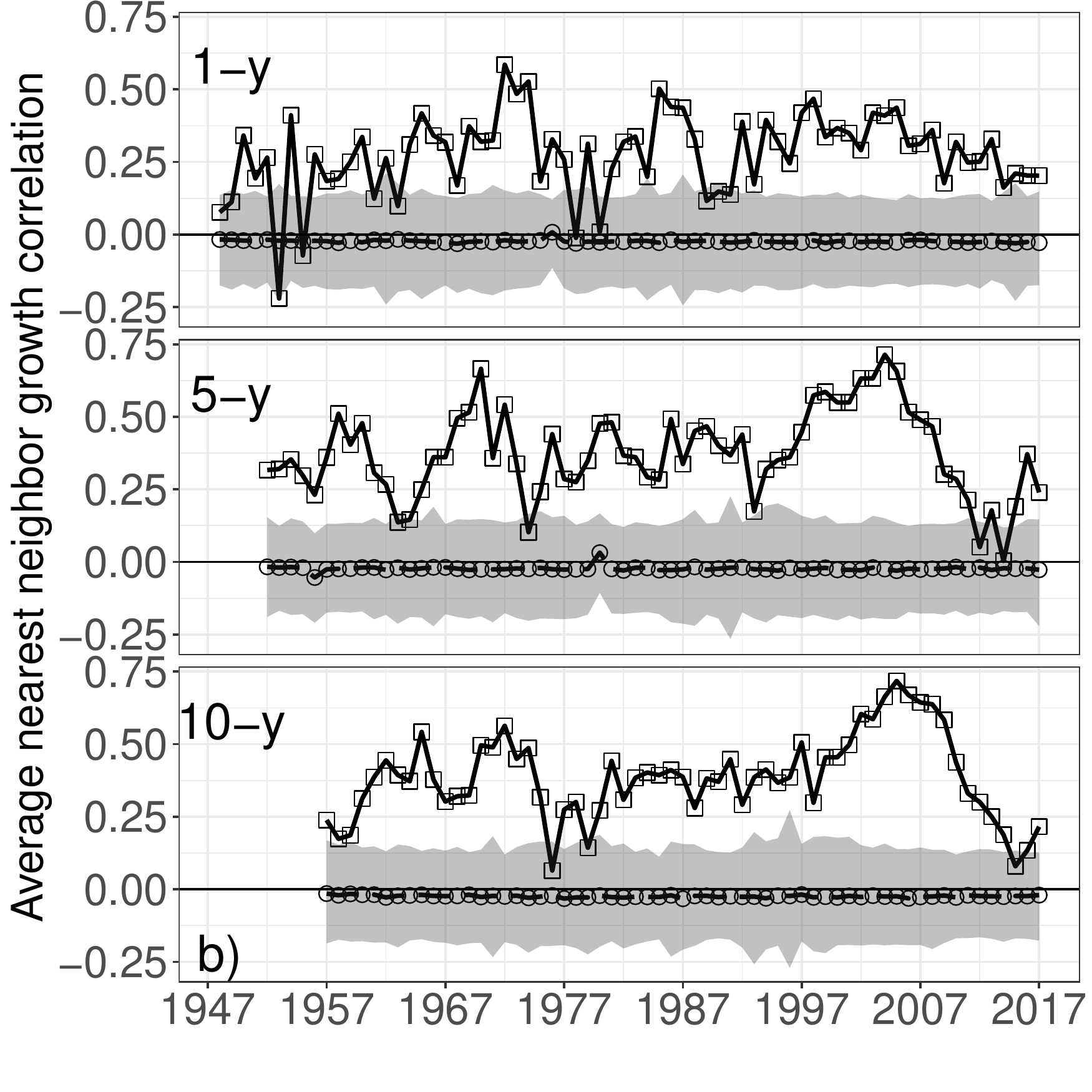} \\
	\includegraphics[width=.44\textwidth]{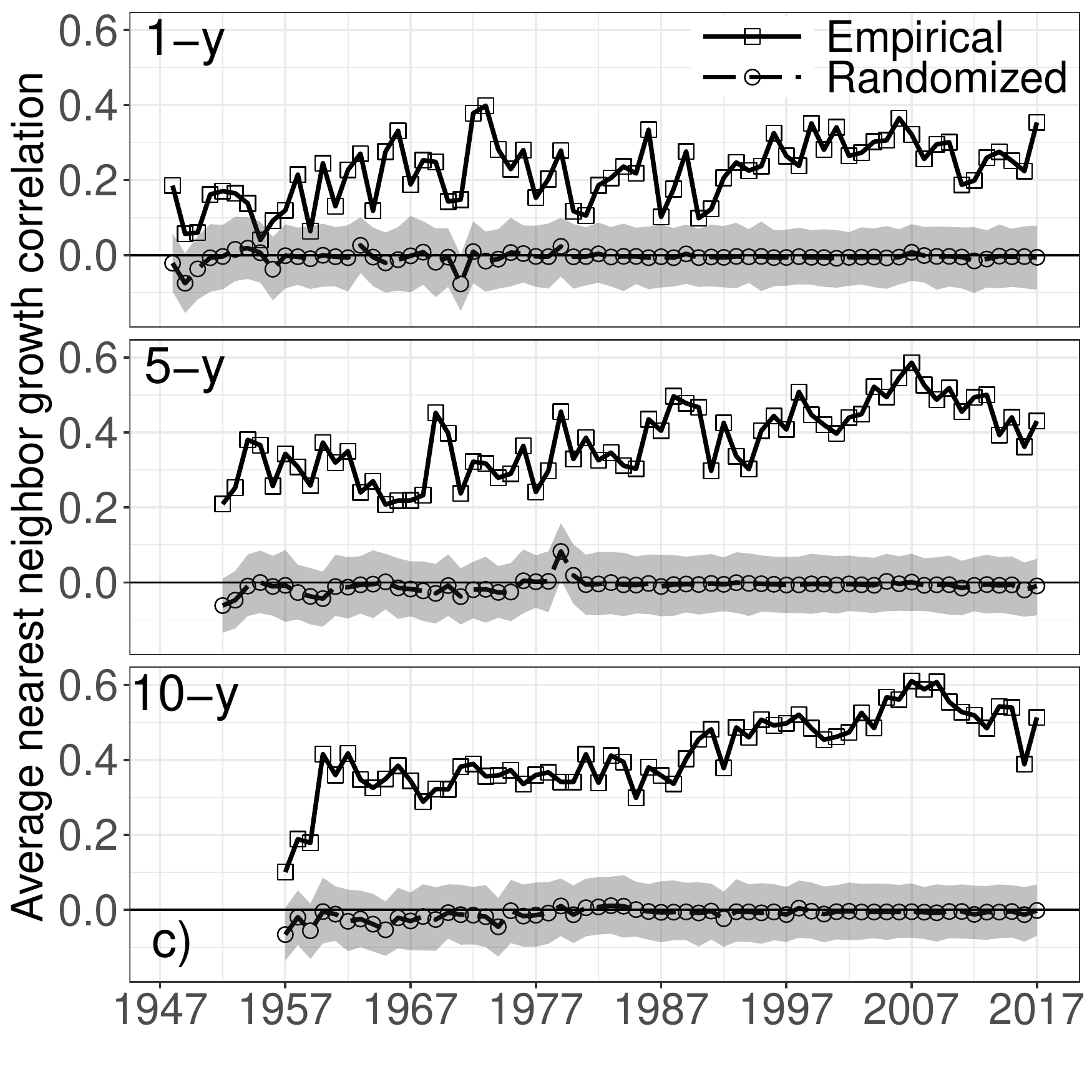}
	\includegraphics[width=.44\textwidth]{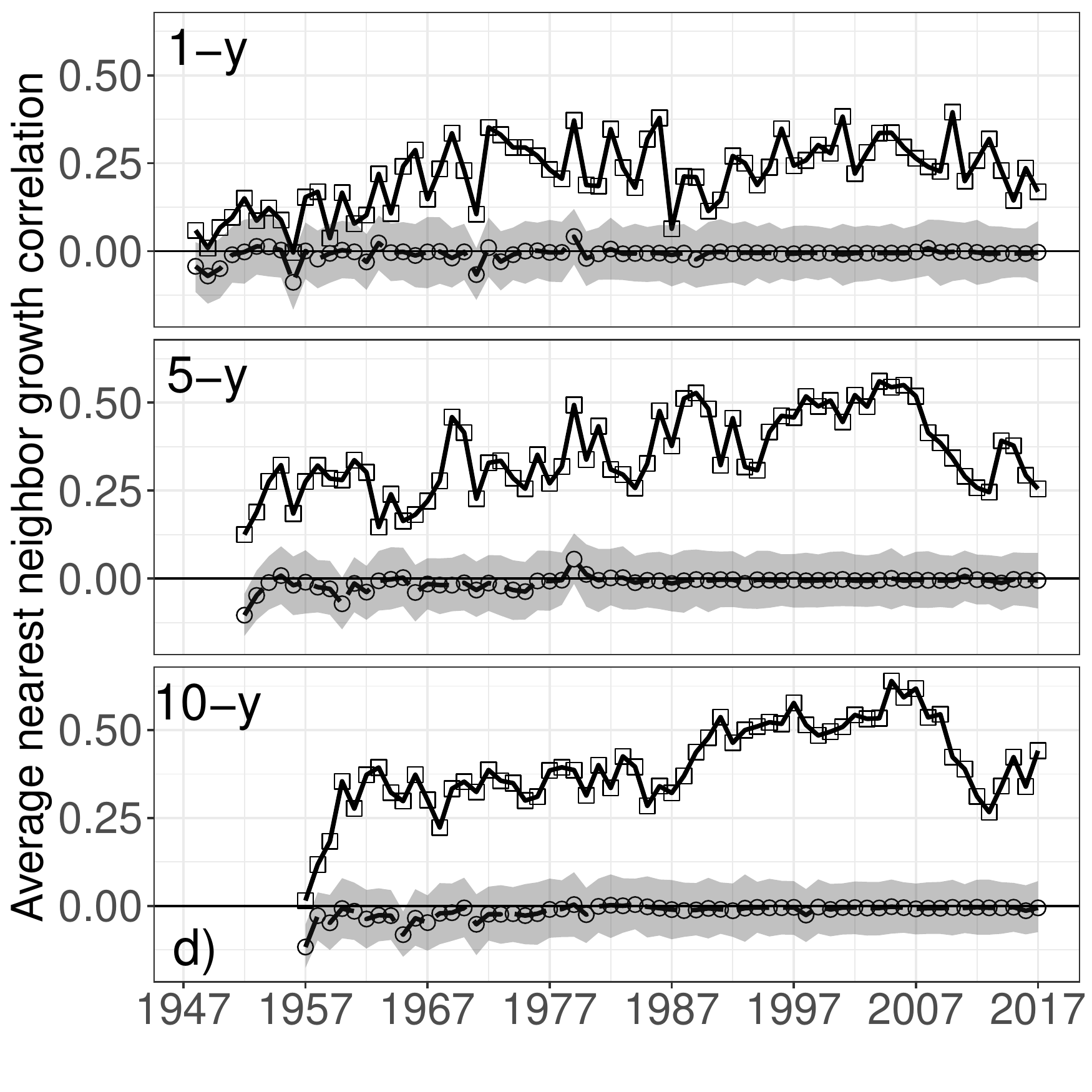}
	\caption{Average nearest neighbor growth rates correlation across time for networks based on the a) 3-digits CPC codes, b) 3-digits IPC codes, c) 4-digits CPC codes and d) 4-digits IPC codes. The shaded area is the 5-95\% inter-quantile range of results based on 1,000 randomized networks (described in the main text).  }	
	\label{fig:assortativity}
\end{figure*}

\begin{table}[!ht]
	\centering
	\begin{tabular}{|l|l|cccc|}
		\hline
		& Lag & CPC 3 & CPC 4 & IPC 3 & IPC 4 \\ 
		\hline
		\multirow{3}{*}{Time average}
		& 1-year & 0.27 & 0.22 & 0.28 & 0.22 \\ 
		& 5-year & 0.39 & 0.38 & 0.38 & 0.35 \\ 
		& 10-year & 0.40 & 0.42 & 0.39 & 0.40 \\ 
		\hline
		\multirow{3}{*}{Time std. deviation}
		& 1-year & 0.15 & 0.08 & 0.14 & 0.10 \\ 
		& 5-year & 0.13 & 0.10 & 0.15 & 0.11 \\ 
		& 10-year & 0.15 & 0.10 & 0.15 & 0.12 \\ 
		\hline
	\end{tabular}
	\caption{Average nearest neighbor growth rate correlations. Time average and time standard deviation of the average nearest neighbor growth rates correlation for distinct technological classification schemes.}
\label{tab:avnng}
\end{table}

\section{Econometric estimation}

\subsection{Econometric network model and SAR}
In the main text we have presented results based on the econometric model
\begin{equation}\label{eq:econometric_model}
g_{i,t} = \frac{ a_{i} }{1- \beta W_{ii,t}} + \frac{\beta }{1- \beta W_{ii,t}} \sum_{j =1}^N W_{ij,t} g_{j,t}(1-\delta_{ij}) + \epsilon_{i,t},
\end{equation}
where $\epsilon_{i,t} \sim N(0,\sigma^2)$.
Note that the econometric model is a generalization of the spatial autoregressive (SAR) model for panel data. In case of a zero diagonal matrix, $W_{ii} = 0$, the model reduces to the fixed-effects SAR with time-varying spatial component
\begin{equation}\label{eq:sar}
g_{i,t} = a_{i}  + \beta \sum_{j =1}^N W_{ij,t} g_{j,t} + \epsilon_{i,t},
\end{equation}
discussed by \cite{wang2015estimation}. Moreover, if the network is fixed in time, $W_t = W$, this model reduces to the standard fixed-effects SAR which can be estimated as outlined in \cite{elhorst2014spatial}.

\subsection{Time-varying network model}
We estimate different variations of these econometric models to see how robust the results in the main text are. First, we calibrate the model as discussed in the main text to all four classification schemes and report the results in Table \ref{t:nw_regress_all}. The results are qualitatively similar for different aggregations.

The econometric estimation also allows use to make the magnitude of network effects in patenting growth rates explicit. Let us define the average direct and average total impacts of research effort growth on patenting growth by
\begin{align}
\bar{I}_{direct,t} &:= \alpha \frac{\sum_{i=1}^N  L_{ii,t} }{N}, \\
\bar{I}_{total,t} &:= \alpha \frac{\sum_{i=1}^N \sum_{j=1}^N  L_{ij,t} }{N},
\end{align}
respectively, yielding the relative network impact of 
\begin{equation}
I_{network,t} := 1- \frac{\bar{I}_{direct}}{\bar{I}_{total}} = 1- \frac{\sum_{i=1}^N L_{ii,t}}{\sum_{i=1}^N  \sum_{j=1}^N L_{ij,t}}.
\end{equation}

Fig. \ref{fig:relnetimpact} shows the relative network impacts $I_{network,t}$ over time for all four classifications. The network effects are growing in time which is not surprising given the increase in network density we have observed. The figure shows that we would estimate higher network impacts in a more aggregated network, i.e. when based on 3-digits classifications. We also find that network effects are slightly larger for IPC codes than for CPC codes, if compared on the same digit-level. But the general trend of rising relative network impacts over time holds for all four classification schemes.

\subsection{Static network model}
We next investigate how the estimated parameters change if we use fixed network specifications. Patent networks are naturally dynamic, but the vast majority of the spatial econometrics literature does not allow for time-varying networks. We thus test whether our results also hold if we keep the technology network fixed.

We do this by estimating the model for every network snapshot separately, i.e. by setting $W_t =W$ in Eq. (\ref{eq:econometric_model}) where $W$ represents a particular network in the sample. We then check how the parameter $\beta$ changes if we use a different network snapshot instead. 
It should be pointed out that there might be an endogeneity issue if more recent networks are included in the model, since the network formation itself could depend on the growth of technological domains. Moreover, this estimation uses future information on the right-hand side of Eq. (\ref{eq:econometric_model}) to explain past left-hand side variables, for example, when the network of 2017 is used as $W$. Nevertheless, we estimate the model with a fixed, potentially future, network to see how results change from the time-varying to the static specification.

In Fig. \ref{fig:fixdenetwork} we plot the coefficient $\beta$ as a function of the used network $W_t$. We show the results for CPC 3-digits and CPC 4-digits classes only, since results are similar for the IPC counterparts. The figure depicts qualitatively different results for the two aggregation levels. For the CPC 3-digits case, we find that $\beta$ exhibits a slight downward trend as a function of the time index. In the CPC 4-digits case, on the other hand, the coefficient is smaller in earlier and larger for more recent networks. Additionally, from the mid-70s on the parameter seems to stabilize around the value of 0.84 which we have estimated in the original time-varying specification.

\begin{table}[b]
	\centering
	\resizebox{\textwidth}{!}{		
		\begin{tabular}{lcccccccccccc}
			\multicolumn{4}{l}{\textit{Dependent variable: patenting growth}} \\ 
			\\[-1.8ex]\hline 
			\hline \\[-1.8ex] 
			& \multicolumn{3}{l}{\textit{CPC3}} & \multicolumn{3}{l}{\textit{CPC4}} & \multicolumn{3}{l}{\textit{IPC3}} & \multicolumn{3}{l}{\textit{IPC4}}\\ 
			\cline{2-3} \cline{5-6} \cline{8-9} \cline{11-12}
			\\[-1.8ex] & (1-y) & (5-y) & (10-y) & (1-y) & (5-y) & (10-y) & (1-y) & (5-y) & (10-y) & (1-y) & (5-y) & (10-y)\\ 
			\hline \\[-1.8ex]
			average $\hat{a}_i$ & 0.01 & 0.03 &0.05  & 0.01 & 0.03 & 0.04 & 0.01 & 0.02 &0.05 &0.01 & 0.03 &0.05\\[1ex] 
			$\hat{\beta}$ & 0.92$^{***}$ & 0.94$^{***}$ &0.94$^{***}$ & 0.85$^{***}$ & 0.93$^{***}$ & 0.94$^{***}$ & 0.92$^{***}$ &0.94$^{***}$ &0.92$^{***}$ & 0.87$^{***}$ & 0.93$^{***}$ &0.93$^{***}$\\ 
			& (0.006) & (0.010) &(0.015) & (0.006) & (0.006) & (0.008) & (0.006) & (0.011) &(0.019) & (0.006) & (0.007) & (0.009) \\[1ex] 
			$\hat{\sigma}^2$ & 0.03 & 0.08 &0.13 & 0.11 & 0.20 & 0.30 & 0.03 & 0.09 &0.14 & 0.11 & 0.23 &0.31\\ 
			%& 0 & 0 &0 & 0.0007 &  0.0030 & 0.0062 & 0 & 0 &0 & 0 & 0 &0\\[1ex] 
			\hline \\[-1.8ex]
			Log-likelihood & 2,525 & -277 &-364 & -13,387 &-5,474  &-3,625  &2,867 &-384 &-395 &-13,431 &-6,049 &-3,677\\ 
			Observations & 8,610 & 1,720 &858 & 43,651 & 8,734 & 4,348  &8,349 &1,669 &834 &42,688 &8,528 &4,427\\
			\hline 
			\hline \\[-1.8ex] 
			\textit{Note:}  & \multicolumn{12}{r}{$^{***}$p$<10^{-16}$}
		\end{tabular}
	\caption{Results from estimating the econometric network model presented in the main text (Eq. \ref{eq:econometric_model}) for different classification schemes. The results are shown for 1-, 5- and 10-year growth rates. }
		\label{t:nw_regress_all}
	}	
\end{table}

\begin{figure*}[!ht]
	\centering
	\includegraphics[width=.75\textwidth]{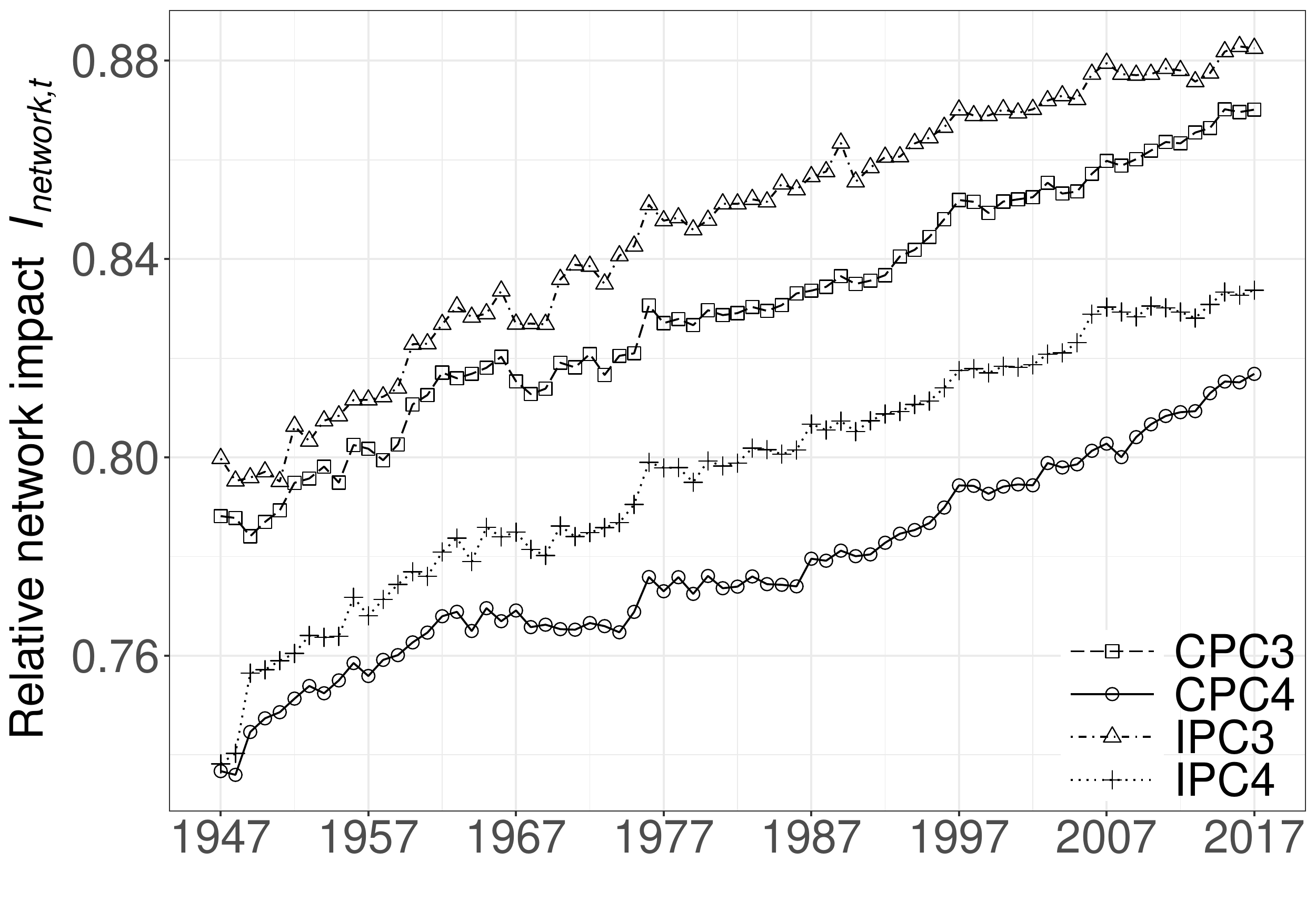}
	\caption{Relative network impacts over time for different classification schemes. Results are based on 1-year growth rates.}	
	\label{fig:relnetimpact}
\end{figure*}

\begin{figure*}[!ht]
	\centering
	\includegraphics[width=.75\textwidth]{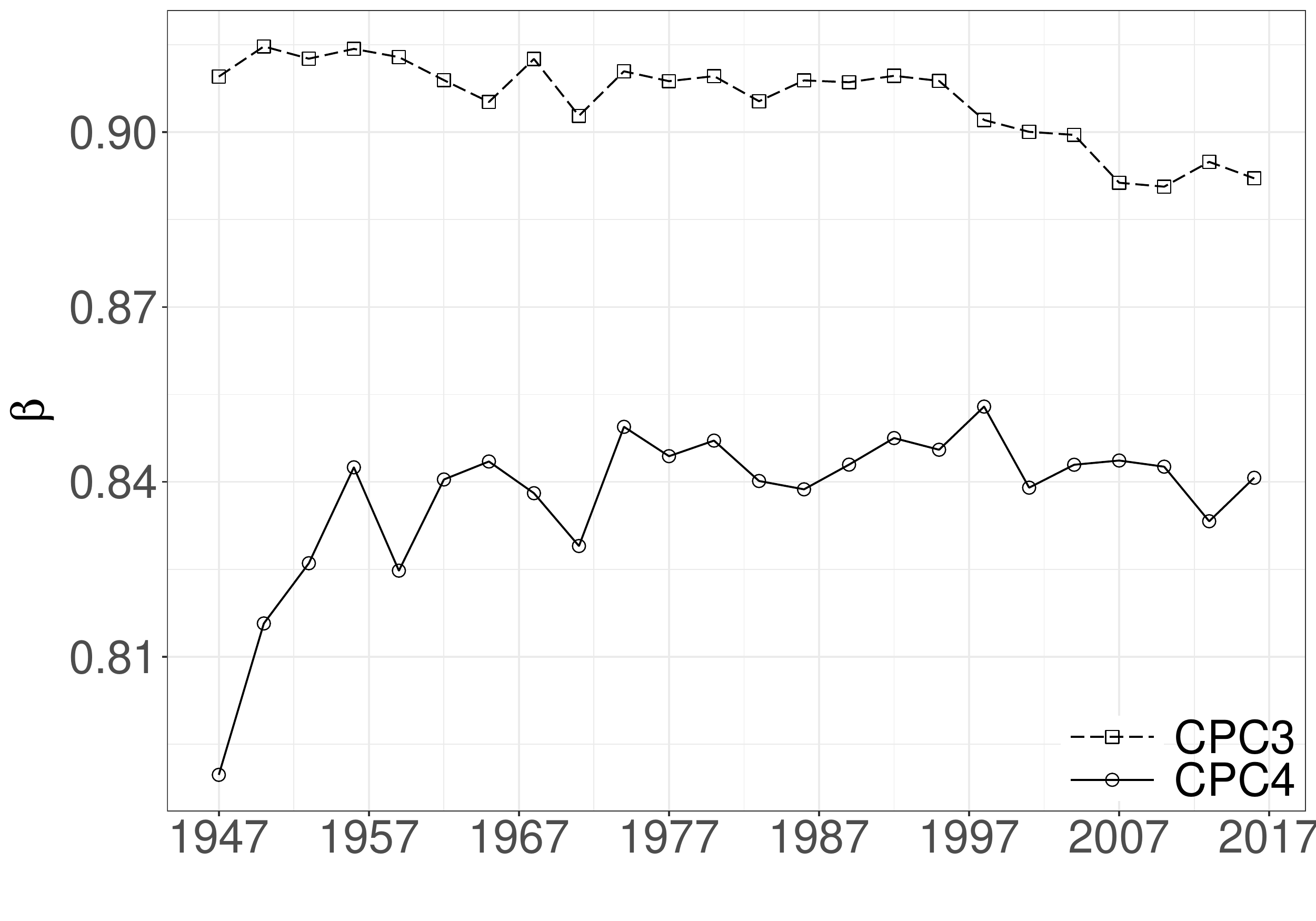}
	\caption{Parameter $\beta$ obtained from estimating Eq. (\ref{eq:econometric_model}), but with a fixed instead of a time-varying network. The x-axis indicates the year of which the network snapshot was obtained from. For example, at 1947 the model was estimated with using $W_t = W_{1947}$ for all $t$. Results are based on 1-year growth rates.} 	
	\label{fig:fixdenetwork}
\end{figure*}

\begin{table}[!b]
	\centering
	\resizebox{\textwidth}{!}{		
		\begin{tabular}{lcccccccccc}
			\multicolumn{11}{r}{\textit{Dependent variable: patenting growth}} \\ 
			\\[-1.8ex]\hline 
			\hline \\[-1.8ex] 
			&(a)  & (b) & (c) & (d) & (e) & (f) & (g) & (h) & (i) & (j)\\[1ex]
			&CPC3  & CPC3 & CPC3 & CPC3 & CPC3 & CPC4 & CPC4 & CPC4 & CPC4 & CPC4\\
			\hline \\[-1.8ex] 
			average $\hat{a}_i$ & 0.011 & 0.023 &0.428  & 0.400 & 0.173 & 0.015 & 0.024 & 0.560 &0.357 &0.127 \\[1ex] 
			%average time effects & & 0.021 &0.428  & 0.400 & 0.171 & & 0.024 & 0.567 &0.357 &0.125	 \\[1ex] 
			$\hat{\beta}$ & 0.830$^{***}$ & 0.509$^{***}$ &0.513$^{***}$ & 0.514$^{***}$ & 0.511$^{***}$ & 0.756$^{***}$ & 0.483$^{***}$ &0.476$^{***}$ &0.483$^{***}$ & 0.482$^{***}$ \\ 
			&(0.009) & (0.018) &(0.017) & (0.018) & (0.018) & (0.008) & (0.105) & (0.09) &(0.105) & (0.105) \\[1ex] 
			Patent rates &  &  &$-0.069^{***}$ &  &  & &  &$-0.128^{***}$ & &  \\ 
			& &  &(0.003) &  &  &  & &(0.002) & &  \\[1ex]
			Patent stock &  &  & & -0.038$^{***}$ &  & & & &$-0.041^{***}$ &  \\ 
			& & & &(0.004) & & &  & &$(0.003)$ &  \\[1ex]
			Discount. pat. stock &  &  & & & $-0.033^{***}$ &  & & & &$-0.035^{***}$ \\ 
			&  &  & & & (0.004) & &  &  & & $(0.003)$ \\[1ex]
			\hline \\[-1.8ex]
			Fixed effects & Yes & Yes & Yes & Yes & Yes  & Yes  & Yes &Yes & Yes & Yes\\ 
			Time effects & No & Yes & Yes & Yes & Yes  &No  & Yes &Yes & Yes & Yes\\
			Log-likelihood & 2,527 & 2,721 & 2,912 &2,763 &2,757  &-13,374  &-12,923 &-11,452 &-12,854 &-12,864\\ 
			Observations & 8,601 & 8,601 &8,601 & 8,601 &8,601 & 43,651  &43,651 &43,651 &43,651 &43,651\\
			\hline 
			\hline \\[-1.8ex] 
			\textit{Note:}  & \multicolumn{10}{r}{$^{***}$p$<10^{-16}$}
		\end{tabular}
		\caption{Results from estimating the SAR with time-varying network component and additional exogenous regressors (Eq. \ref{eq:sar_vector}) for different classification schemes. The first five columns (a-e) show the estimated parameters for the CPC 3-digits aggregation and the last five columns (f-j) for the CPC 4-digits aggregation. The results are based on 1-year growth rates. 
		}
		\label{t:sar_regress_all}
	}	
\end{table}

\subsection{Time-varying spatial autoregressive model}
We also estimated a standard SAR model (zero-diagonal) with time-varying network component (Eq. \ref{eq:sar}) for comparison. The SAR model is convenient for demonstration purposes, since extensive literature renders it a ``standard'' model for studying network effects. Due to its simpler form, it is also straightforward to include further regressors to check for robustness. We therefore estimate Eq. (\ref{eq:sar}) also with additional regressors such as time effects and technological class sizes. %measured as yearly patenting rate or as cumulative patent stock (also including a discount factor). 

The estimation of the SAR model simplifies in comparison to the full network model of Eq. (\ref{eq:econometric_model}). To estimate Eq. (\ref{eq:sar}) including exogenous regressors, we stack all elements $g_{i,t}$ into the $NT$-dimensional vector $\bar{g} := (g_{1,1}, g_{2,1}, ..., g_{N,1}, g_{1,2}, ..., g_{N,T})$. In the same manner we obtain the vector $\bar{\epsilon}$. By defining the $T$-dimensional vector of ones $\iota_T$ and the block-diagonal matrix $W := diag(W_1, ..., W_T)$, we can rewrite Eq. (\ref{eq:sar}) into matrix notation
\begin{equation} \label{eq:sar_vector}
	\bar{g} = (\iota_T \otimes \mathbb{I}_N)a  + \beta W \bar{g} + \bar{X} \delta + \bar{ \epsilon },
\end{equation}
where $\bar{X}$ is the $NT \times K$ matrix of exogenous regressors and $\delta$ the corresponding $K$-dimensional vector of coefficients.
After subtracting the time average, Eq. (\ref{eq:sar_vector}) simplifies to 
\begin{equation} \label{eq:sar_vector_demean}
g = \beta W g + X \delta + \epsilon,
\end{equation}
where $g$, $X$ and $\epsilon$ are the time-demeaned versions of $\bar{g}$, $\bar{X}$ and $\bar{ \epsilon }$, respectively. 
The log-likelihood of the time-demeaned model then reads
\begin{equation} \label{eq:loglik}
\text{LogL} = -\frac{NT}{2} \ln(2 \pi \sigma^2)  + \sum_{t=1}^{T} \ln |\mathbb{I} - \beta {W}_t| - \frac{1}{2 \sigma^2}  \epsilon^\top \epsilon,
\end{equation}
with $\epsilon = (\mathbb{I}_{NT} - \beta   {W}) g - X \delta$. The parameter $\beta$ is obtained by solving
\begin{align*}
\underset{\beta}{\text{argmax}} \quad
& constant  + \sum_{k=1}^{NT} \ln (1 - \beta e_k) \\
&-  \frac{NT}{2} \ln  ( g^\top g - 2 \beta g^\top W g  - \beta^2 g^\top W^\top W g  - A^\top X  (X^\top X)^{-1} X^\top Ag,
\end{align*}
where $A := (\mathbb{I}_{NT} - \beta W )$ and $e_k$ represents eigenvalues of matrix $W$. Under standard regularity assumptions, we derive the variance-covariance matrix by inverting the Fisher information matrix which yields
\begin{equation}
var(\sigma^2, \beta, \delta) = 
\begin{bmatrix}
\frac{NT}{\sigma^4} & -\mathbb{E} \left[ \frac{\partial^2 LogL}{ \partial \sigma^2 \partial \beta } \right] &0_K^\top \\
 \cdot & -\mathbb{E} \left[ \frac{\partial^2 LogL}{ \partial \beta^2 } \right] & \frac{1}{\sigma^2} X^\top W A^{-1} X \delta \\
 \cdot& \cdot & \frac{1}{\sigma^2} X^\top X \\
\end{bmatrix}^{-1},
\end{equation} 
where $-\mathbb{E} \left[ \frac{\partial^2 LogL}{ \partial \delta^2 } \right] = trace(A^{-1}W A^{-1}W) + \frac{1}{2\sigma^2} trace \left( [A^{-1} X \delta \delta^\top X^\top A^{\top -1} + \sigma^2 A^{-1}  A^{\top -1}] W^\top W \right)$ and 
$-\mathbb{E} \left[ \frac{\partial^2 LogL}{ \partial \sigma^2 \partial \beta } \right] = 
\frac{1}{\sigma^2} trace \left( A^\top W [ A^{-1} X \delta \delta^\top X^\top A^{\top -1} + \sigma^2 A^{-1}  A^{\top -1}]  + \delta^\top X^\top W A^{-1}X\delta \right) $. $0_K$ denotes a $K$-dimensional vector of zeros. Matrix elements indicated by the dots are not reported since they are trivially filled due to symmetry.

In Table \ref{t:sar_regress_all} we show the results of estimating the SAR with time-varying network component for the CPC 3-digits, columns (a) to (e), and CPC 4-digits aggregation, columns (f) to (j). We find that the network coefficient $\beta$ is estimated to be slightly larger, if the more aggregated networks are used. We also include time effects in the regression which reduces the network parameter in the CPC 4-digits case from 0.76 to roughly 0.48. When estimating the model for higher growth rate lags (not shown here), including time effects has less impact on the coefficient.
The coefficient is fairly robust against adding further regressors such as including logged patenting rates, columns (c) and (h), logged cumulative patent stock, (d) and (i), or logged discounted cumulative patent stock, (e) and (j), where we have used a 15\% yearly discount factor as suggested by \cite{hall2005market}. The exogenous regressors are highly significant and have a negative sign, indicating that higher patenting rates and larger knowledge stocks at time point $t$ come with smaller patenting rate growth in the following year. The average magnitude of the fixed effects increases as a consequence of including exogenous regressors.

\section{Predicting innovation dynamics}

In the main text we have shown that predictions of patenting rates in technological classes can be significantly improved if network effects are taken into account.
In this section we present further details on predicting patenting rates with the introduced model. Results shown in the main text are based on the CPC 4-digits codes. Here, we show how the model predictions change when using alternative aggregation levels. Moreover, we discuss further aspects of the presented results and investigate alternative forecast benchmarks.

\subsection{ARIMA forecasts}
We first focus on the forecasts obtained from the ARIMA models which we state again for the ease of reference:
\begin{equation} \label{eq:arima}
g_{i,t} = \nu_i + \sum_{s=0}^{p} \phi_{i,s} g_{i,t-s} + \sum_{s=0}^{q} \psi_{i,s} u_{i,t-s} + u_{i,t},
\end{equation}
where $\phi_{i,0} = \psi_{i,0} = 0$. 
As indicated in the main text, we have fitted every $(p,q)$ combination up to order (5,5) of the ARIMA($p$,1,$q$) model for each (logged) time series in the training set (1947-1987) and used the parameter fits to forecast the time series in the validation set (1988-2002).  We then refitted the ARIMA($p$,1,$q$) which yielded the smallest median absolute percentage error in the validation set to the data up to 2002 to forecast the test set from 2003 to 2017. Table \ref{t:arima} gives an overview of the best ($p$,$q$) combinations for different aggregation methods.
In the CPC 4-digit case discussed in the main text, almost 25\% of model specifications use less than two AR and MA terms, with the the geometric random walk ($p=q=0$) alone yielding the best forecasts already in 12\% percent of all cases. But we also find time series where higher-order models are preferred. Similarly for alternative aggregations, the geometric random walk model is the most frequent choice.

\begin{table}[!b]
	\begin{minipage}{.5\linewidth}
		\centering
				\begin{tabular}{|c|r|rrrrrr|}
			\hline 
			\multicolumn{2}{|c|}{CPC3}& \multicolumn{6}{c|}{MA order $q$} \\
			\cline{1-8}
			\multirow{8}*{\rotatebox{90}{AR order $p$}}		
			& & 0 & 1 & 2 & 3 & 4 & 5 \\ 
			\cline{3-8}
			&0 &0.15 &0.01 &0.03 &0.03 &0.02 &0.02 \\
			&1 &0.04 &0.00 &0.01 &0.00 &0.02 &0.02 \\
			&2 &0.03 &0.02 &0.02 &0.03 &0.02 &0.04 \\
			&3 &0.03 &0.02 &0.03 &0.00 &0.02 &0.02 \\
			&4 &0.02 &0.02 &0.01 &0.02 &0.03 &0.07 \\
			&5 &0.02 &0.02 &0.01 &0.03 &0.05 &0.07 \\
			\hline  
		\end{tabular}
		\begin{tabular}{|c|r|rrrrrr|}
			\hline 
			\multicolumn{2}{|c|}{IPC3}& \multicolumn{6}{c|}{MA order $q$} \\
			\cline{1-8}
			\multirow{8}*{\rotatebox{90}{AR order $p$}}		
			& & 0 & 1 & 2 & 3 & 4 & 5 \\ 
			\cline{3-8}
			&0 &0.10 &0.03 &0.04 &0.02 &0.03 &0.02 \\
			&1 &0.01 &0.03 &0.02 &0.00 &0.03 &0.03 \\
			&2 &0.08 &0.00 &0.02 &0.01 &0.01 &0.04 \\
			&3 &0.03 &0.00 &0.02 &0.04 &0.02 &0.03 \\
			&4 &0.04 &0.03 &0.01 &0.03 &0.03 &0.03 \\
			&5 &0.01 &0.05 &0.02 &0.02 &0.08 &0.05 \\
			\hline  
		\end{tabular}
	\end{minipage}%
	\begin{minipage}{.5\linewidth}
		\centering
					\begin{tabular}{|c|r|rrrrrr|}
			\hline 
			\multicolumn{2}{|c|}{CPC4}& \multicolumn{6}{c|}{MA order $q$} \\
			\cline{1-8}
			\multirow{8}*{\rotatebox{90}{AR order $p$}}		
			& & 0 & 1 & 2 & 3 & 4 & 5 \\ 
			\cline{3-8}
			&0 & 0.12 & 0.04 & 0.03 & 0.02 & 0.03 & 0.03 \\
			&1 & 0.06 & 0.02 & 0.01 & 0.02 & 0.02 & 0.04 \\ 
			&2 & 0.02 & 0.02 & 0.02 & 0.03 & 0.03 & 0.03 \\ 
			&3 & 0.03 & 0.01 & 0.02 & 0.01 & 0.03 & 0.03 \\ 
			&4 & 0.03 & 0.02 & 0.03 & 0.01 & 0.03 & 0.03 \\ 
			&5 & 0.03 & 0.03 & 0.01 & 0.03 & 0.03 & 0.02 \\ 
			\hline  
		\end{tabular}	
		\begin{tabular}{|c|r|rrrrrr|}
			\hline 
			\multicolumn{2}{|c|}{IPC4}& \multicolumn{6}{c|}{MA order $q$} \\
			\cline{1-8}
			\multirow{8}*{\rotatebox{90}{AR order $p$}}		
			& & 0 & 1 & 2 & 3 & 4 & 5 \\ 
			\cline{3-8}
			&0 &0.13 &0.05 &0.02 &0.02 &0.02 &0.04 \\
			&1 &0.05 &0.02 &0.01 &0.02 &0.02 &0.04 \\
			&2 &0.02 &0.02 &0.02 &0.02 &0.03 &0.03 \\
			&3 &0.02 &0.00 &0.01 &0.03 &0.02 &0.03 \\
			&4 &0.03 &0.01 &0.01 &0.02 &0.03 &0.03 \\
			&5 &0.04 &0.04 &0.03 &0.03 &0.04 &0.03 \\
			\hline  
		\end{tabular}
	\end{minipage} 
\caption{Distribution of $(p,q)$ pairs which minimize the median absolute percentage error in the validation set for the ARIMA(p,1,q) forecasts. $p$ is the number of autoregressive terms and $q$ the number of moving average terms in the ARIMA model. For example, in the CPC3 case we have 0.15 for the combination $(p,q)=(0,0)$, meaning that for 15\% of all cases the geometric random walk with drift yielded the best predictions in the validation set. }
\label{t:arima}
\end{table}

\subsection{Network model forecasts}
To obtain the unconditional forecasts, a similar model selection procedure is applied. Here, we first calibrate the network model to the data in the training set and choose, as discussed in the main text, the parameter $k'$ such that the median absolute percentage error is minimized in the validation set. The model is then refitted to the data, including training and validation set (1947-2002), to predict patenting in the test set from 2003 to 2017 with given $k'$. When doing this for the four different technology aggregations, we obtain the following results: $k'=9$ for CPC 3-digits, $k'=3$ for CPC 4-digits, $k'=6$ for IPC 3-digits and $k'=2$ for IPC 4-digits. Thus, the applied model selection procedure always prefers $k'>0$, i.e. the procedure always suggests to include network effects in the predictions.

\subsection{Performance evaluation}
As in the main text, we compare forecasts based on the predictability gain measure 
\begin{align}
PG1_{i,t} &=  \frac{ | {P}_{i,t} - \hat{P}_{i,t}^{ARIMA} | - | {P}_{i,t} - \hat{P}_{i,t}^{network} | }{ |{P}_{i,t} | }, \label{eq:pg}
\end{align} 
i.e. taking the difference between ARIMA and network model absolute percentage errors (APE) for every predicted value.
We also check the predictability gains based on the absolute errors (AE) instead of the absolute percentage errors by computing
\begin{align}
PG2_{i,t} &=  | {P}_{i,t} - \hat{P}_{i,t}^{ARIMA} | - | {P}_{i,t} - \hat{P}_{i,t}^{network} |.
\label{eq:pg_ae}
\end{align} 
We then take averages, standard deviations, medians and interquartile ranges for every year and plot the results
for all four aggregation schemes in Fig. \ref{fig:pg_cpc3} - \ref{fig:pg_ipc4}. 
In every figure, panel a) shows the average predictability gains based on APE, Eq. (\ref{eq:pg}). Panel b) gives the corresponding yearly median values and interquartile ranges of the APE-based predictability gains.
Averages and standard errors of AE predictability gains, Eq. (\ref{eq:pg_ae}), are shown in panel c) and the corresponding medians and interquartile ranges are visualized in panel d).
We discuss the results for each aggregation scheme separately.\\[0.1cm]
 
\noindent
\textbf{CPC 3-digit codes (Fig. \ref{fig:pg_cpc3}):} \\
Similarly as the results shown in the main text, predictions can be substantially improved in the conditional forecasts where the growth rates of a focal technology's neighborhood is known. This holds true for averages and medians, regardless of the considered performance metric. 
In contrast to the result in the main text, we find that the unconditional forecasts are not able to beat the ARIMA model systematically. 
The main reason why in this scenario the unconditional predictions underperform is that the optimal tuning parameter $k'$ is not stable between the validation and test set. While a large $k'=9$ is chosen based on the validation set forecasts, this choice is suboptimal when predicting patenting rates in the test set. This is largely a result of temporal instability of coefficients. Estimating the model from earlier years yields higher fixed effects and a slightly lower $\beta$ than when estimating it from more recent data, leading to an overprediction of patenting levels in the test set. A potential way to tackle this problem in future research is to allow for time-varying coefficients. \\[0.1cm]

\noindent
\textbf{CPC 4-digit codes (Fig. \ref{fig:pg_cpc4}):} \\
In all four panels predictability gains of conditional forecasts are substantial. 
There is more variation in the unconditional forecasts.
%, predictability gains are always positive on average for t, panel a) and c). When taking the median, we find positive median predictability gains in the first eleven years, ranging from 1.3\% to 30\% and negative predictability gains in the last four years ($-5$ to $-13$\%) when using $PG1$.
Panel a) is the same plot as presented in the main text which plots average $PG1$ (solid lines) and twice the standard errors (shaded areas) over time. In both prediction exercises, the network model comes with substantial positive predictability gains.
In panel b) the median predictability gains, as well as the interquartile range are shown. In the unconditional forecasts, we see positive median predictability gains in the first eleven years, ranging from 1.3\% to 30\% and negative predictability gains in the last four years ($-5$\% to $-13$\%) when using $PG1$. The conditional forecasts always exhibit positive median predictability gains.
Panels c) shows the average predictability gains based on absolute errors. On average, the network model outperforms the ARIMA forecasts in both prediction exercises, with the conditional forecasts yielding positive prediction gains in every year and the unconditional forecasts in 11 out of 15 years. Panel d) plots the median predictability gains based on the corresponding interquartile range for $PG2$ which yields qualitatively similar results as panel b). \\[0.1cm]

\noindent
\textbf{IPC 3-digit codes (Fig. \ref{fig:pg_ipc3}):} \\
The average predictability gains over the ARIMA models in the conditional forecast setting reach almost 60\% in 2009 when measured in percentages, panel a), or roughly 800 patents when looking at absolute errors, panel b). In the unconditional scenario, average predictability gains based on $PG1$ are less pronounced (although positive in every single year), but are substantial when considering the absolute error based $PG2$ measure. 
The median predictability gain based on the absolute percentage error in the conditional forecasts scenario lies between 1.6\% and 55\% as can be seen in panel c). In the unconditional scenario, the median gains are smaller, but always positive.  Panel d) reveals similar patterns as the ones observed in panel c). 
\\[0.1cm]

\noindent
\textbf{IPC 4-digit codes (Fig. \ref{fig:pg_ipc4}):}\\
For the IPC 4-digit codes we find average percentage predictability gains between 4\% in 2003 and 74\% in 2013 when considering the conditional network forecasts; see panel a). The predictability gains for the unconditional forecasts can get as high as 29\% in the year 2016. Panel b) shows that the median predictability gains are always positive for the conditional forecasts (ranging from 5\% to 68\%) and positive for the unconditional forecasts in 13 out of 15 forecasts. Only in the last two years, the median predictability gains are slightly negative (less than one percent).
When benchmarking the forecasts with absolute errors, Eq. (\ref{eq:pg_ae}), we find that predictability gains are always positive on average and tend to increase in time for both prediction scenarios; panel c). Qualitatively, panel d) resembles panel b) where median predictability gains are always positive, except for the last two years in the unconditional forecasts.

\begin{figure}
	\caption{Prediction results based on \textbf{CPC 3-digit codes}.
		a) Average predictability gain based on $PG1$, Eq. (\ref{eq:pg}. Shaded areas indicate twice the standard error. b) Median predictability gain based on $PG1$. Shaded areas indicate the interquartile range.
		c) Average predictability gain based on $PG2$, Eq. (\ref{eq:pg_ae}. Shaded areas indicate twice the standard error. d) Median predictability gain based on $PG2$. Shaded areas indicate the interquartile range.
	}
	\includegraphics[width=\textwidth]{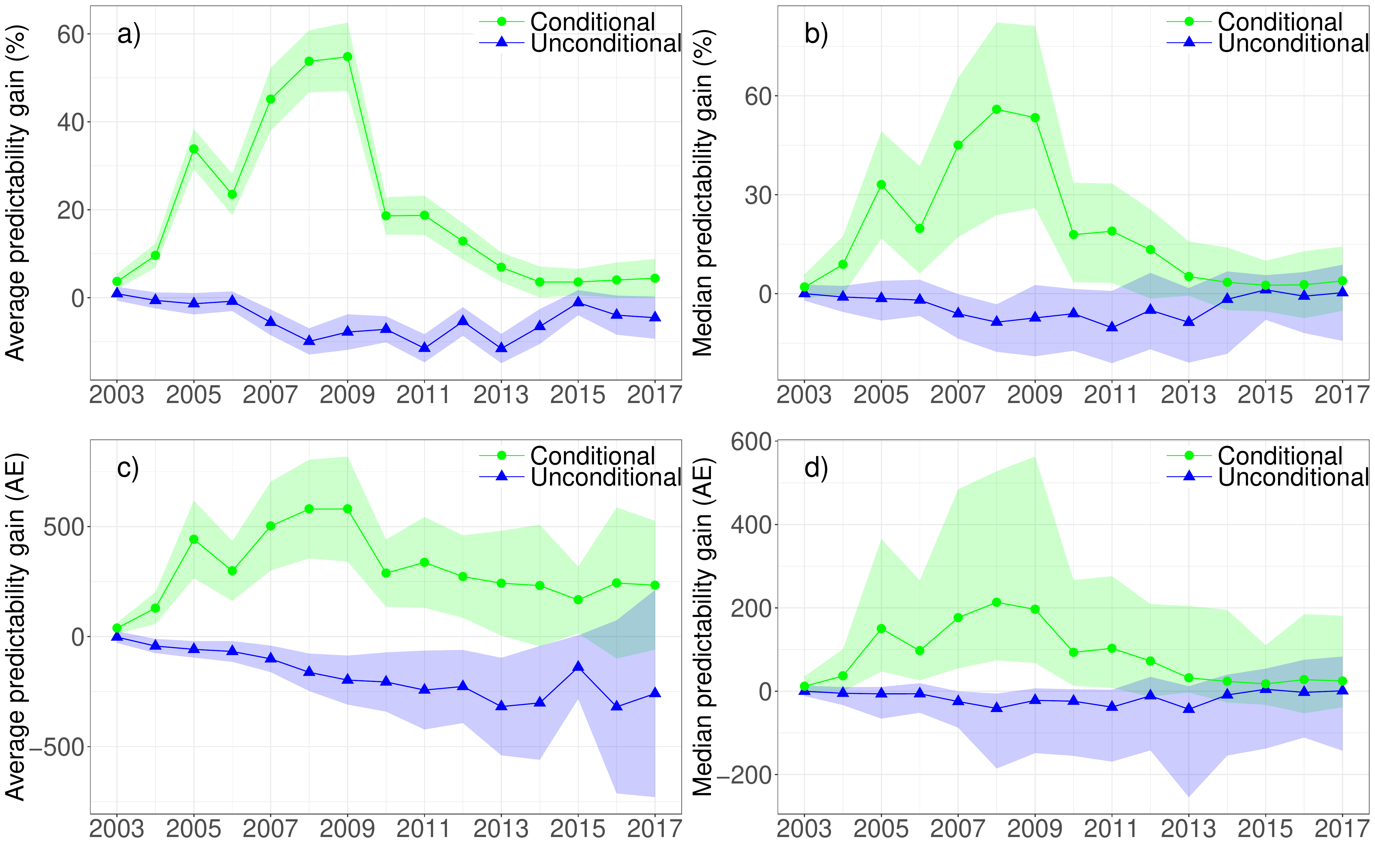}\label{fig:pg_cpc3}
	\caption{Prediction results based on \textbf{CPC 4-digit codes}.
		a) Average predictability gain based on $PG1$, Eq. (\ref{eq:pg}. Shaded areas indicate twice the standard error. b) Median predictability gain based on $PG1$. Shaded areas indicate the interquartile range.
		c) Average predictability gain based on $PG2$, Eq. (\ref{eq:pg_ae}. Shaded areas indicate twice the standard error. d) Median predictability gain based on $PG2$. Shaded areas indicate the interquartile range.}
	\includegraphics[width=\textwidth]{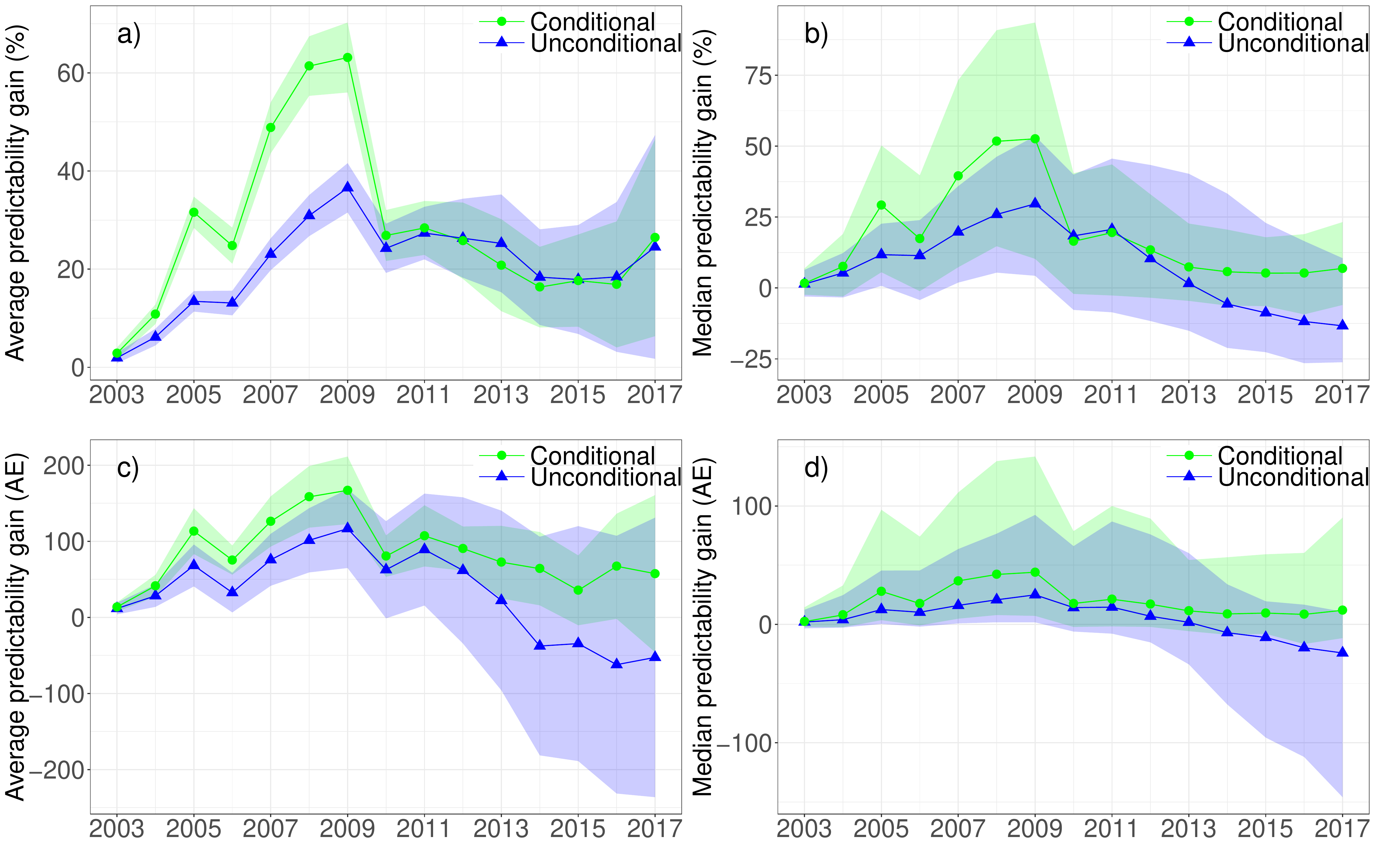}\label{fig:pg_cpc4}
\end{figure}

\begin{figure}
	\caption{Prediction results based on \textbf{IPC 3-digit codes}.
		a) Average predictability gain based on $PG1$, Eq. (\ref{eq:pg}. Shaded areas indicate twice the standard error. b) Median predictability gain based on $PG1$. Shaded areas indicate the interquartile range.
		c) Average predictability gain based on $PG2$, Eq. (\ref{eq:pg_ae}. Shaded areas indicate twice the standard error. d) Median predictability gain based on $PG2$. Shaded areas indicate the interquartile range.
	}
	\includegraphics[width=\textwidth]{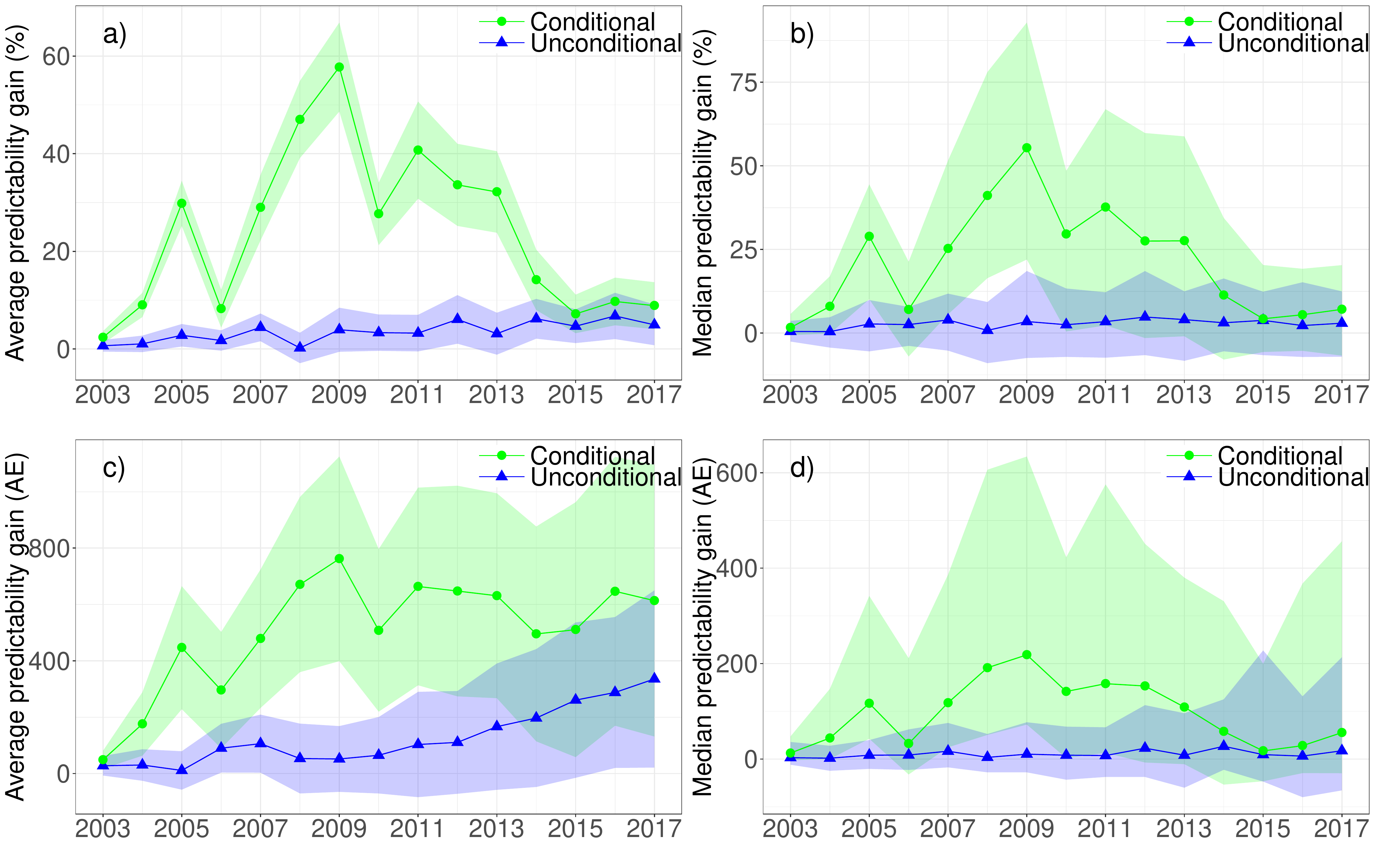}\label{fig:pg_ipc3}
	\caption{Prediction results based on \textbf{IPC 4-digit codes}.
		a) Average predictability gain based on $PG1$, Eq. (\ref{eq:pg}. Shaded areas indicate twice the standard error. b) Median predictability gain based on $PG1$. Shaded areas indicate the interquartile range.
		c) Average predictability gain based on $PG2$, Eq. (\ref{eq:pg_ae}. Shaded areas indicate twice the standard error. d) Median predictability gain based on $PG2$. Shaded areas indicate the interquartile range.}
	\includegraphics[width=\textwidth]{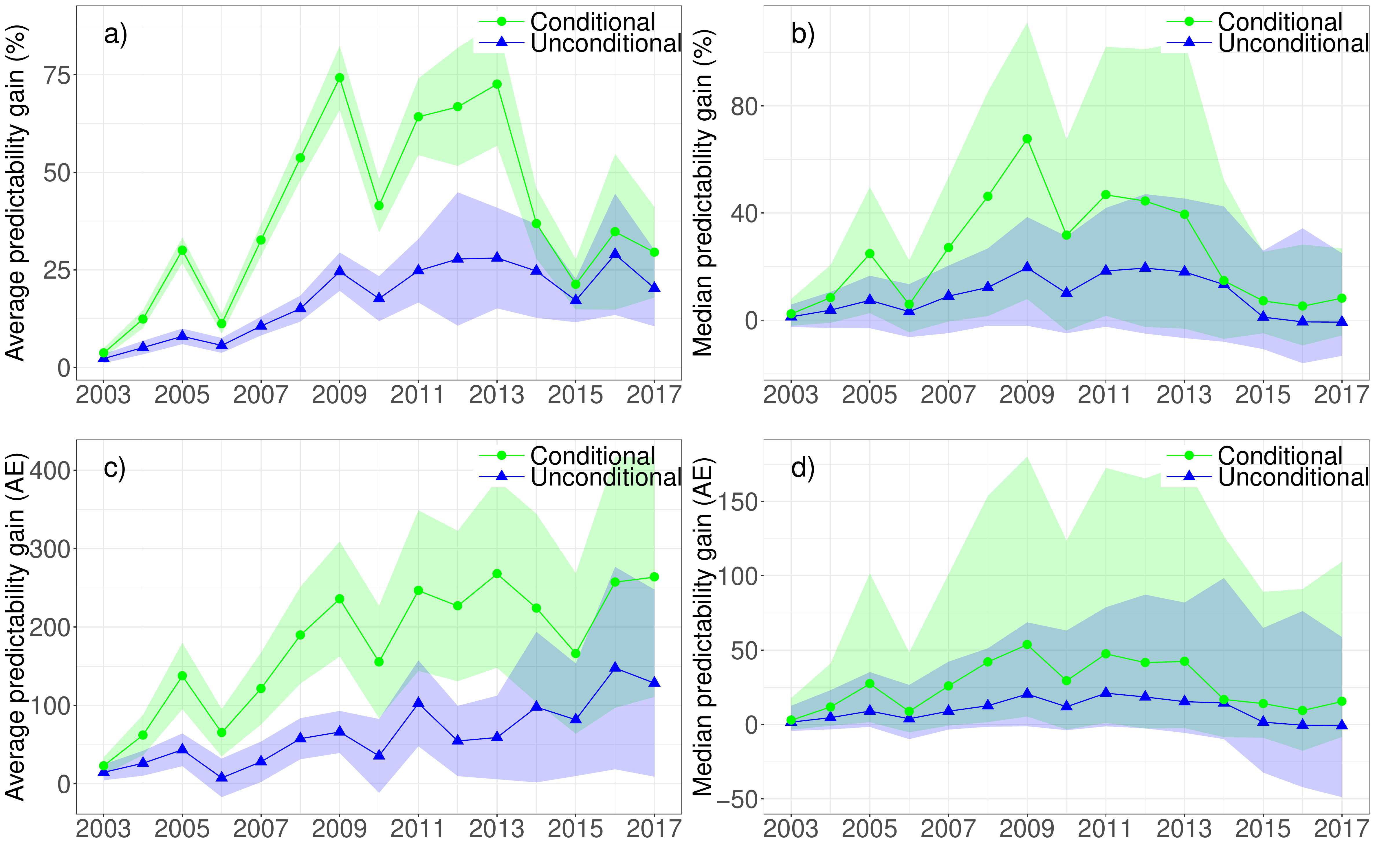}\label{fig:pg_ipc4}
\end{figure}

\clearpage

\bibliography{../tech_ref} % refers to example.bib
\bibliographystyle{abbrvnat} % or try abbrvnat or unsrtnat

% TTLEPAGE

\begin{center}
\huge{\textbf{Supplementary Information}}
\end{center}
\vspace{0.05cm}

\begin{center}
\LARGE{Innovation dynamics in the technological ecosystem}
\end{center}
\vspace{0.05cm}

\begin{center}
\large Anton Pichler, Fran\c{c}ois Lafond, J. Doyne Farmer
\end{center}

\tableofcontents
\newpage

%-----------------------------------------------------------------------------------------
\section{Data sources and preprocessing}

Our database contains all US utility patents granted between 1836 and 2018, except for rare exceptions. These patents are classified in two different classification systems: Cooperative Classification Scheme (CPC) and International Patent Classification (IPC) (up to 2015, they were also classified in the US Patent Classification System (USPC); we use this only in Section \ref{section:acemogluetal2016} where we discuss the results of \cite{acemoglu2016innovation}. Since we use their data we do not discuss this further here.) 

CPC classifications are retrieved from the Master Classification File available from the USPTO/Reed Tech (version 2019-04-01). IPC classifications and patent dates are retrieved from Google Patents Public Data, provided by IFI CLAIMS Patent Services, accessed June 2018. 

Citation data, we merge the citation database of \cite{kogan2017technological} with the citation data collected by \cite{lafond2019long} and with the database provided by the USPTO (patentsview.org/download, accessed: 14/05/2019).

\section{Robustness to alternative classification systems}
We check the robustness of the results by applying our method to two different technology classification schemes, at two different levels in each case. In the main text, we discuss results based on the Cooperative Classification Scheme (CPC) 4-digits codes. Here, we also show results based on the CPC 3-digits, International Patent Classification (IPC) 3-digits and IPC 4-digits codes scheme. In the following, we describe the data in detail and how we have processed it for the presented analysis.

Figure \ref{fig:patenting_classes}a) shows total patenting rates in the US from 1836 to 2018 for distinct classification codes of patents. While we find that CPC codes for almost every patent back to 1836 (the solid and the dotted line are almost indistinguishable in the plot), reclassification of IPC codes seems to be less rigorous for patents before 1920. For 2018 the data is not yet entirely updated, indicated by the sharp drop in patenting rates at the end of the time series. We therefore exclude the year 2018 in further analysis. Figure \ref{fig:patenting_classes}b) shows the citations counts per year. The figure illustrates that reliable data on citations made is available only from 1947 on, but we have information on citations received by earlier patents.

\begin{figure*}[!ht]
	\centering
	\includegraphics[width=.65\textwidth]{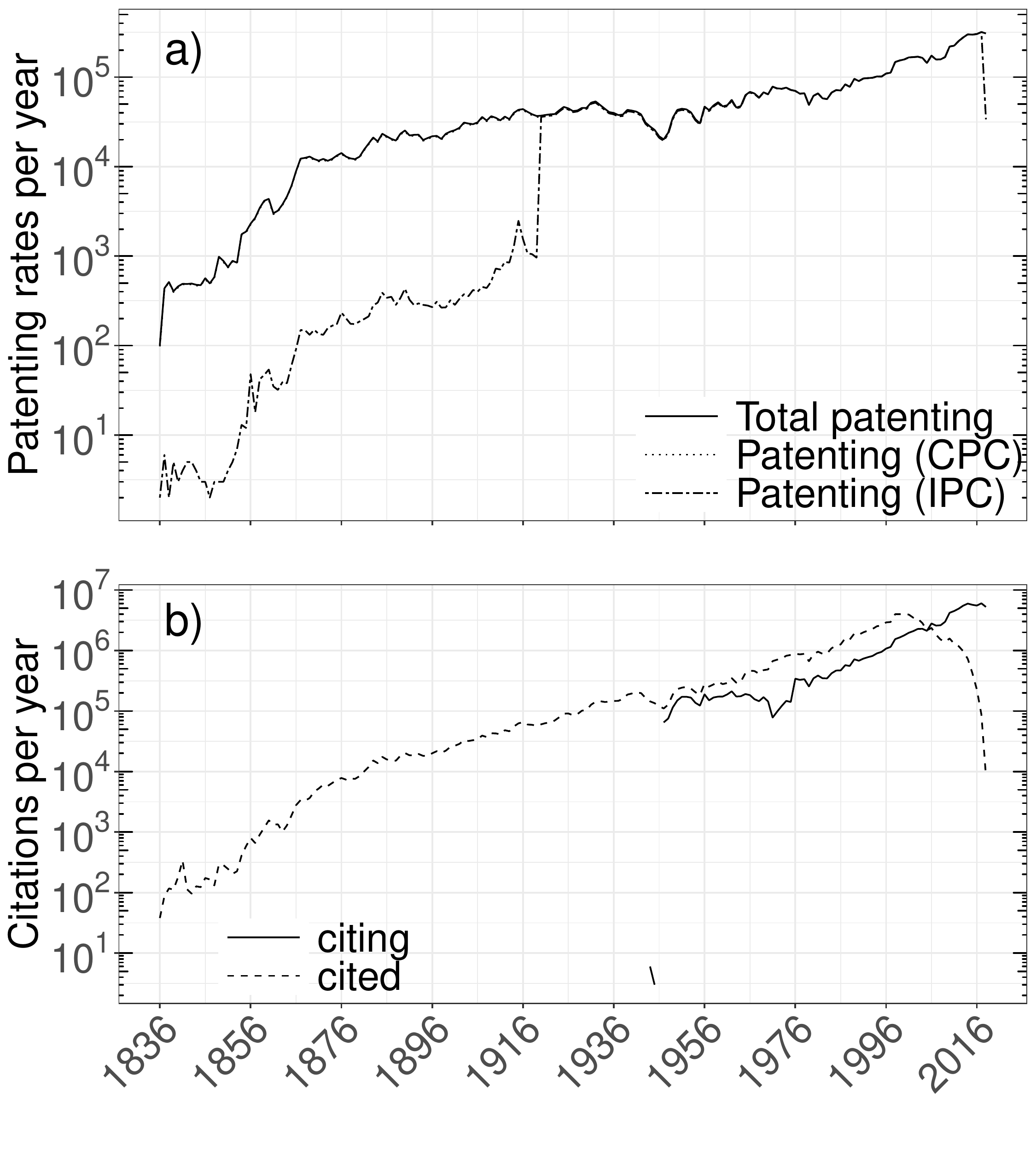} 
	\caption{a) Time series of US patents granted between 1836 to 2019. The solid line is the number of total granted patents, the dotted line is the number of patents with CPC attributes and the dotdashed line the number of patents having IPC attributes. 
	b) Time series of citations per year in the data set. \textit{Citing} refers to the number of citations made in a given year and \textit{cited} refers to the number of citations received.
}	
	\label{fig:patenting_classes}
\end{figure*}

We map patents onto their CPC and IPC codes and eliminate duplicates if the same patent was classified multiple times with the same 3- or 4-digits code. We only include technological classes which were at least ten times assigned to patents between 1836 and 2000. This removes no class at all in the CPC 3-digits case and only a single class in the CPC 4-digits case (A61P with only a single patent up to 2000). The cleaning is more effective in the IPC regimes where there are several classes which are very infrequently used. Introducing this threshold eliminates 143 IPC 3-digits and 389 IPC 4-digits classes. For the CPC classes, we also eliminate all Y-codes which are tags rather than separate technology classes. Table \ref{tab:classes} summarizes some basic statistics of the used data, after excluding the year 2018.

\begin{table}[!ht]
	\centering
	\begin{tabular}{lrrr}
		Classification & \# classes & \# observations & \# patents \\ 
		\hline
		CPC 3-digits & 125  & 13,006,073 & 9,828,035\\ 
		CPC 4-digits & 649  & 14,537,098 & 9,828,035\\ 
		IPC 3-digits & 121  & 12,000,519 & 8,450,379\\ 
		IPC 4-digits & 633  & 13,697,093 & 8,450,379\\ 
		\hline
	\end{tabular}
	\caption{ Descriptive statistics of used data after initial cleaning.}
	\label{tab:classes}
\end{table}

We obtain a temporal network for each of the four classification schemes. To avoid links between technological domains which one would expect by chance, we apply the significance sampling procedure \citep{alstott2017mapping} described in the main text. As can be seen in Figure \ref{fig:density_classes}, the density is reduced substantially after applying the significance sampling procedure, but is also increasing in time. Unsurprisingly, the network density is higher when using the more aggregate 3-digits classification. 

\begin{figure*}[!ht]
	\centering
	\includegraphics[width=\textwidth]{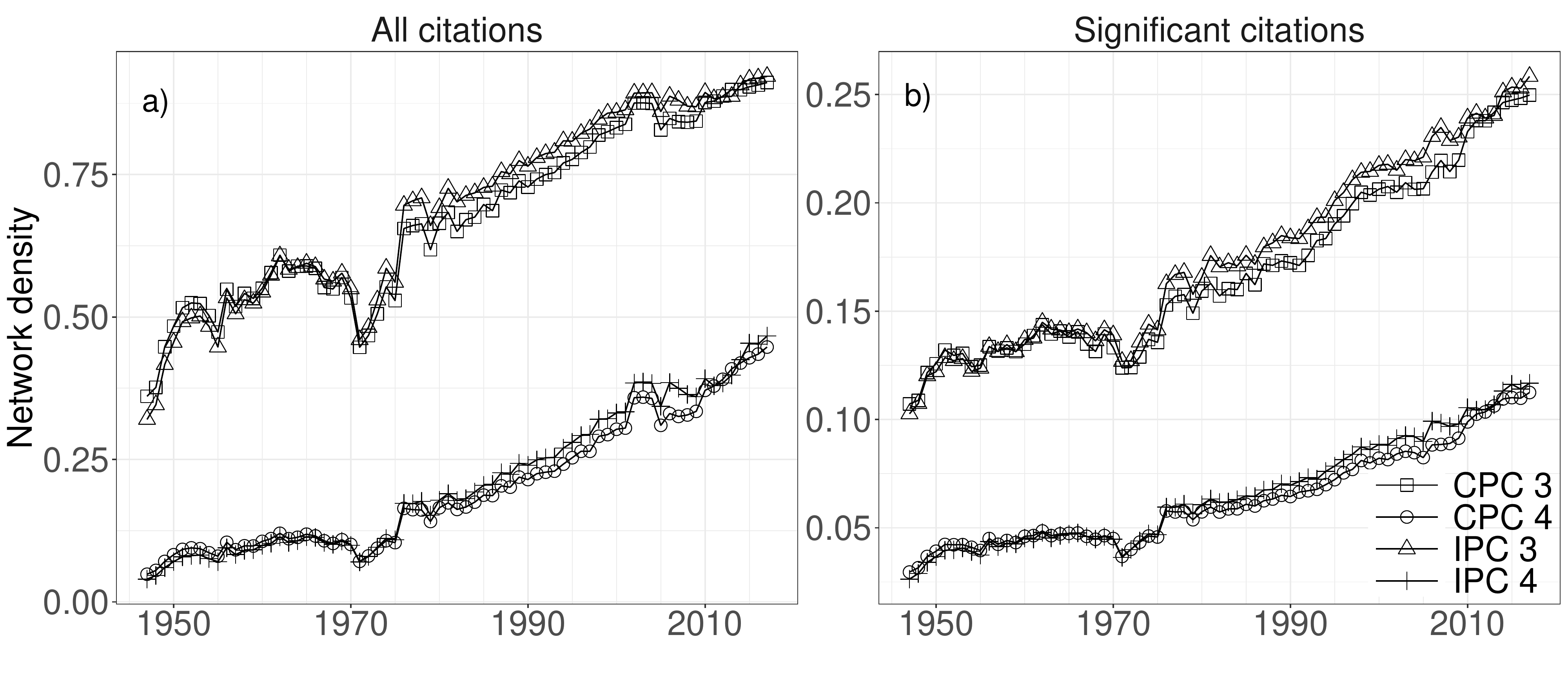} 
	\caption{a) The network densities over time for different classification schemes when all citations are included. b) The network densities after removing non-significant links. }	
	\label{fig:density_classes}
\end{figure*}

\section{Temporal network stability}
As a measure of temporal stability, Figure \ref{fig:nw_stability} plots the average cosine similarity between knowledge bases at $t$ and $t-l$, or in more mathematical terms
\begin{equation*}
\text{similarity}_{t,t-l}^\text{av.} =
\frac{1}{N} 
\sum_{i=1}^N
\frac{\sum_j W_{ij,t} W_{ij,t-l}}{\sqrt{\sum_j W_{ij,t}^2} \sqrt{ \sum_j W_{ij,t-l}^2} }.
\end{equation*}
We observe that knowledge bases tend to be fairly stable in time, i.e. the future knowledge base of a given technology is likely to be similar to its current one. After the temporal stability of the network was increasing for most parts of the second half of the last century, it became weaker in the last 20 years. As one would expect, we find that the average knowledge base similarity decreases for higher lags, although there is still a pronounced positive relationship between knowledge inputs over a 10-year horizon.
A qualitative difference between 3-digits and 4-digits classification schemes is apparent. The networks based on the more aggregated 3-digits classifications exhibit higher correlations in time for all shown lag choices. Moreover, the differences in temporal correlations between different time lags become less pronounced. This suggests that it is important to use more aggregated technology classification if the research is interested on the temporal evolution and structural change of the technological ecosystem.
\begin{figure*}[!ht]
	\centering
	\includegraphics[width=.44\textwidth]{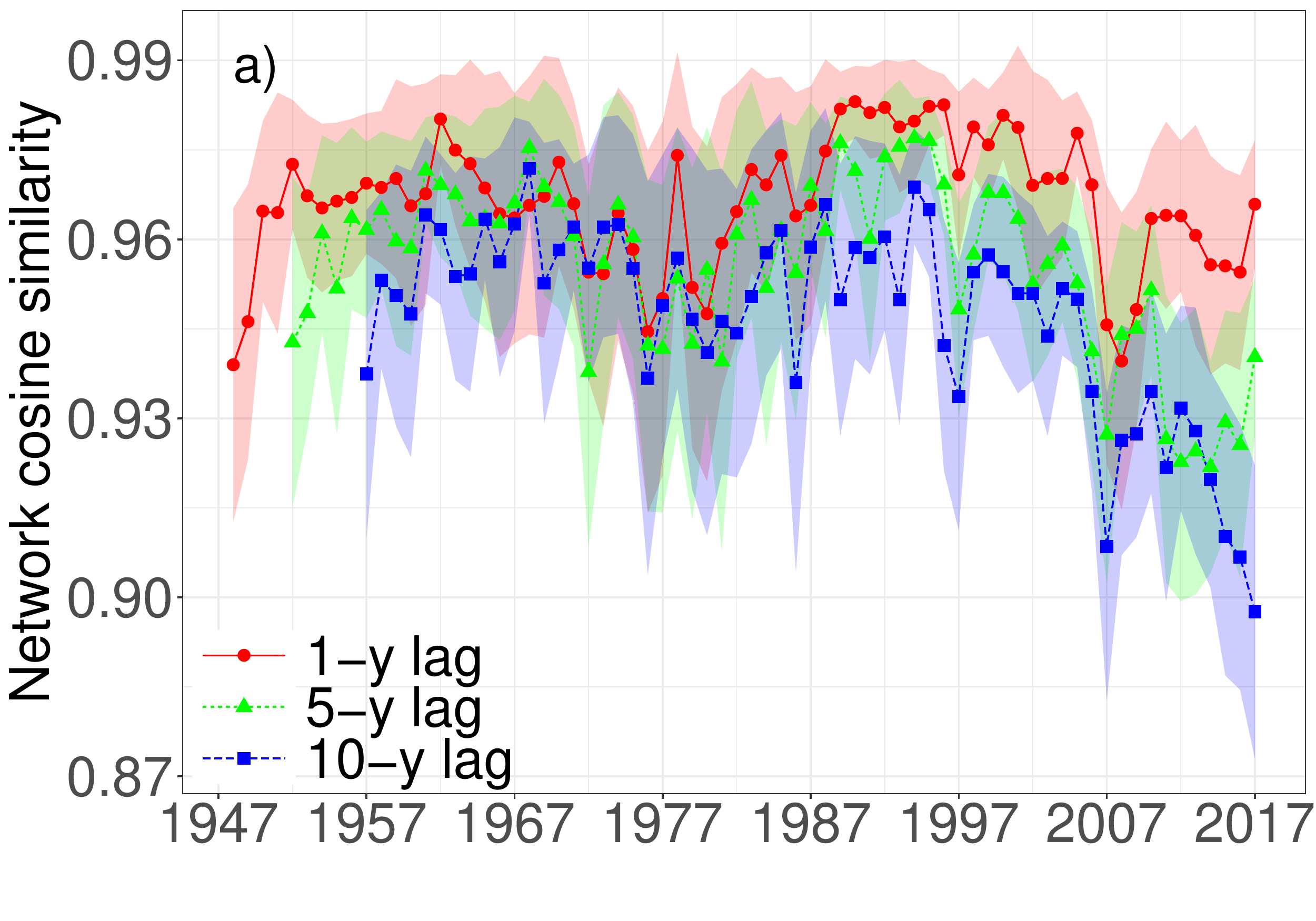}
	\includegraphics[width=.44\textwidth]{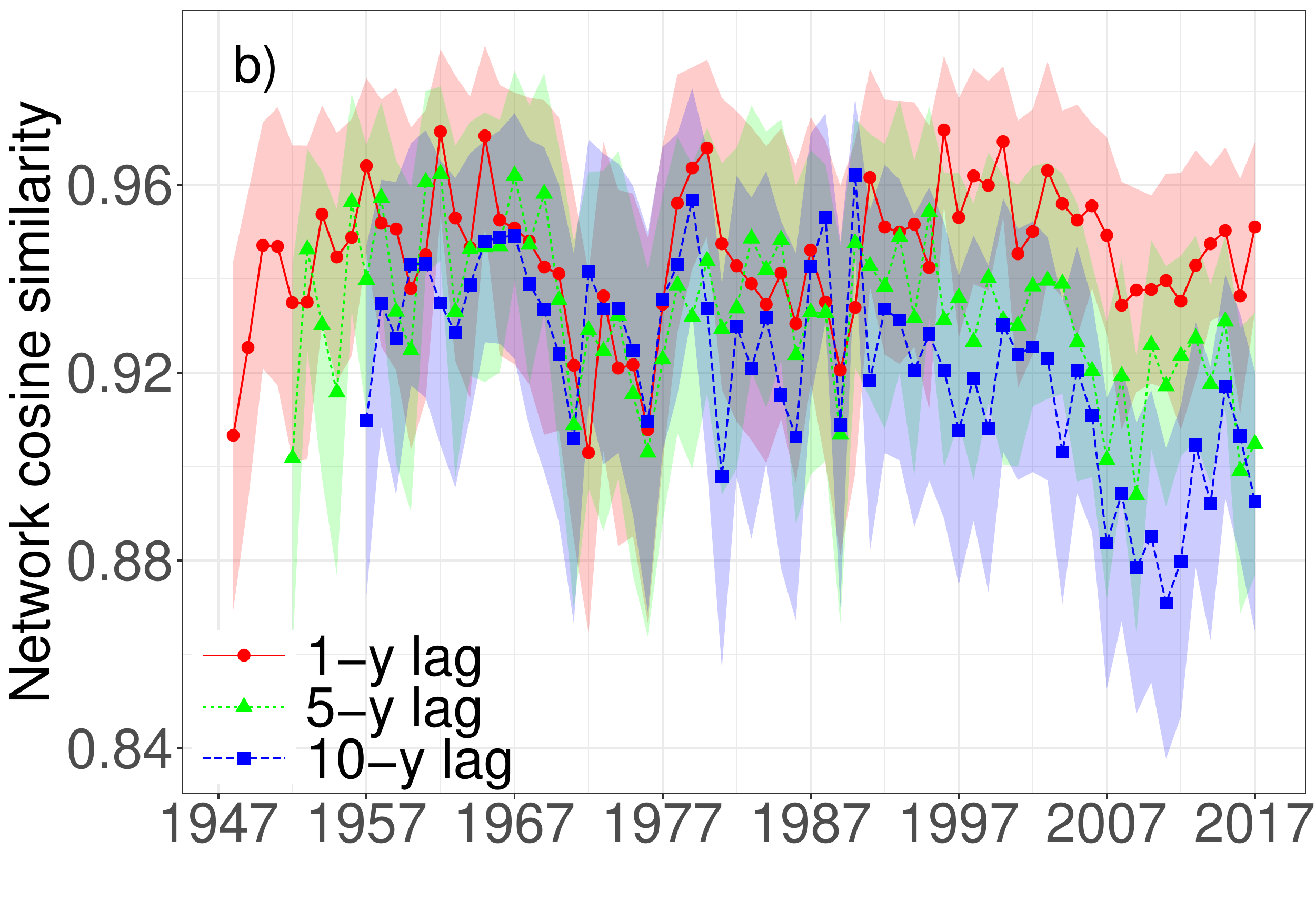} \\
	\includegraphics[width=.44\textwidth]{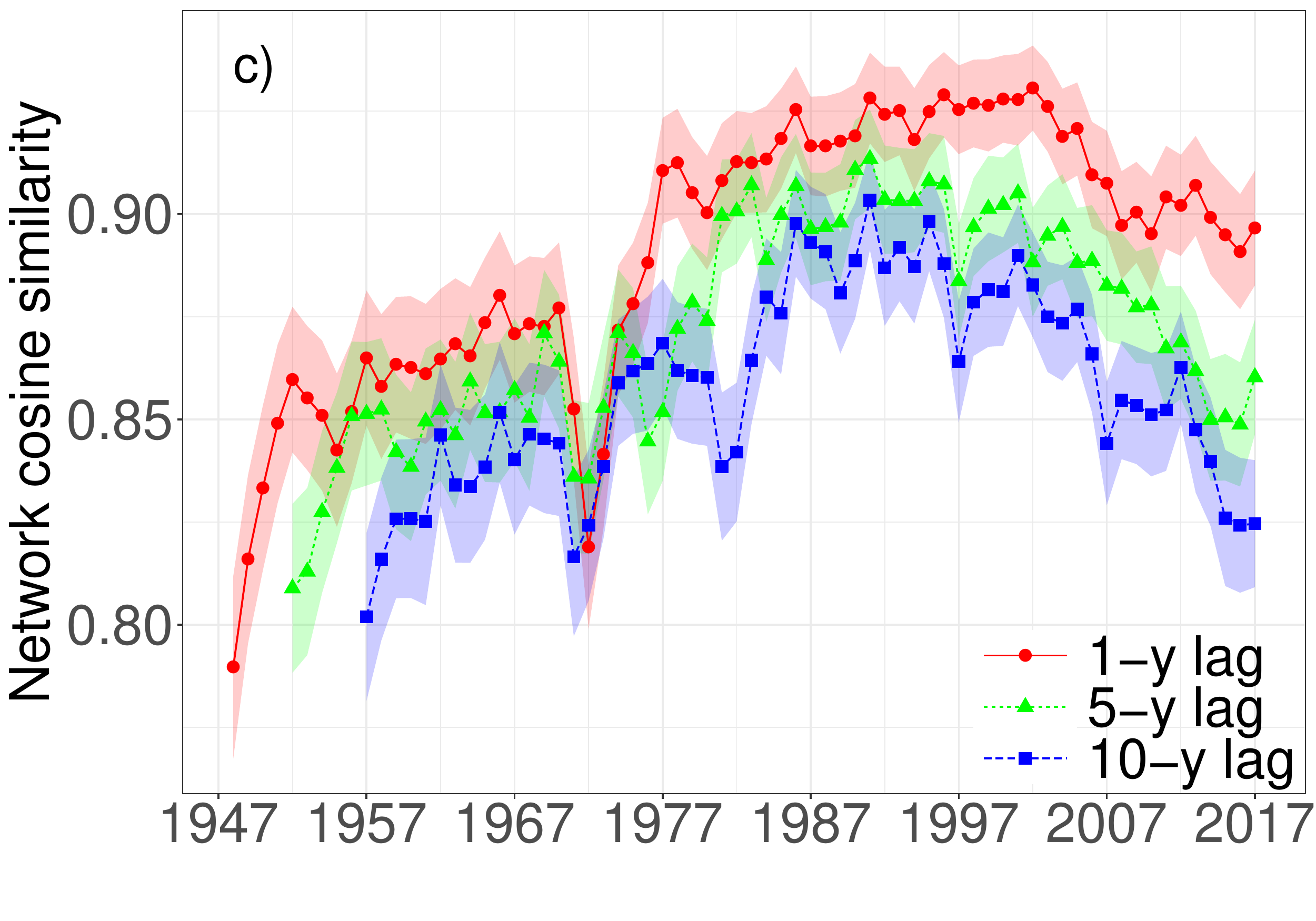}
	\includegraphics[width=.44\textwidth]{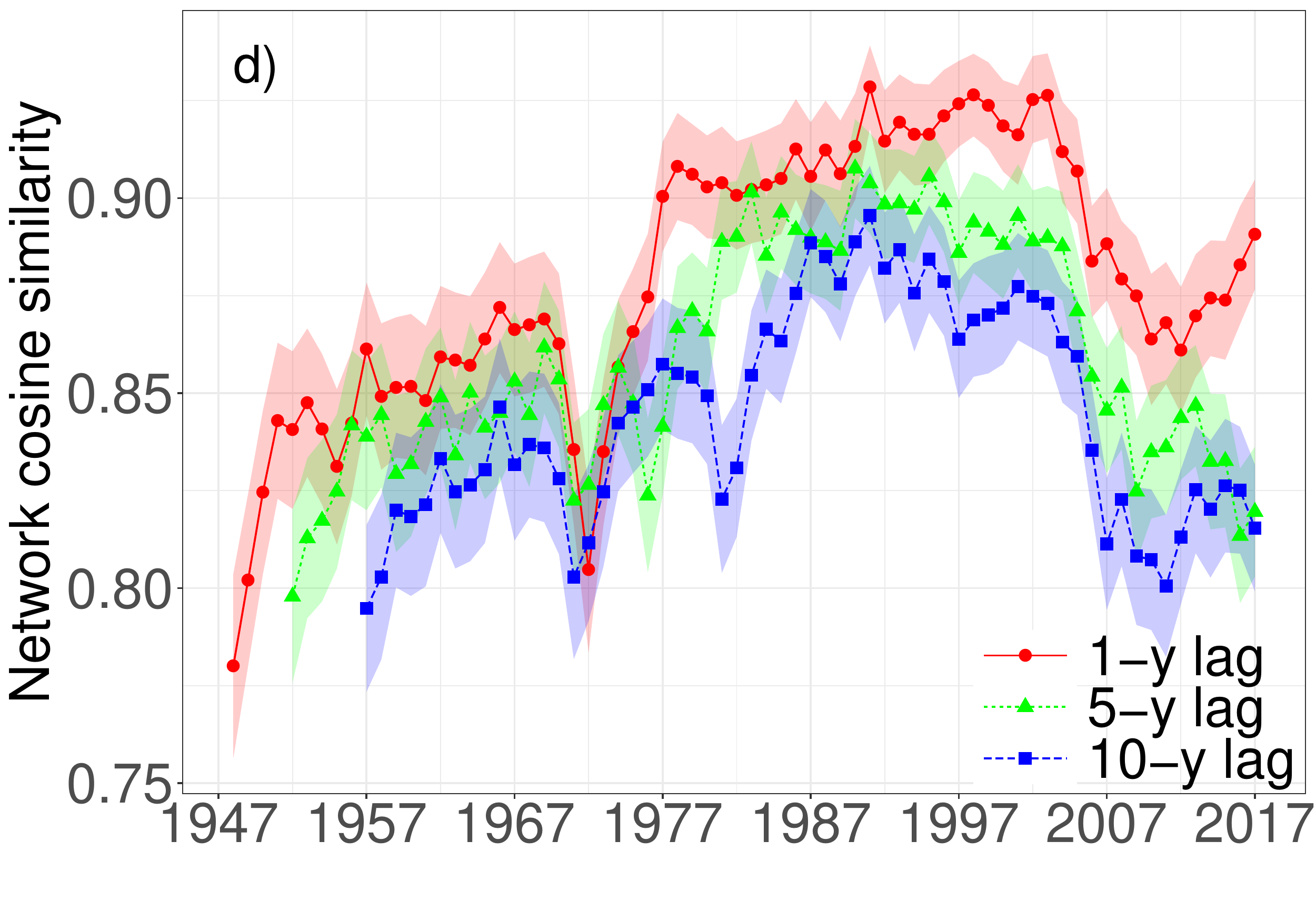}
	\caption{Average temporal cosine similarity between knowledge inputs over time. The shaded area shows the range of the average plus/minus twice the standard errors. The results are shown for networks based on the a) 3-digits CPC codes, b) 3-digits IPC codes and c) 4-digits CPC codes and d) 4-digits IPC codes. }	
	\label{fig:nw_stability}
\end{figure*}

\section{Nearest neighbor growth correlations}
In the main text we have shown that the technology network is highly assortative with respect to growth rates, suggesting that patenting growth rates are similar if the technologies are connected. Figure \ref{fig:assortativity} plots the average nearest neighbor growth rates correlation over time for the differently aggregated networks. The positive assortativity patterns is more noisy in the more aggregated networks, panel a) and b), but still pronounced for most of the cases. The average nearest neighbor growth rates correlation is relatively insensitive with respect to different growth rate lengths choices when consider the more aggregated networks, again pointing at the importance of using more granular networks for discerning temporal effects in the technological evolution.

Table \ref{tab:avnng} shows that the correlation magnitude is similar for different classification schemes, but systematically less noisy when using the more granular 4-digits classification. Data aggregation often reduces noise by averaging out extrema. Thus, it is not obvious a priori that the more fine-grained network representations exhibit less variation in their average nearest neighbor growth rates correlations.

\begin{figure*}[!ht]
	\centering
	\includegraphics[width=.44\textwidth]{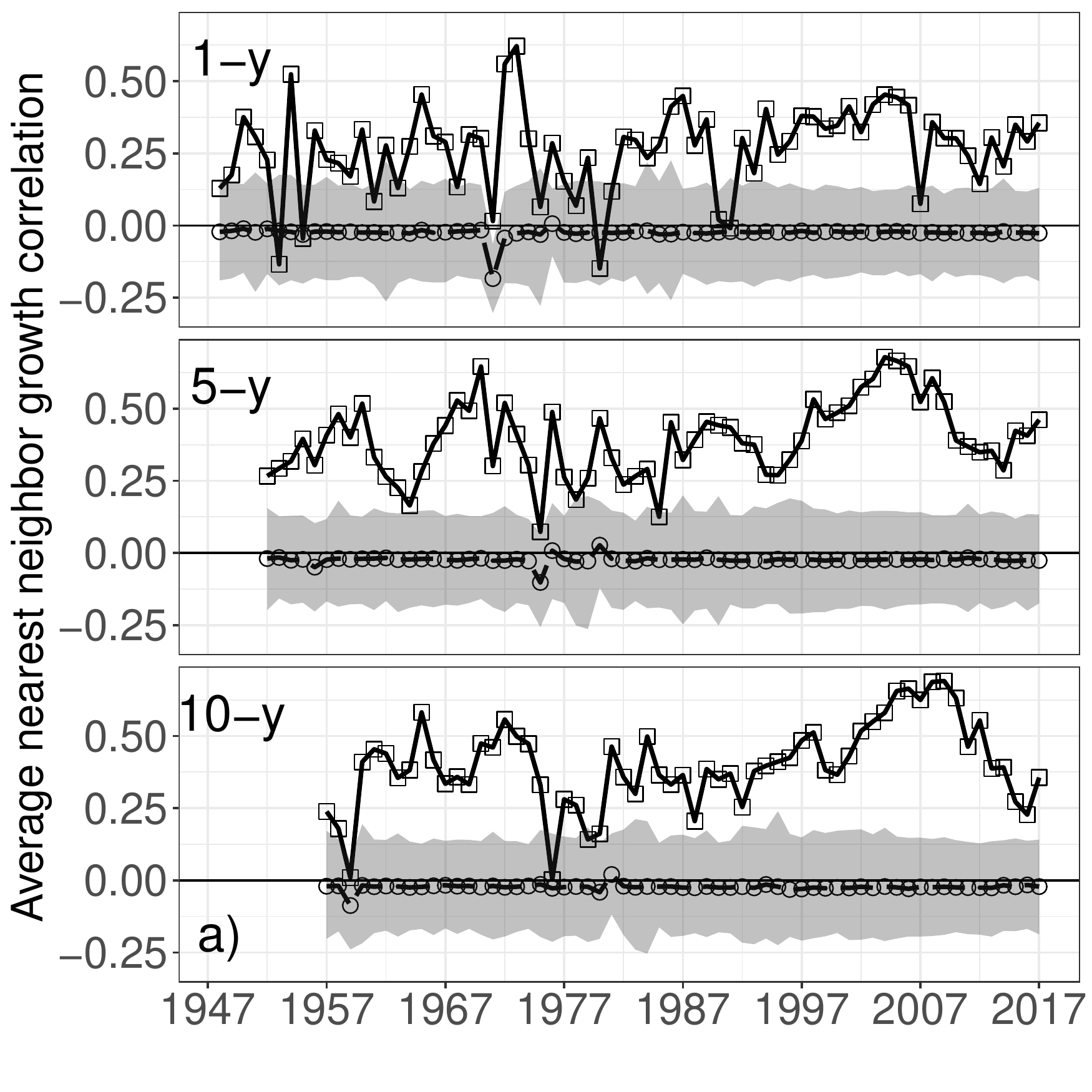}
	\includegraphics[width=.44\textwidth]{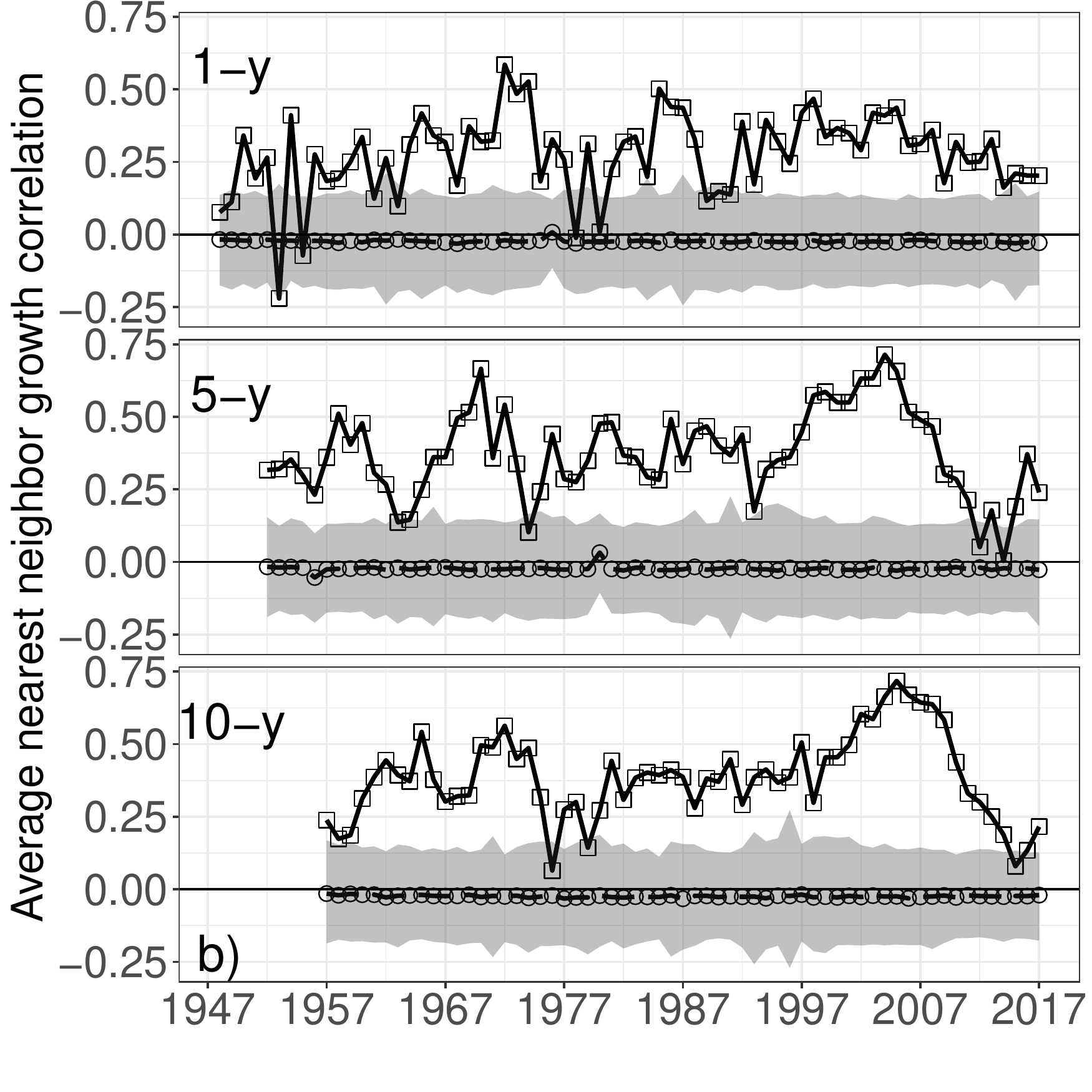} \\
	\includegraphics[width=.44\textwidth]{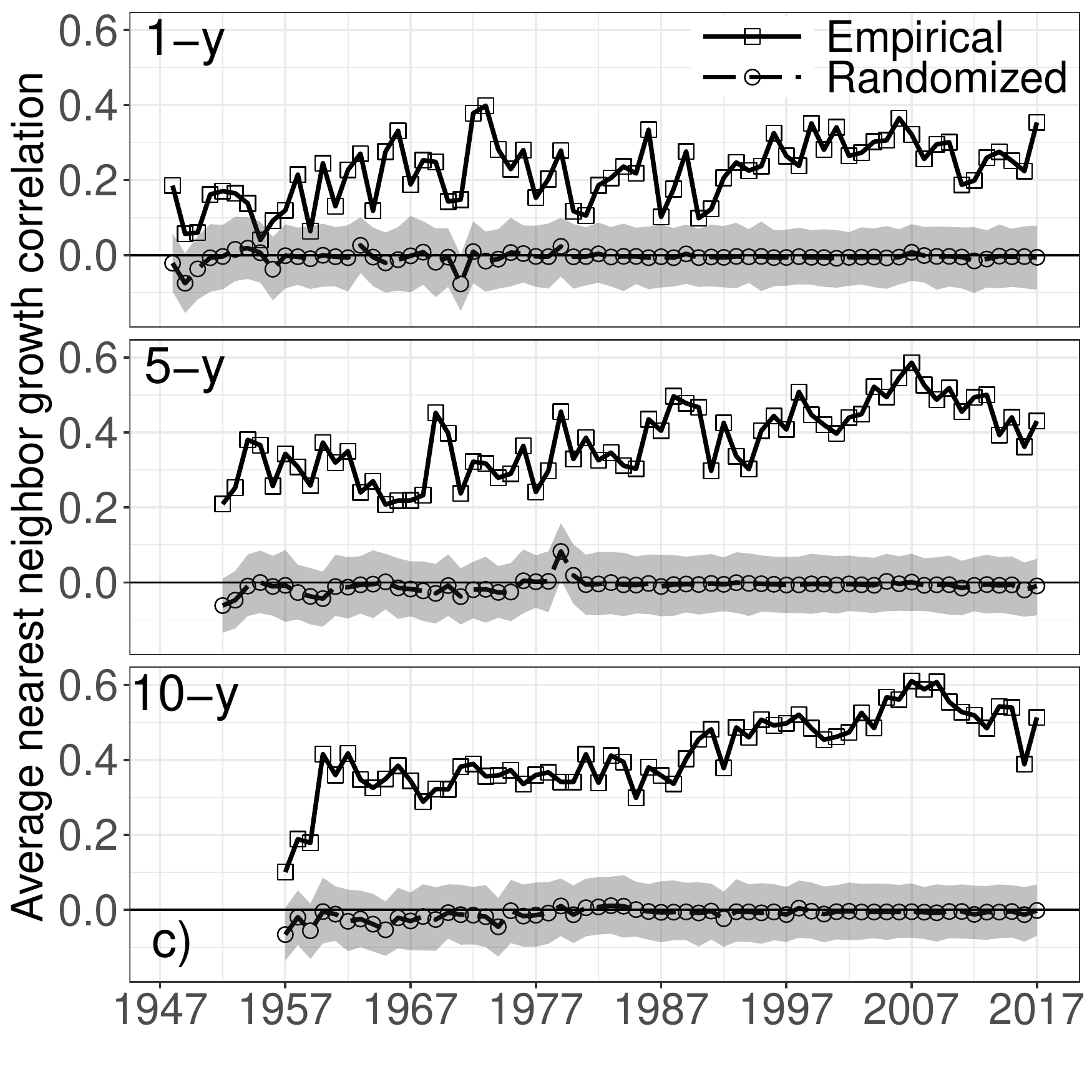}
	\includegraphics[width=.44\textwidth]{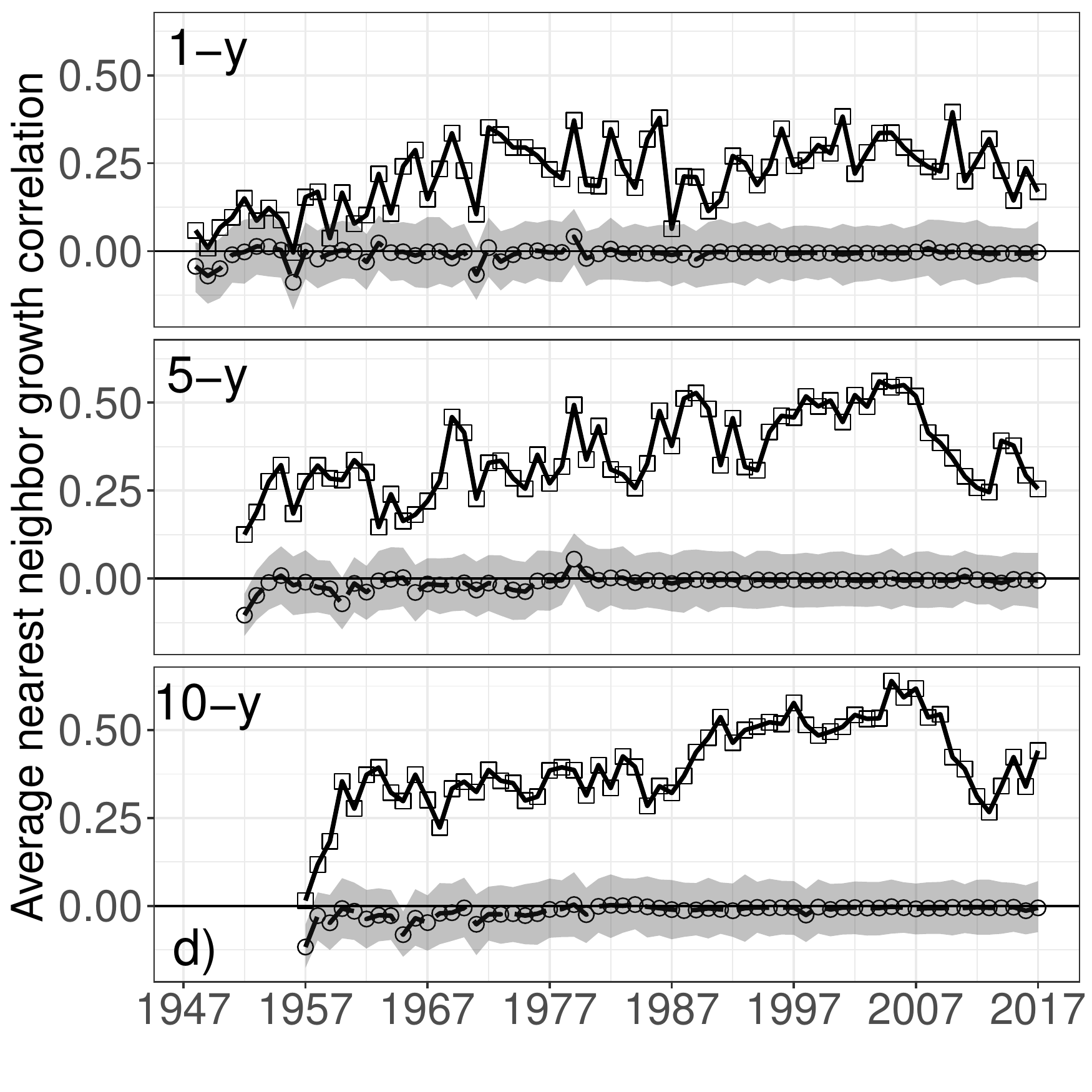}
	\caption{Average nearest neighbor growth rates correlation across time for networks based on the a) 3-digits CPC codes, b) 3-digits IPC codes, c) 4-digits CPC codes and d) 4-digits IPC codes. The shaded area is the 5-95\% inter-quantile range of results based on 1,000 randomized networks (described in the main text).  }	
	\label{fig:assortativity}
\end{figure*}

\begin{table}[!ht]
	\centering
	\begin{tabular}{|l|l|rrrr|}
		\hline
		& Lag & CPC 3 & CPC 4 & IPC 3 & IPC 4 \\ 
		\hline
		\multirow{3}{*}{Time average}
		& 1-year & 0.25 & 0.21 & 0.26 & 0.20 \\ 
		& 5-year & 0.38 & 0.37 & 0.36 & 0.35 \\ 
		& 10-year & 0.44 & 0.42 & 0.41 & 0.41 \\ 
		\hline
		\multirow{3}{*}{Time std. deviation}
		& 1-year & 0.16 & 0.09 & 0.15 & 0.10 \\ 
		& 5-year & 0.19 & 0.09 & 0.19 & 0.11 \\ 
		& 10-year & 0.20 & 0.10 & 0.20 & 0.11 \\ 
		\hline
	\end{tabular}
	\caption{Average nearest neighbor growth rate correlations. Time average and time standard deviation of the average nearest neighbor growth rates correlation for distinct technological classification schemes.}
\label{tab:avnng}
\end{table}

\section{Knowledge base similarities}
We conduct another test to check whether innovation dynamics of technologies are influenced by their knowledge base. If the knowledge base affects a technology's patenting rates, we should expect technological domains with similar knowledge bases to exhibit similar innovation dynamics. We test this hypothesis in two ways.
First, we ask whether technologies with similar knowledge bases in the year 2000 have grown similarly over the course of the next ten years.
To quantify the similarity of knowledge bases, we compute the cosine similarity for each pair of input vectors (i.e. each pair of rows in the matrix $W_t$). 
As a measure of growth rate similarity, we take the product of each pair of 10-year growth rates. A growth rate product will be positive if both technologies have growth rates of the same sign and negative otherwise.   

\begin{figure}
	\begin{minipage}{\textwidth}
			\centering
	\includegraphics[height=.19\textheight]{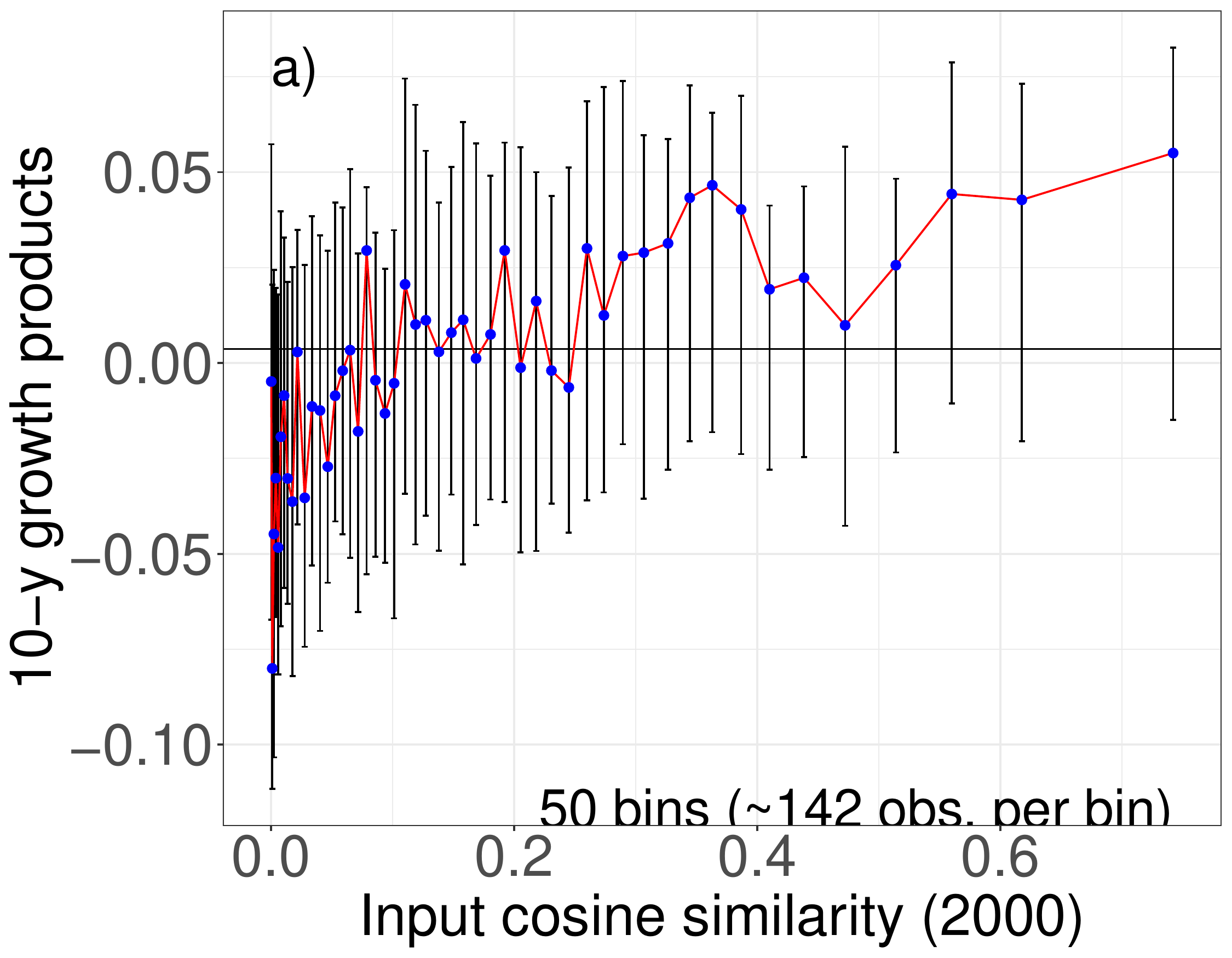}
\includegraphics[height=.19\textheight]{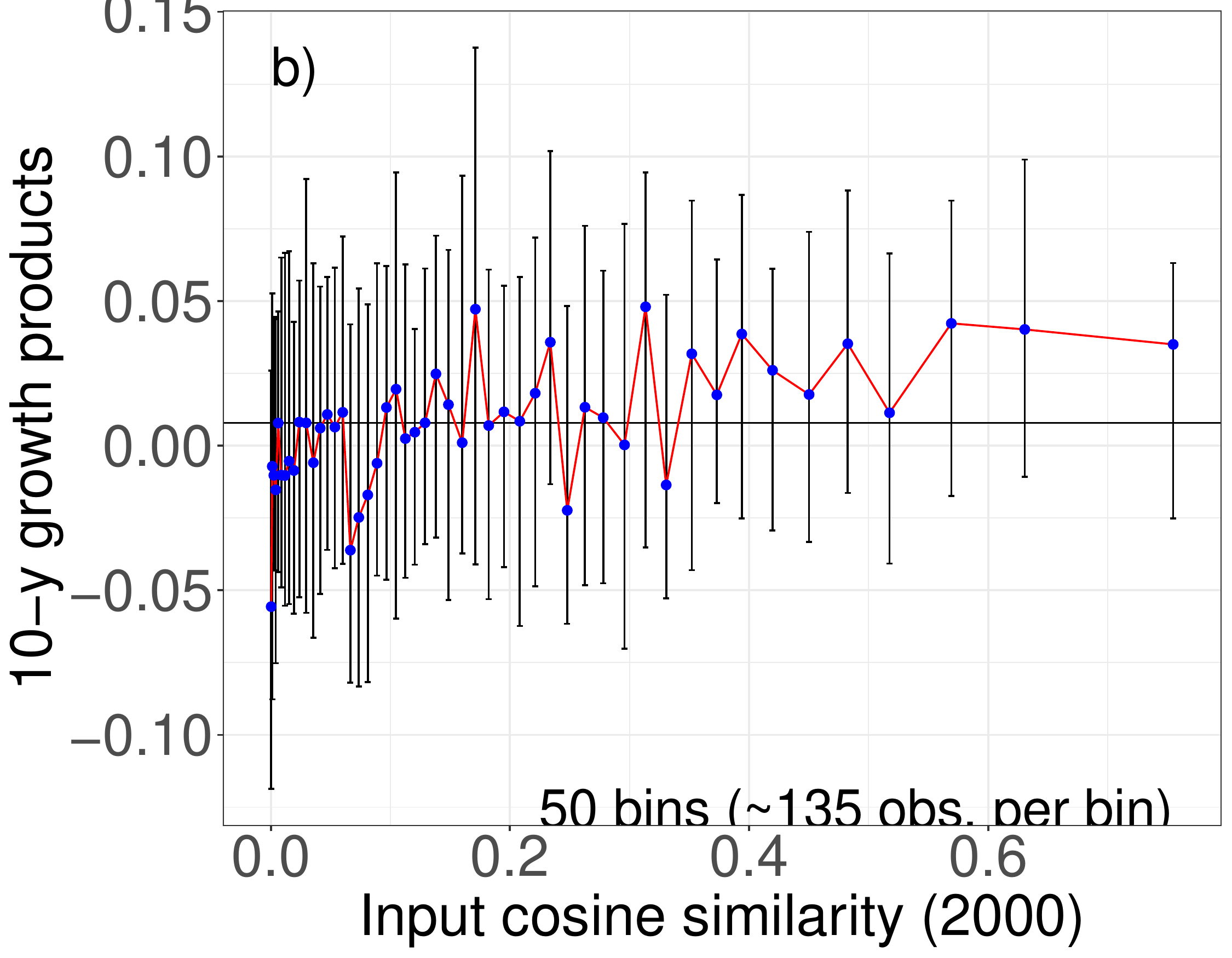} \\
\includegraphics[height=.19\textheight]{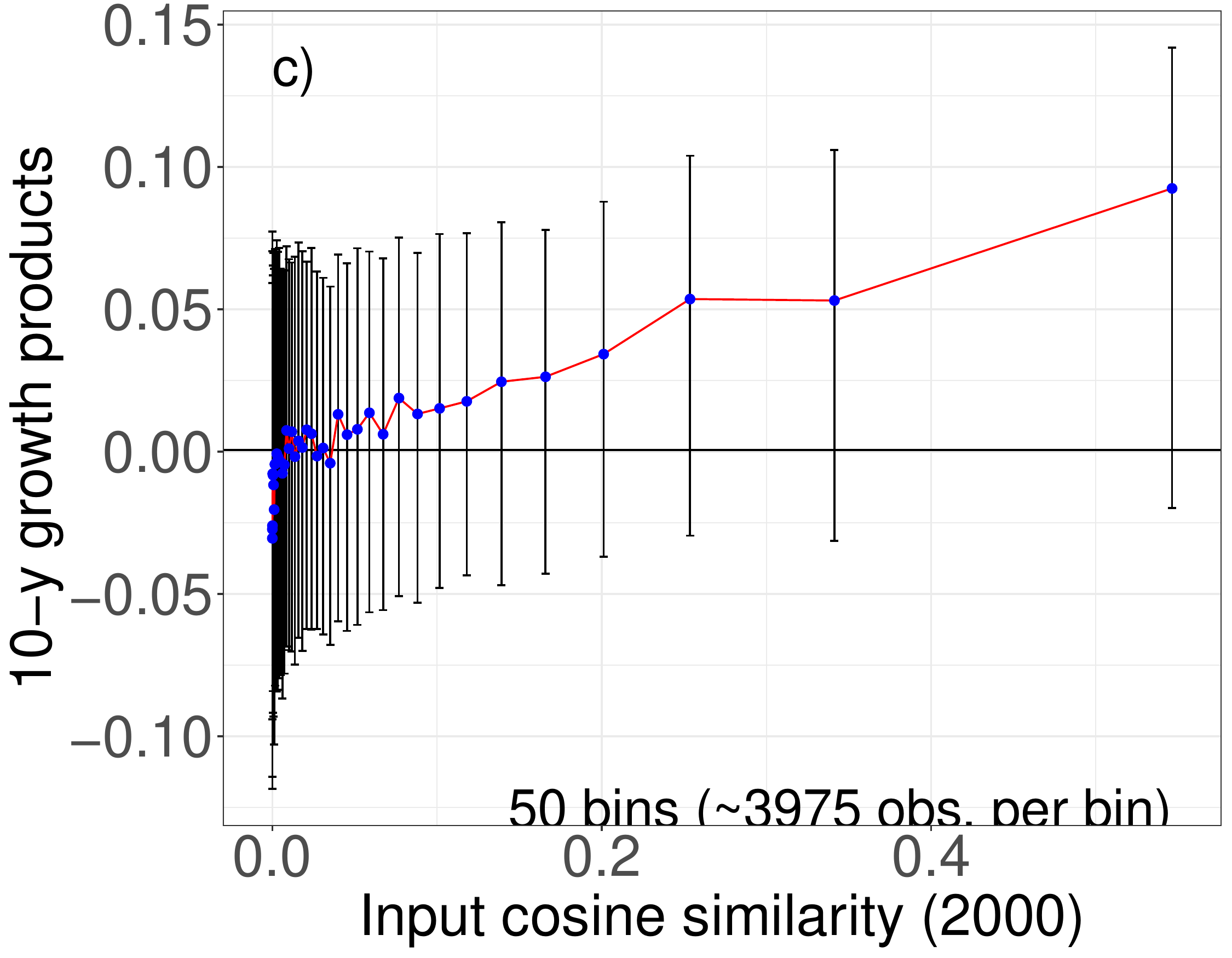}
\includegraphics[height=.19\textheight]{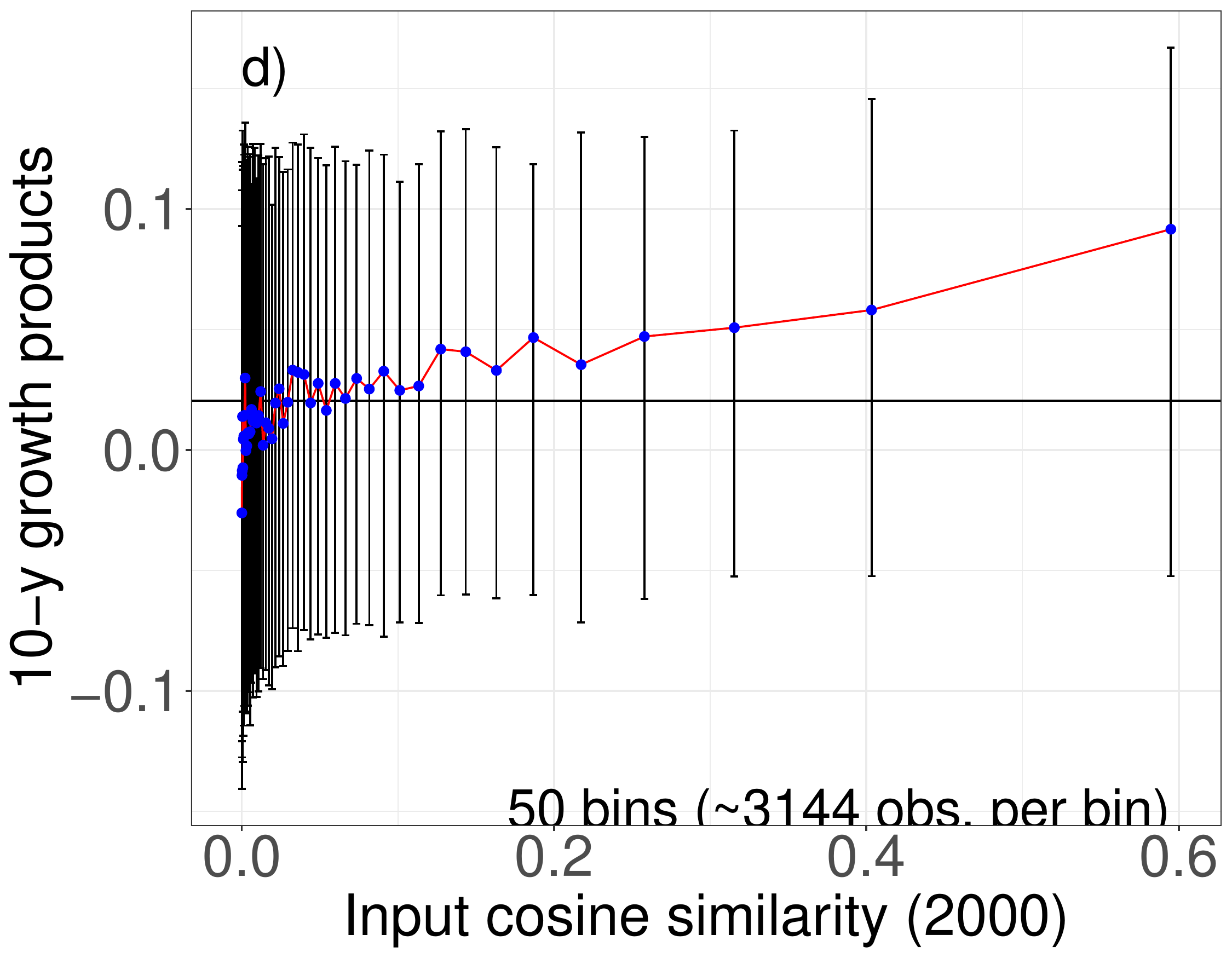}
	\caption{Pairwise growth rate products as a function of input cosine similarity for networks based on the a) 3-digits CPC codes, b) 3-digits IPC codes, c) 4-digits CPC codes and d) 4-digits IPC codes. Input cosine similarities are taking from the year 2000 and growth products are computed based on the 10-year growth rates from 2000 to 2010. The blue points indicate the bin average and the black bars denote the interquartile range of the bin. The horizontal black line indicates the sample average. }	
\label{fig:growth_products}
	\end{minipage}
	\begin{minipage}{\textwidth}
			\centering
	\includegraphics[height=.19\textheight]{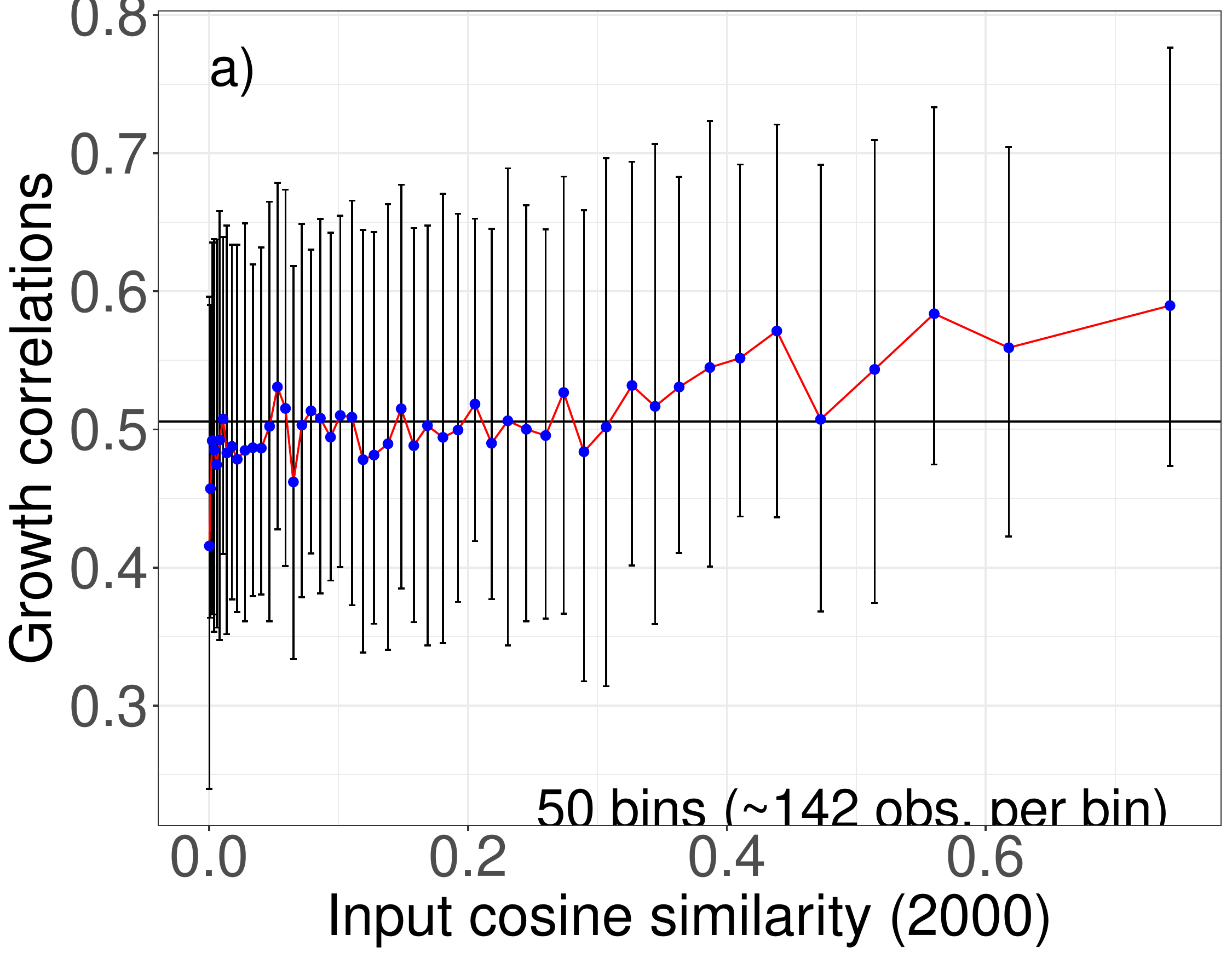}
\includegraphics[height=.19\textheight]{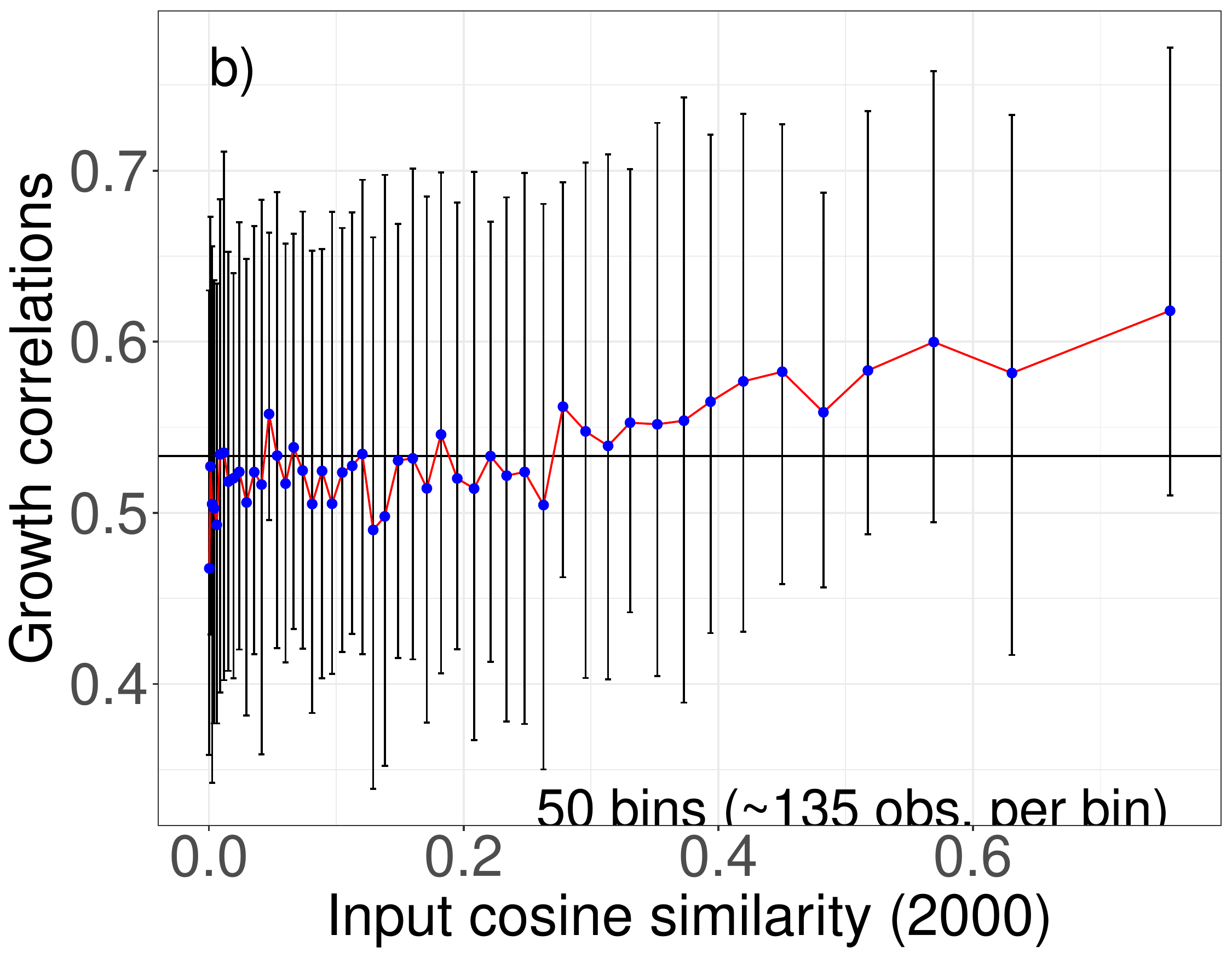} \\
\includegraphics[height=.19\textheight]{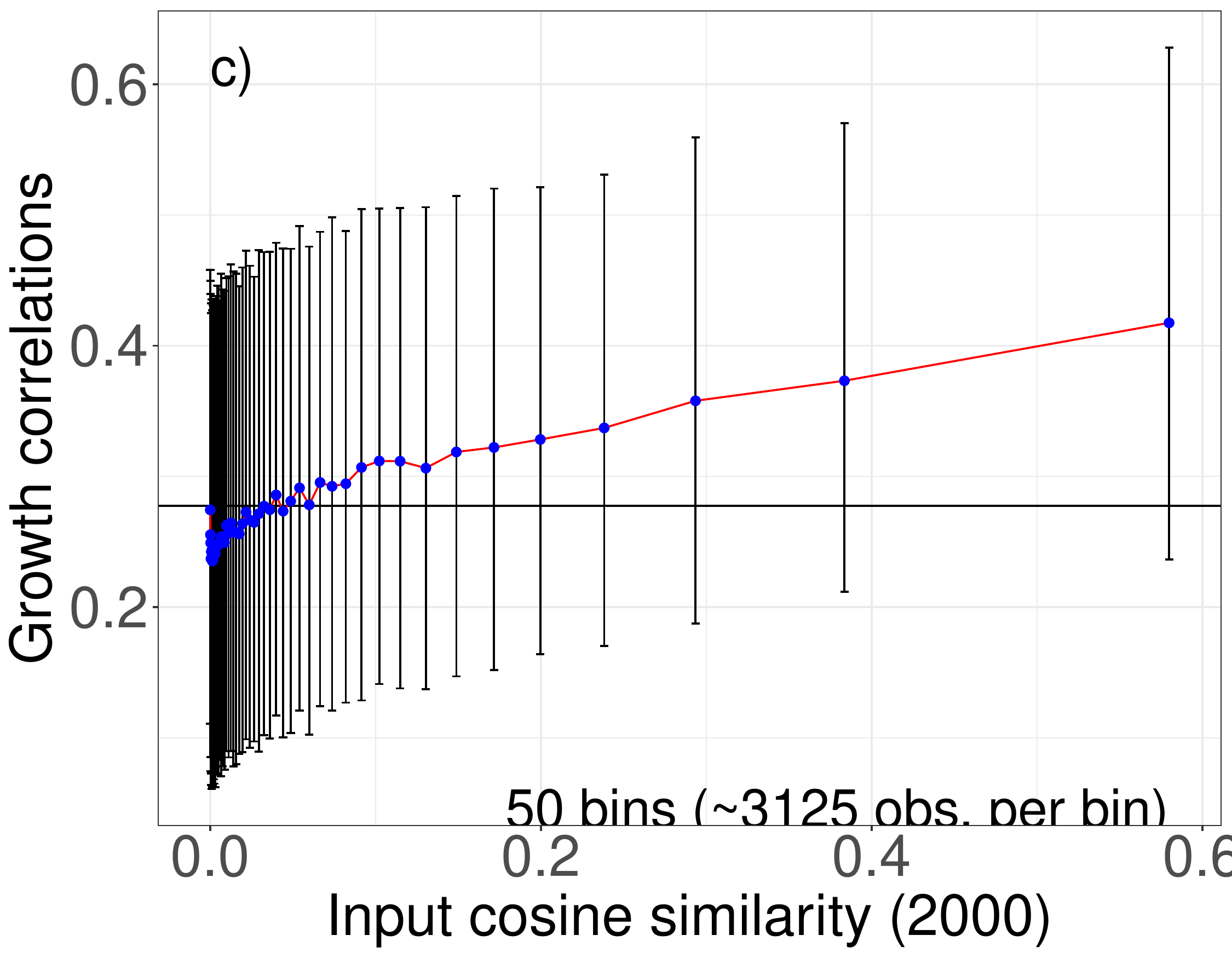}
\includegraphics[height=.19\textheight]{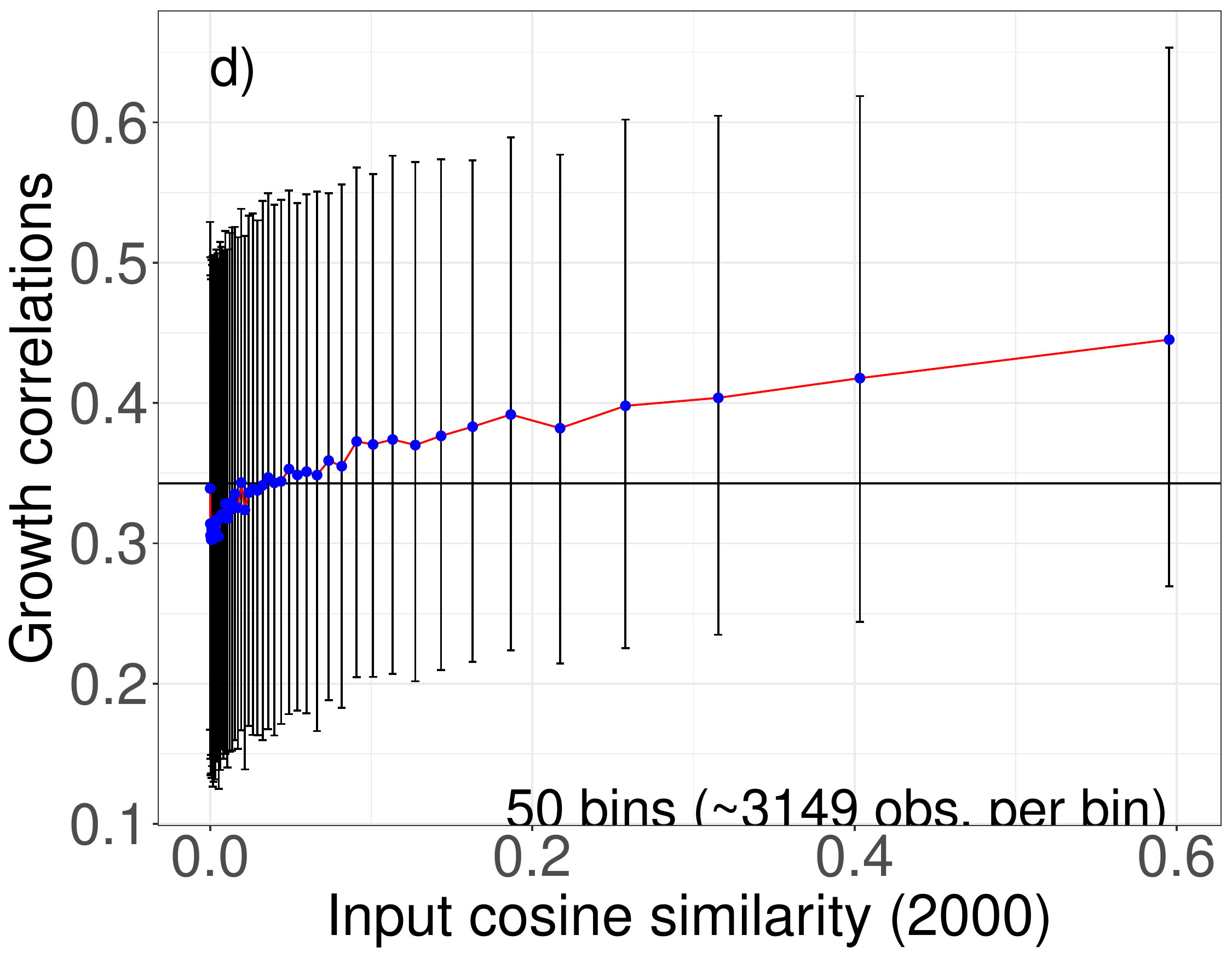}
	\end{minipage}
	\caption{Growth rate correlations as a function of input cosine similarity for networks based on the a) 3-digits CPC codes, b) 3-digits IPC codes, c) 4-digits CPC codes and d) 4-digits IPC codes. Input cosine similarities are taking from the year 2000. Correlations are computed for 1-year growth rates, every year between 2000 and 2017 (i.e. at most 17 observations per pair). The blue points indicate the bin average and the black bars denote the interquartile range of the bin. The horizontal black line indicates the sample average. }	
\label{fig:growth_correlations}
\end{figure}

Figure \ref{fig:growth_products} plots the growth rate products versus the knowledge input similarities. 
Since this pairwise comparison yields around 200,000 observations for the 4-digits classifications, we have binned the data and report bin averages and the interquartile range in the visualization. As expected, we  find that growth rate products are increasing as a function of knowledge base similarity. 
Figure \ref{fig:growth_products}c) depicts the result for the CPC 4-digit case. Here, the average of growth products over the entire sample is close to zero, but rises to 0.09 when considering only pairs with knowledge base similarity above 0.41. 
Regressing the growth rate products against input similarities yields a highly significant coefficient of 0.21, but the effect is extremely noisy ($R^2 = 0.004$).
The result is similar for the IPC 4-digit codes, Figure \ref{fig:growth_products}d). For the 3-digit examples, the relationship is somewhat less stable in the more disaggregated networks, see panel a) and b), but still positive.

In a second test we check if technologies with similar knowledge bases have correlated growth trajectories. To do this, we plot the pairwise knowledge base similarities in the year 2000 against the 1-year growth rate correlations of the subsequent years (2001 to 2017).
Figure \ref{fig:growth_correlations} confirms that there is a positive relationship between knowledge base similarity and growth rate correlations. 
Growth rates of patent classes tend to be correlated, with a positive average correlation of 0.26 for the CPC 4-digit classes; see Figure \ref{fig:growth_correlations}c). But when looking at the bin of technology pairs with highest knowledge base similarities, we find a substantially higher average growth rate correlation of 0.41. Thus, if two technologies rely on a similar set of knowledge, they do not only tend to experience similar growth rates, but also exhibit correlated growth trajectories over time.
The result is again similar for the IPC 4-digit case, Figure \ref{fig:growth_correlations}d), and less pronounced for the 3-digit codes, panel a) and b), which exhibit substantially higher average growth rate correlations.
While the positive relationship holds for all four cases, it should be noted that the effect is extremely noisy.

\begin{table}[!htbp] \centering 
	\caption{Results for regressing pairwise growth products against input cosine similarities. Input cosine similarities are taking from the year 2000 and growth products are computed based on the 10-year growth rates from 2000 to 2010.} 
	\label{tab:reg_products} 
	\begin{tabular}{@{\extracolsep{5pt}}lcccc} 
		\\[-1.8ex]\hline 
		\hline \\[-1.8ex] 
		& \multicolumn{4}{c}{\textit{Dependent variable:} Growth products} \\ 
		\cline{2-5} 
		\\[-1.8ex] &CPC3 &CPC4 & IPC3 & IPC4  \\ 
		\hline \\[-1.8ex] 
		Input similarity &0.117$^{***}$  &0.209$^{***}$ &0.071$^{***}$ & 0.142$^{***}$ \\ 
		&(0.011) &(0.007) & (0.011) & (0.008)  \\ 
		&  & & & \\ 
		Constant &$-$0.018$^{***}$ &$-$0.010$^{***}$ &$-$0.006$^{*}$ & 0.011$^{***}$ \\ 
		&(0.003) &(0.001) &(0.003) & (0.001) \\ 
		&  & & & \\ 
		\hline \\[-1.8ex] 
		Observations &7,117 &198,765 &6,740 & 157,198  \\ 
		R$^{2}$ &0.016 &0.004 &0.006 & 0.002  \\ 
		\hline 
		\hline \\[-1.8ex] 
		\textit{Note:}  & \multicolumn{4}{r}{$^{*}$p$<$0.1; $^{**}$p$<$0.05; $^{***}$p$<$0.01} \\ 
	\end{tabular} 
\end{table}

\begin{figure*}[!ht]
	\centering
	 \includegraphics[width=.44\textwidth]{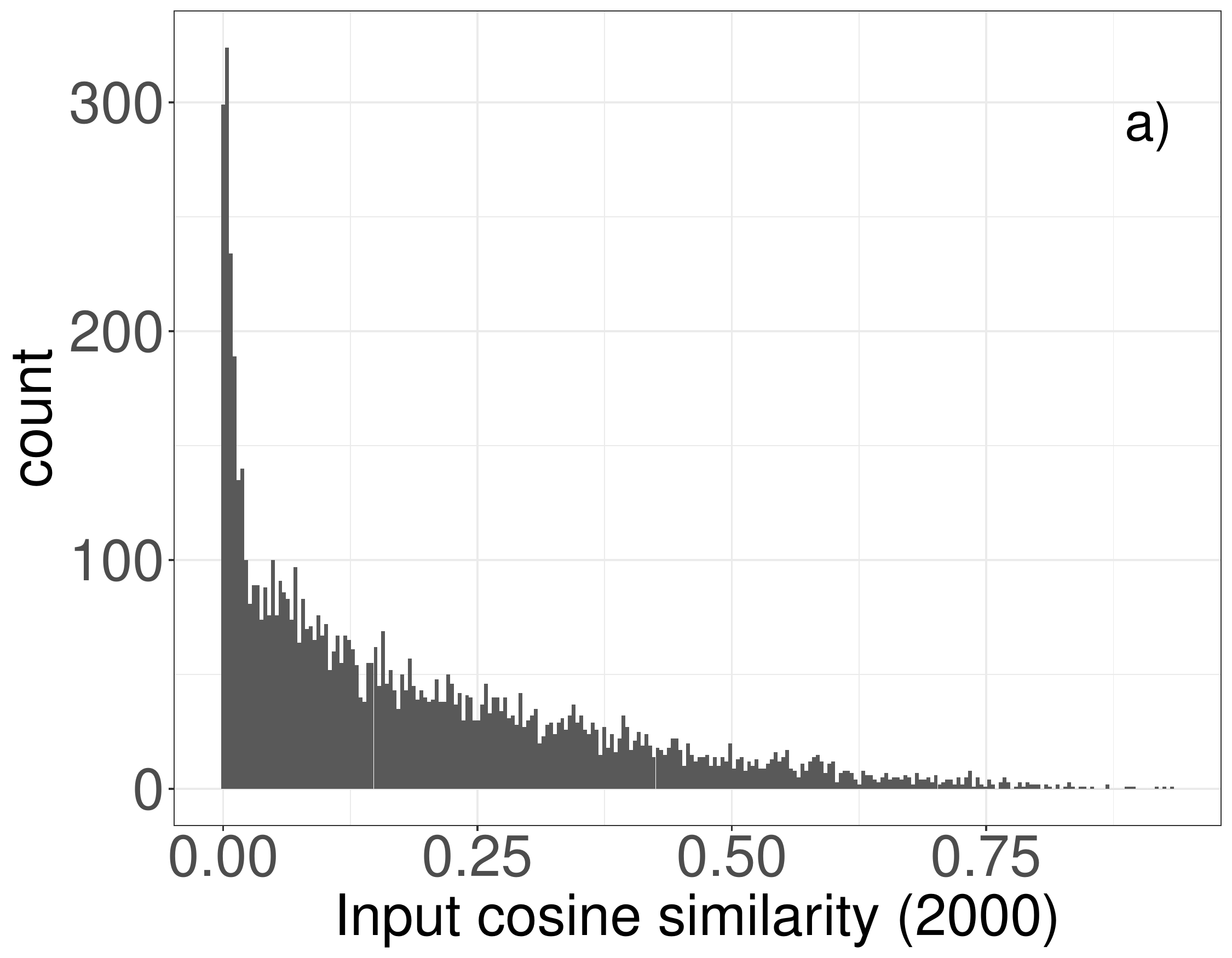}
	 \includegraphics[width=.44\textwidth]{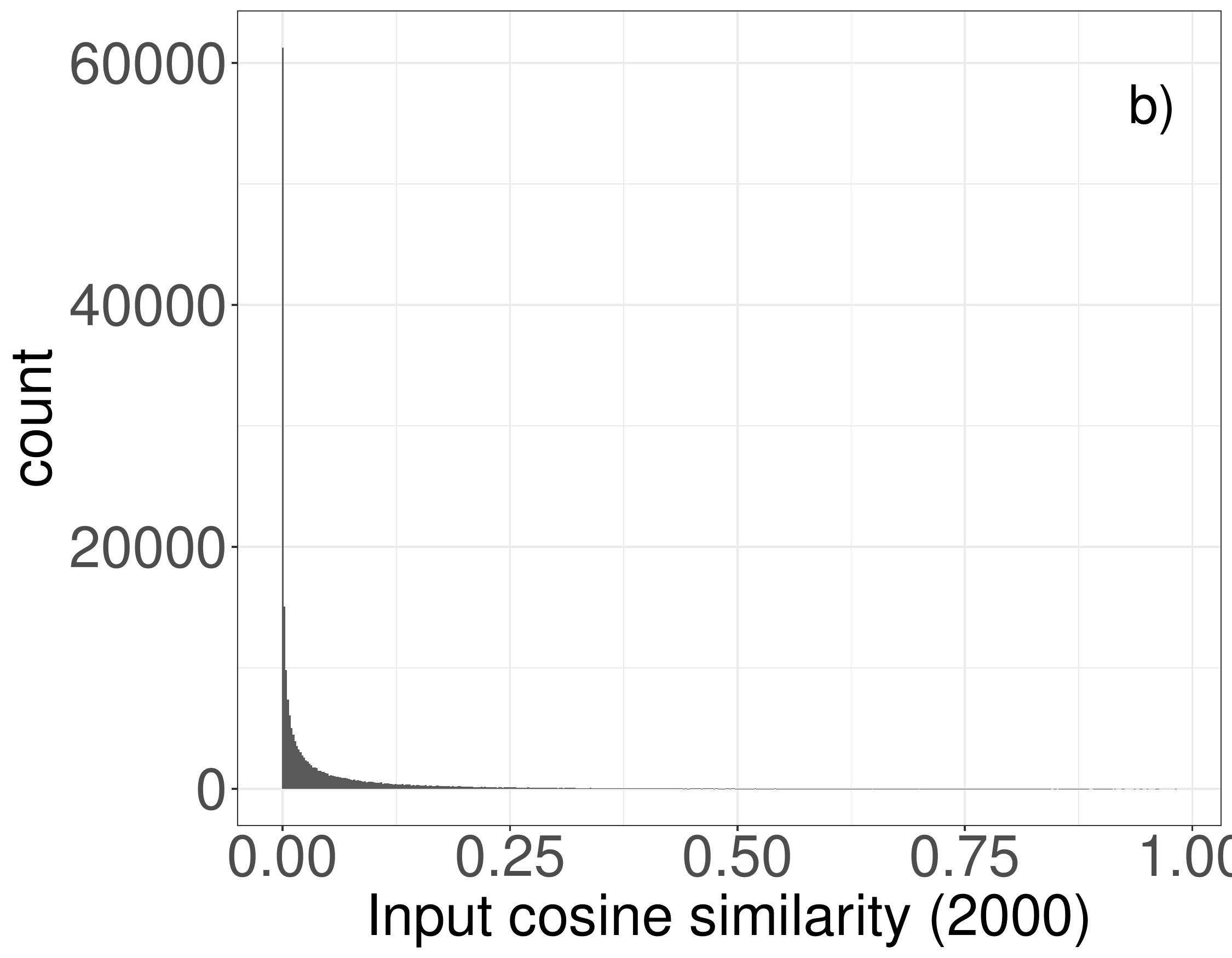}
	 \includegraphics[width=.44\textwidth]{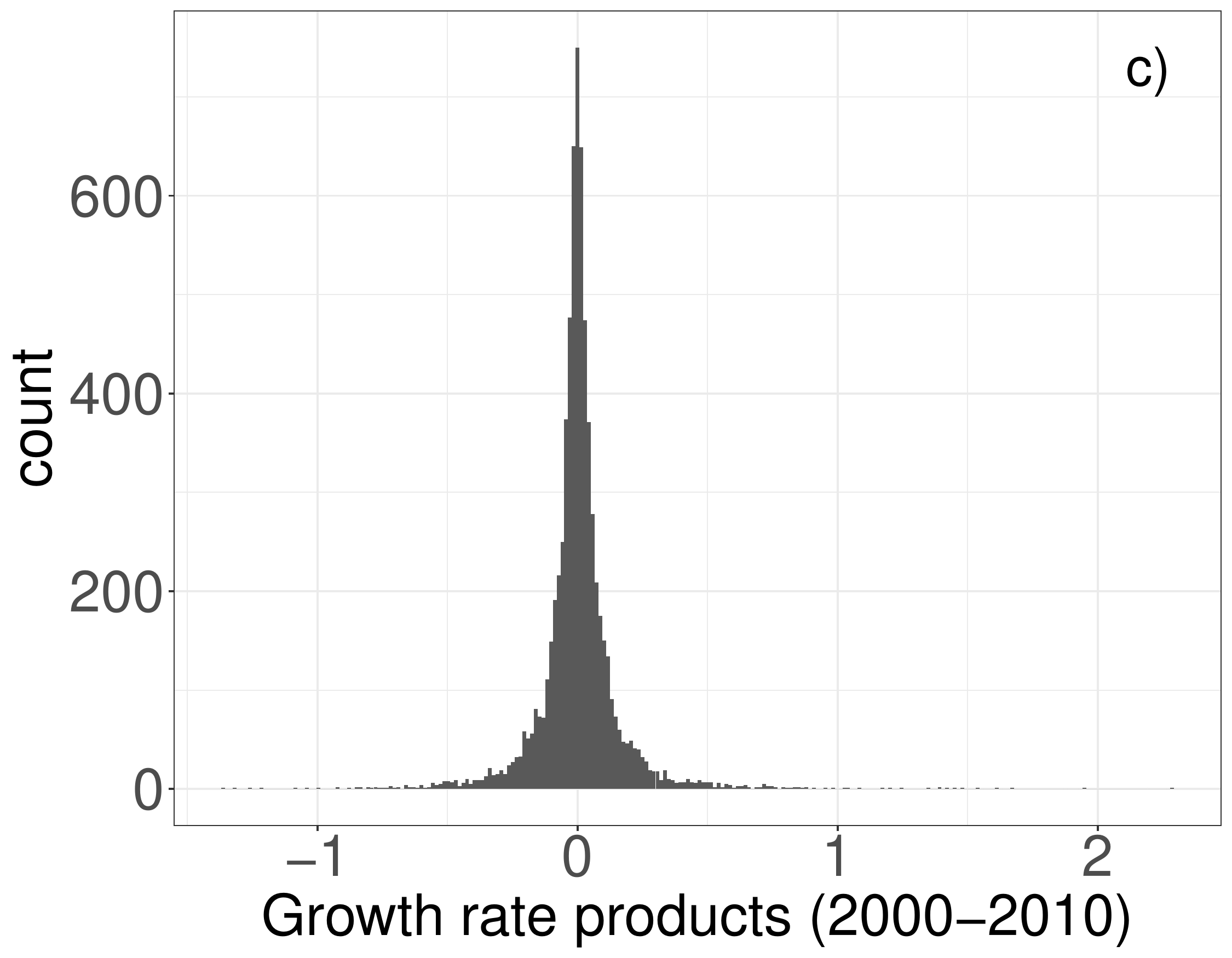}
	 \includegraphics[width=.44\textwidth]{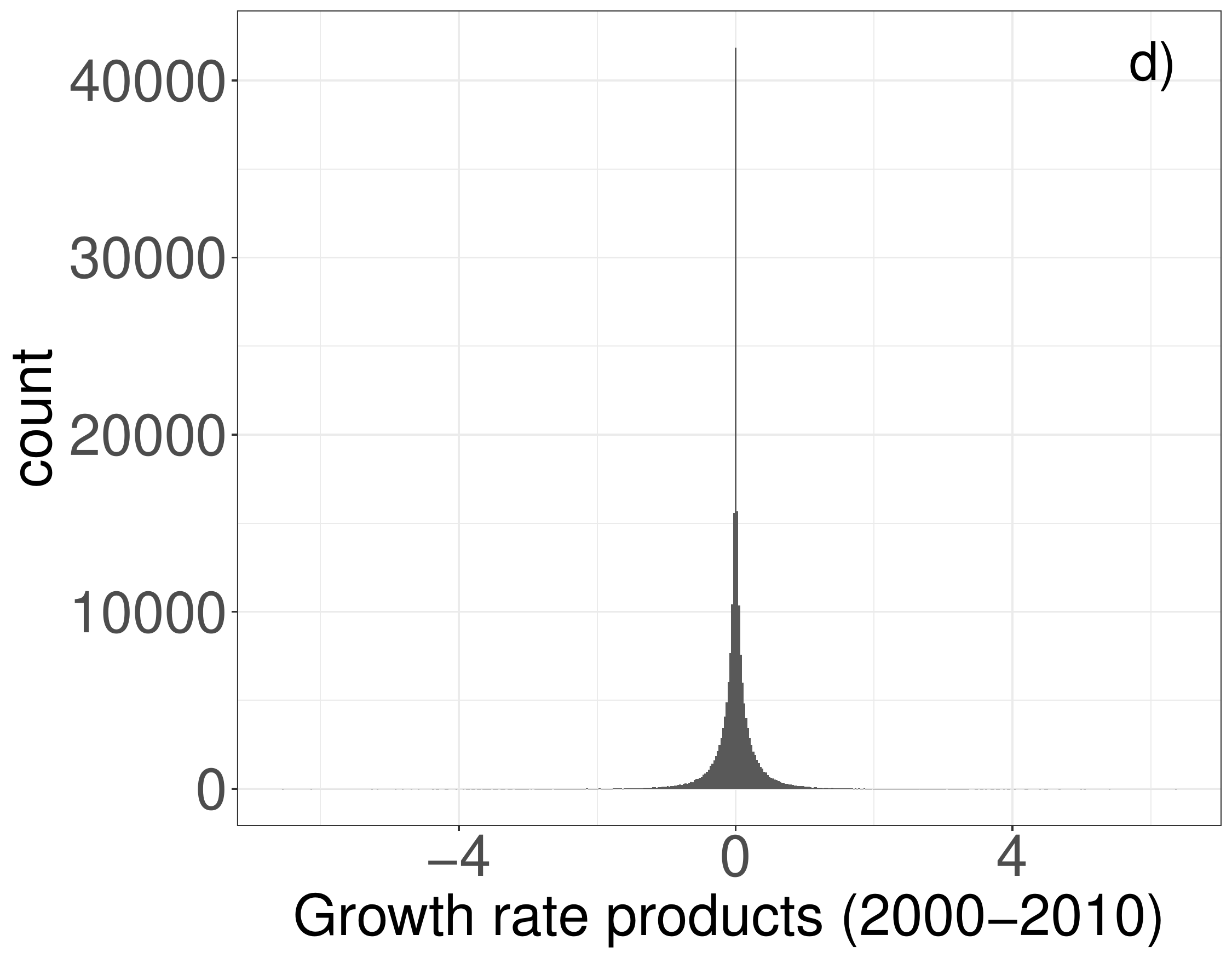}
	 \includegraphics[width=.44\textwidth]{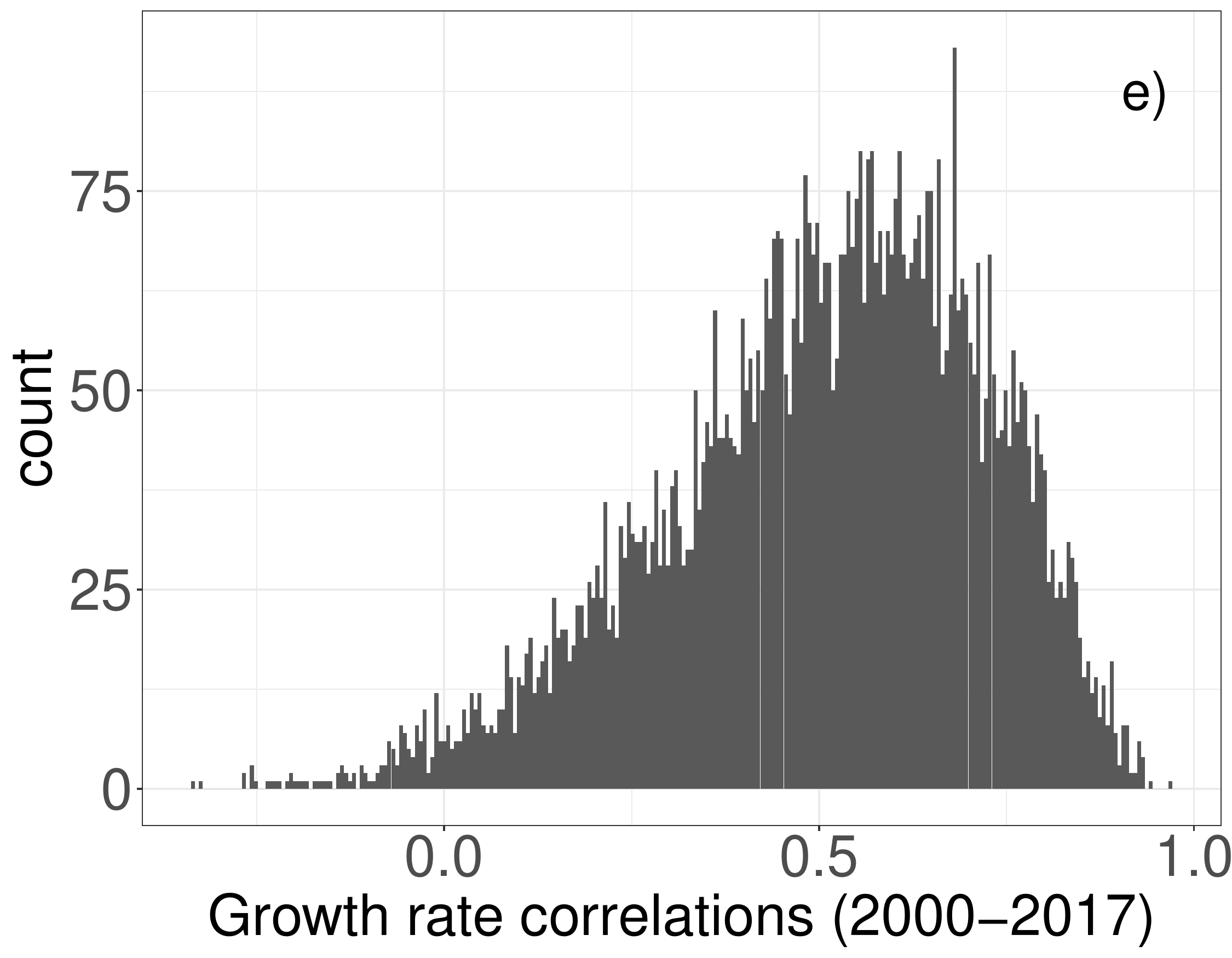}
	 \includegraphics[width=.44\textwidth]{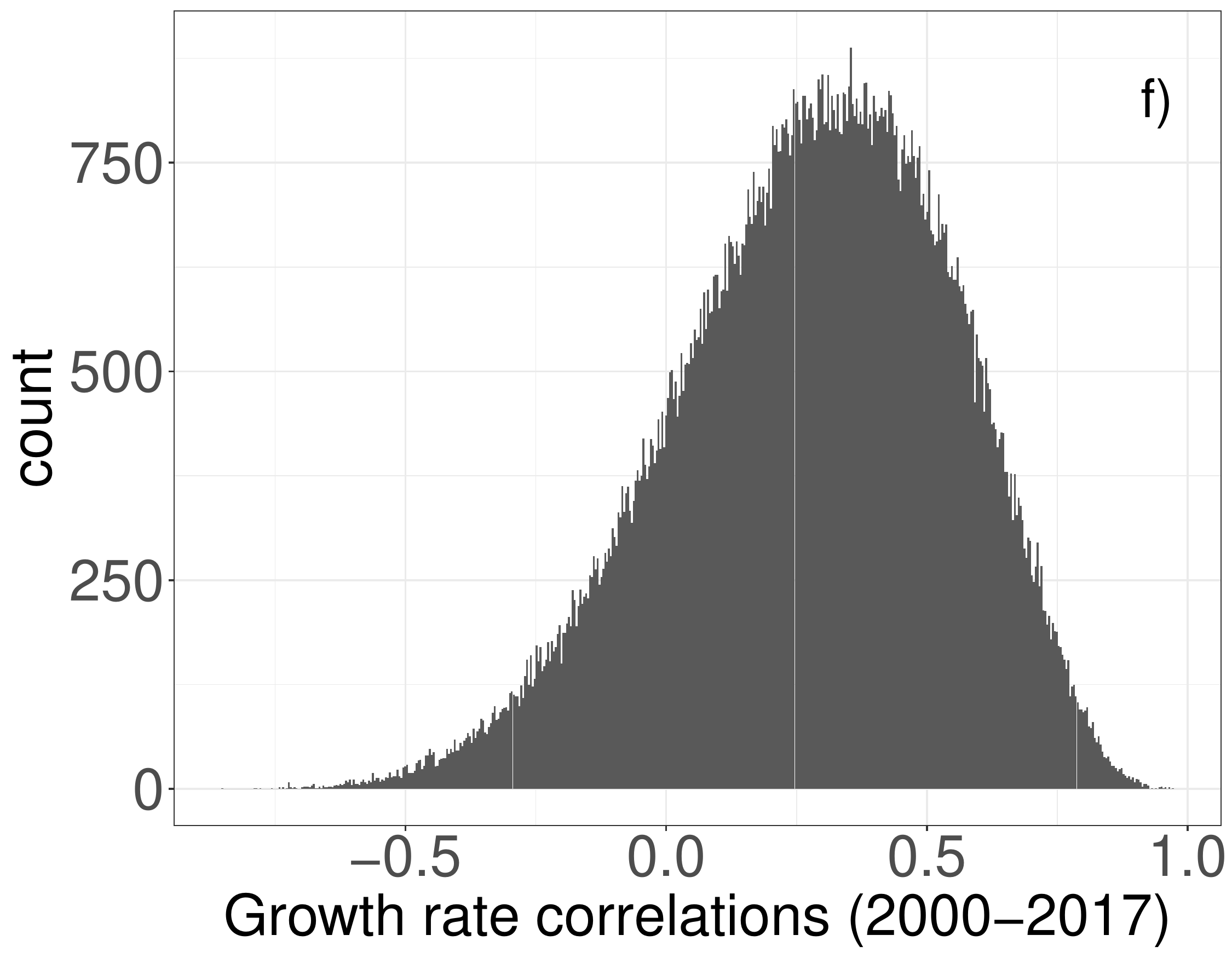}
	\caption{Histograms of input cosine similarities for a) CPC 3-digits and b) CPC 4-digits codes.
		 Histograms of growth rate products for c) CPC 3-digits and d) CPC 4-digits codes.
	  	 Histograms of growth rate correlations for e) CPC 3-digits and f) CPC 4-digits codes.
  	 }	
	\label{fig:histograms}
\end{figure*}

Figure \ref{fig:histograms} shows histograms of the knowledge input similarities, growth rate products and growth rate correlations, respectively. We show the histograms only for CPC codes, since the plots look very different for the 3- and 4-digits aggregation, but very similar for IPC and CPC codes when compared on the same aggregation level.

\clearpage

\section{Econometric estimation}

In the main text we have presented results based on the econometric model
\begin{equation}\label{eq:econometric_model}
g_{i,t} = \frac{ a_{i} }{1- \beta W_{ii,t}} + \frac{\beta }{1- \beta W_{ii,t}} \sum_{j =1}^N W_{ij,t} g_{j,t}(1-\delta_{ij}) + \epsilon_{i,t},
\end{equation}
where $\epsilon_{i,t} \sim N(0,\sigma^2)$.
Note that the econometric model is a generalization of the spatial autoregressive (SAR) model for panel data. In case of a zero diagonal matrix, $W_{ii} = 0$, the model reduces to the fixed-effects SAR with time-varying spatial component
\begin{equation}\label{eq:sar}
g_{i,t} = a_{i}  + \beta \sum_{j =1}^N W_{ij,t} g_{j,t} + \epsilon_{i,t},
\end{equation}
discussed by \cite{wang2015estimation}. Moreover, if the network is fixed in time, $W_t = W$, this model reduces to the standard fixed-effects SAR which can be estimated as outlined in \cite{elhorst2014spatial}.

In the following we will estimate different variations of these econometric models to see how robust the results in the main text are. Table \ref{t:nw_regress_all} shows the estimated parameters when fitting Equation \ref{eq:econometric_model} to the data for all four different classification schemes. When using the more aggregated technology codes (3-digits IPC and CPC), we find slightly higher ``network parameters'' $\beta$, but overall, the results are qualitatively similar for different aggregations.
 
Figure \ref{fig:relnetimpact} shows the relative network impacts $I_{network,t}$ over time for all four classifications. The figure shows that we would estimate higher network impacts in a more aggregated network, i.e. when based on 3-digits classifications. We also find that network effects are slightly larger for IPC codes than for CPC codes, if compared on the same digit-level. But the general trend of rising relative network impacts over time holds for all four classification schemes.

We next investigate how the estimated parameters change if we use fixed network specifications. Patent networks are naturally dynamic, but the vast majority of literature on network econometrics do not allow for time-varying networks. Thus, we test how the time-varying nature of our model influences the network parameter $\beta$. 
We do this by estimating the model for every network snapshot separately. We then check how the parameter $\beta$ changes as a function of the included time-fixed network. It should be pointed out that there might be an endogeneity issue if more recent networks $W_t$ are included in the model, since the network formation itself could depend on the growth of technological domains. Moreover, this estimation uses future information on the right-hand side of Equation \ref{eq:econometric_model} to explain past left-hand side variables, for example, when $W_t = W_{2017}$.
In Figure \ref{fig:fixdenetwork} we plot the coefficient $\beta$ as a function of the used network $W_t$. We show the results for CPC 3-digits and CPC 4-digits classes only, since results are similar for the IPC counterparts. The figure depicts qualitatively different results for the two aggregation levels. For the CPC 3-digits case, we find that $\beta$ exhibits a slight downward trend as a function of the time index. In the CPC 4-digits case, on the other hand, the coefficient is smaller in the early years and larger for more recent networks. Additionally, from the mid-70s on the parameter seems to stabilize around the value of 0.84 which we have estimated in the original time-varying specification.

\begin{table}[b]
	\centering
	\resizebox{\textwidth}{!}{		
		\begin{tabular}{lcccccccccccc}
			\multicolumn{4}{l}{\textit{Dependent variable: patenting growth}} \\ 
			\\[-1.8ex]\hline 
			\hline \\[-1.8ex] 
			& \multicolumn{3}{l}{\textit{CPC3}} & \multicolumn{3}{l}{\textit{CPC4}} & \multicolumn{3}{l}{\textit{IPC3}} & \multicolumn{3}{l}{\textit{IPC4}}\\ 
			\cline{2-3} \cline{5-6} \cline{8-9} \cline{11-12}
			\\[-1.8ex] & (1-y) & (5-y) & (10-y) & (1-y) & (5-y) & (10-y) & (1-y) & (5-y) & (10-y) & (1-y) & (5-y) & (10-y)\\ 
			\hline \\[-1.8ex]
			average $\hat{a}_i$ & 0.01 & 0.03 &0.05  & 0.01 & 0.03 & 0.07 & 0.01 & 0.03 &0.06 &0.01 & 0.04 &0.08\\[1ex] 
			$\hat{\beta}$ & 0.91$^{***}$ & 0.94$^{***}$ &0.93$^{***}$ & 0.84$^{***}$ & 0.90$^{***}$ & 0.89$^{***}$ & 0.92$^{***}$ &0.93$^{***}$ &0.92$^{***}$ & 0.85$^{***}$ & 0.90$^{***}$ &0.86$^{***}$\\ 
			& (0.006) & (0.010) &(0.015) & (0.006) & (0.008) & (0.011) & (0.006) & (0.011) &(0.019) & (0.006) & (0.09) & (0.014) \\[1ex] 
			$\hat{\sigma}^2$ & 0.03 & 0.08 &0.13 & 0.11 & 0.20 & 0.30 & 0.03 & 0.09 &0.14 & 0.11 & 0.24 &0.32\\ 
			%& 0 & 0 &0 & 0.0007 &  0.0030 & 0.0062 & 0 & 0 &0 & 0 & 0 &0\\[1ex] 
			\hline \\[-1.8ex]
			Log-likelihood & 2,526 & -277 &-364 & -13,383 &-5,464  &-3,612  &2,867 &-384 &-395 &-13,427 &-6,039 &-3,657\\ 
			Observations & 8,610 & 1,720 &858 & 43,651 & 8,734 & 4,348  &8,349 &1,669 &834 &42,688 &8,528 &4,427\\
			\hline 
			\hline \\[-1.8ex] 
			\textit{Note:}  & \multicolumn{12}{r}{$^{***}$p$<10^{-16}$}
		\end{tabular}
	\caption{Results from estimating the econometric network model presented in the main text (Equation \ref{eq:econometric_model}) for different classification schemes. The results are shown for 1-, 5- and 10-year growth rates. }
		\label{t:nw_regress_all}
	}	
\end{table}

\begin{figure*}[!ht]
	\centering
	\includegraphics[width=.75\textwidth]{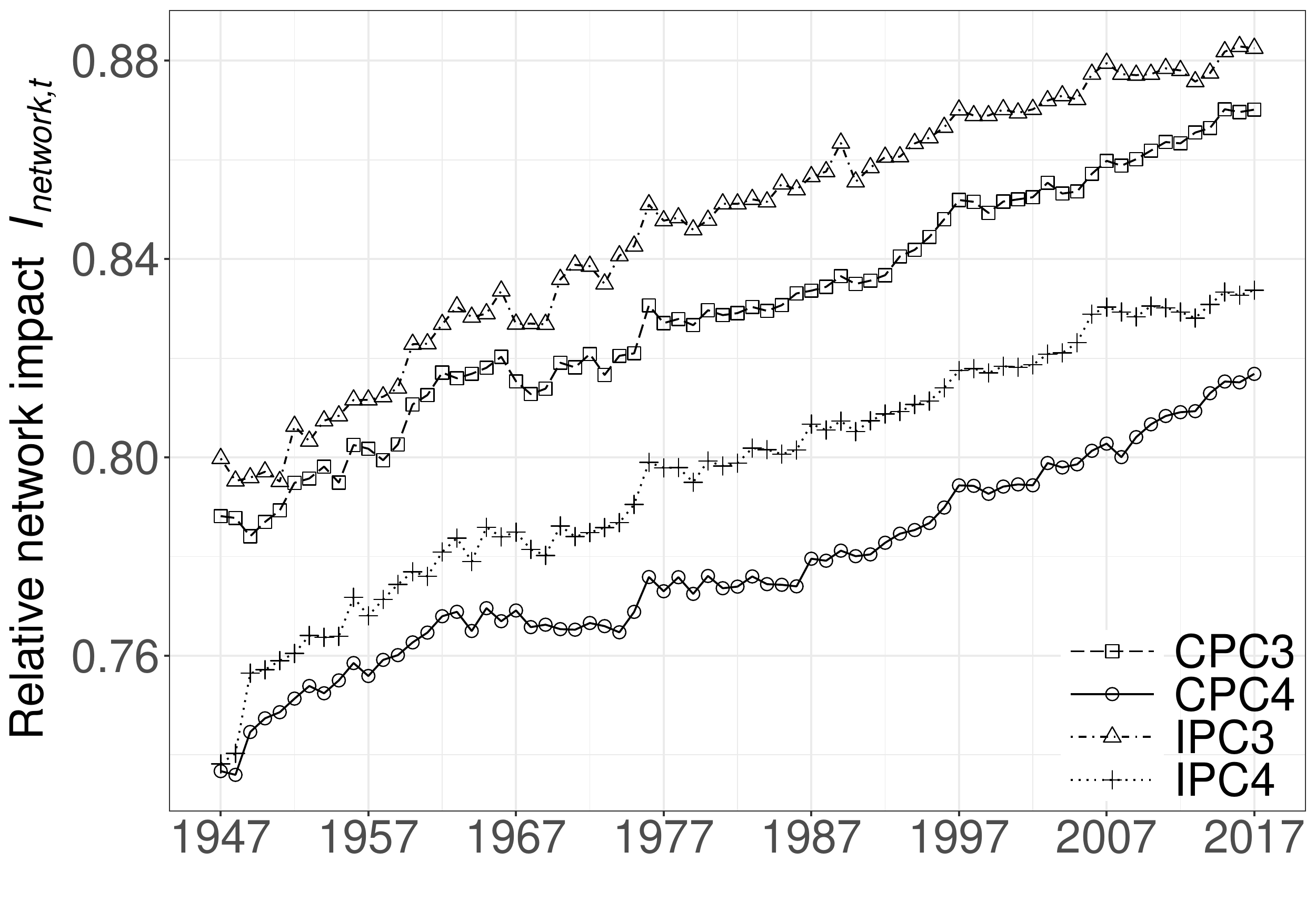}
	\caption{Relative network impacts over time for different classification schemes. Results are based on 1-year growth rates.}	
	\label{fig:relnetimpact}
\end{figure*}

\begin{figure*}[!ht]
	\centering
	\includegraphics[width=.75\textwidth]{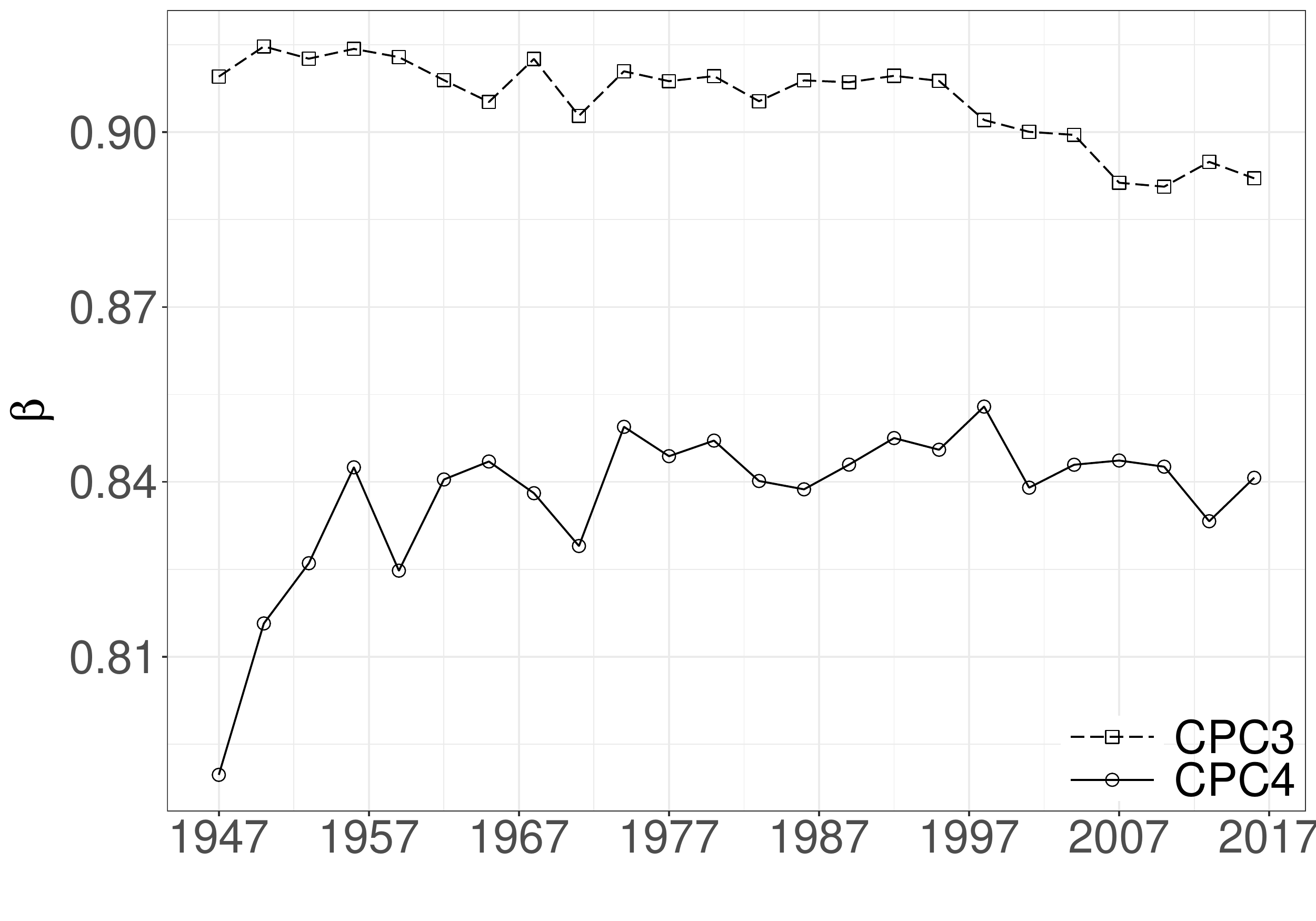}
	\caption{Parameter $\beta$ when estimated with fixed network. The x-axis is the time index of the used network $W_t$. For example, if $t=1947$, the model was estimated with using $W_{1947}$. Results are based on 1-year growth rates.} 	
	\label{fig:fixdenetwork}
\end{figure*}

\begin{table}[!b]
	\centering
	\resizebox{\textwidth}{!}{		
		\begin{tabular}{lcccccccccc}
			\multicolumn{11}{r}{\textit{Dependent variable: patenting growth}} \\ 
			\\[-1.8ex]\hline 
			\hline \\[-1.8ex] 
			&(a)  & (b) & (c) & (d) & (e) & (f) & (g) & (h) & (i) & (j)\\[1ex]
			&CPC3  & CPC3 & CPC3 & CPC3 & CPC3 & CPC4 & CPC4 & CPC4 & CPC4 & CPC4\\
			\hline \\[-1.8ex] 
			average $\hat{a}_i$ & 0.011 & 0.023 &0.428  & 0.400 & 0.173 & 0.015 & 0.024 & 0.560 &0.357 &0.127 \\[1ex] 
			%average time effects & & 0.021 &0.428  & 0.400 & 0.171 & & 0.024 & 0.567 &0.357 &0.125	 \\[1ex] 
			$\hat{\beta}$ & 0.830$^{***}$ & 0.509$^{***}$ &0.513$^{***}$ & 0.514$^{***}$ & 0.511$^{***}$ & 0.756$^{***}$ & 0.483$^{***}$ &0.476$^{***}$ &0.483$^{***}$ & 0.482$^{***}$ \\ 
			&(0.009) & (0.018) &(0.017) & (0.018) & (0.018) & (0.008) & (0.105) & (0.09) &(0.105) & (0.105) \\[1ex] 
			Patent rates &  &  &$-0.069^{***}$ &  &  & &  &$-0.128^{***}$ & &  \\ 
			& &  &(0.003) &  &  &  & &(0.002) & &  \\[1ex]
			Patent stock &  &  & & -0.038$^{***}$ &  & & & &$-0.041^{***}$ &  \\ 
			& & & &(0.004) & & &  & &$(0.003)$ &  \\[1ex]
			Discount. pat. stock &  &  & & & $-0.033^{***}$ &  & & & &$-0.035^{***}$ \\ 
			&  &  & & & (0.004) & &  &  & & $(0.003)$ \\[1ex]
			\hline \\[-1.8ex]
			Fixed effects & Yes & Yes & Yes & Yes & Yes  & Yes  & Yes &Yes & Yes & Yes\\ 
			Time effects & No & Yes & Yes & Yes & Yes  &No  & Yes &Yes & Yes & Yes\\
			Log-likelihood & 2,527 & 2,721 & 2,912 &2,763 &2,757  &-13,374  &-12,923 &-11,452 &-12,854 &-12,864\\ 
			Observations & 8,601 & 8,601 &8,601 & 8,601 &8,601 & 43,651  &43,651 &43,651 &43,651 &43,651\\
			\hline 
			\hline \\[-1.8ex] 
			\textit{Note:}  & \multicolumn{10}{r}{$^{***}$p$<10^{-16}$}
		\end{tabular}
		\caption{Results from estimating the SAR with time-varying network component and additional exogenous regressors (Equation \ref{eq:sar_vector}) for different classification schemes. The first five columns (a-e) show the estimated parameters for the CPC 3-digits aggregation and the last five columns (f-j) for the CPC 4-digits aggregation. The results are based on 1-year growth rates. 
		}
		\label{t:sar_regress_all}
	}	
\end{table}

We also estimated the SAR model with time-varying network component (Equation \ref{eq:sar}) for comparison. The SAR model is convenient for demonstration purposes, since extensive literature make it a ``standard'' model for studying network effects. Due to its simpler form, it is also straightforward to include further regressors to check for robustness. We therefore estimate Equation \ref{eq:sar} also with additional regressors such as time effects and technological class size measured as yearly patenting rate or as cumulative patent stock (also including a discount factor).

The estimation of the SAR model simplifies in comparison to the full network model of Equation \ref{eq:econometric_model}. To estimate Equation \ref{eq:sar} including exogenous regressors, we stack all elements $g_{i,t}$ into the $NT$-dimensional vector $\bar{g} := (g_{1,1}, g_{2,1}, ..., g_{N,1}, g_{1,2}, ..., g_{N,T})$. In the same manner we obtain the vector $\bar{\epsilon}$. By defining the $T$-dimensional vector of ones $\iota_T$ and the block-diagonal matrix $W := diag(W_1, ..., W_T)$, we can rewrite Equation 2 into matrix notation
\begin{equation} \label{eq:sar_vector}
	\bar{g} = (\iota_T \otimes \mathbb{I}_N)a  + \beta W \bar{g} + \bar{X} \delta + \bar{ \epsilon },
\end{equation}
where $\bar{X}$ is the $NT \times K$ matrix of exogenous regressors and $\delta$ the corresponding $K$-dimensional vector of coefficients.
After subtracting the time average, Equation \ref{eq:sar_vector} simplifies to 
\begin{equation} \label{eq:sar_vector_demean}
g = \beta W g + X \delta + \epsilon,
\end{equation}
where $g$, $X$ and $\epsilon$ are the time-demeaned versions of $\bar{g}$, $\bar{X}$ and $\bar{ \epsilon }$, respectively. 
The log-likelihood of the time-demeaned model then reads
\begin{equation} \label{eq:loglik}
\text{LogL} = -\frac{NT}{2} \ln(2 \pi \sigma^2)  + \sum_{t=1}^{T} \ln |\mathbb{I} - \beta {W}_t| - \frac{1}{2 \sigma^2}  \epsilon^\top \epsilon,
\end{equation}
with $\epsilon = (\mathbb{I}_{NT} - \beta   {W}) g - X \delta$ and the parameter $\beta$ is obtained by solving
\begin{align*}
\underset{\beta}{\text{argmax}} \quad
& constant  + \sum_{k=1}^{NT} \ln (1 - \beta e_k) \\
&-  \frac{NT}{2} \ln  ( g^\top g - 2 \beta g^\top W g  - \beta^2 g^\top W^\top W g  - A^\top X  (X^\top X)^{-1} X^\top Ag,
\end{align*}
where $A := (\mathbb{I}_{NT} - \beta W )$ and $e_k$ represents eigenvalues of matrix $W$. Under standard regularity assumptions, we derive the variance-covariance matrix by inverting the Fisher information matrix which yields
\begin{equation}
var(\sigma^2, \beta, \delta) = 
\begin{bmatrix}
\frac{NT}{\sigma^4} & -\mathbb{E} \left[ \frac{\partial^2 LogL}{ \partial \sigma^2 \partial \beta } \right] &0_K^\top \\
 \cdot & -\mathbb{E} \left[ \frac{\partial^2 LogL}{ \partial \beta^2 } \right] & \frac{1}{\sigma^2} X^\top W A^{-1} X \delta \\
 \cdot& \cdot & \frac{1}{\sigma^2} X^\top X \\
\end{bmatrix}^{-1},
\end{equation} 
where $-\mathbb{E} \left[ \frac{\partial^2 LogL}{ \partial \delta^2 } \right] = trace(A^{-1}W A^{-1}W) + \frac{1}{2\sigma^2} trace \left( [A^{-1} X \delta \delta^\top X^\top A^{\top -1} + \sigma^2 A^{-1}  A^{\top -1}] W^\top W \right)$ and 
$-\mathbb{E} \left[ \frac{\partial^2 LogL}{ \partial \sigma^2 \partial \beta } \right] = 
\frac{1}{\sigma^2} trace \left( A^\top W [ A^{-1} X \delta \delta^\top X^\top A^{\top -1} + \sigma^2 A^{-1}  A^{\top -1}]  + \delta^\top X^\top W A^{-1}X\delta \right) $. $0_K$ denotes a $K$-dimensional vector of zeros and matrix elements indicated by the dots are left out here since they are trivially filled due to symmetry.

In Table \ref{t:sar_regress_all} we show the results of estimating the SAR with time-varying network component for the CPC 3-digits, columns (a) to (e), and CPC 4-digits aggregation, columns (f) to (j). We see again that the network coefficient $\beta$ is estimated to be larger if the more aggregated networks are used. We also include time effects in the regression which reduces the network parameter in the CPC 4-digits case from 0.76 to roughly 0.48. The coefficient is fairly robust against adding further regressors such as including logged patenting rates, columns (c) and (h), logged cumulative patent stock, (d) and (i), or logged discounted cumulative patent stock, (e) and (j), where we have used a 15\% yearly discount factor as suggested by \cite{hall2005market}. The exogenous regressors are highly significant and have a negative sign, indicating that higher patenting rates and larger knowledge stocks at time point $t$ come with smaller patenting rate growth in the following year. The average magnitude of the fixed effects increases as a consequence of including exogenous regressors.

\section{Predicting innovation dynamics}

In the main text we have shown that predictions of patenting rates in technological classes can be significantly improved if network effects are taken into account.
In this section we present further details on predicting patenting rates with the introduced model. Results shown in the main text are based on the CPC 4-digits codes. Here, we show how the model predictions change when using alternative aggregation levels. Moreover, we discuss further aspects of the presented results and investigate alternative forecast benchmarks.

We first focus on the ARIMA forecasts. As outlined in the main text, we have fitted every $(p,q)$ combination up to order (5,5) of the ARIMA($p$,1,$q$) model for each (logged) time series in the training set (1947-1987) and used the parameter fits to forecast the time series in the validation set (1988-2002).  We then refitted the ARIMA($p$,1,$q$) which yielded the smallest median absolute percentage error in the validation set to the data up to 2002 to forecast the test set from 2003 to 2017. Table \ref{t:arima} gives an overview of the best ($p$,$q$) combinations for different aggregation methods.
In the CPC 4-digit case discussed in the main text, almost 25\% of model specifications use less than two AR and MA terms, with the the geometric random walk ($p=q=0$) alone yielding the best forecasts already in 12\% percent of all cases. But we also find time series where higher-order models are preferred. Similarly for alternative aggregations, the geometric random walk model is the most frequent choice.

\begin{table}[!b]
	\begin{minipage}{.5\linewidth}
		\centering
				\begin{tabular}{|c|r|rrrrrr|}
			\hline 
			\multicolumn{2}{|c|}{CPC3}& \multicolumn{6}{c|}{MA order $q$} \\
			\cline{1-8}
			\multirow{8}*{\rotatebox{90}{AR order $p$}}		
			& & 0 & 1 & 2 & 3 & 4 & 5 \\ 
			\cline{3-8}
			&0 &0.15 &0.01 &0.03 &0.03 &0.02 &0.02 \\
			&1 &0.04 &0.00 &0.01 &0.00 &0.02 &0.02 \\
			&2 &0.03 &0.02 &0.02 &0.03 &0.02 &0.04 \\
			&3 &0.03 &0.02 &0.03 &0.00 &0.02 &0.02 \\
			&4 &0.02 &0.02 &0.01 &0.02 &0.03 &0.07 \\
			&5 &0.02 &0.02 &0.01 &0.03 &0.05 &0.07 \\
			\hline  
		\end{tabular}
		\begin{tabular}{|c|r|rrrrrr|}
			\hline 
			\multicolumn{2}{|c|}{IPC3}& \multicolumn{6}{c|}{MA order $q$} \\
			\cline{1-8}
			\multirow{8}*{\rotatebox{90}{AR order $p$}}		
			& & 0 & 1 & 2 & 3 & 4 & 5 \\ 
			\cline{3-8}
			&0 &0.10 &0.03 &0.04 &0.02 &0.03 &0.02 \\
			&1 &0.01 &0.03 &0.02 &0.00 &0.03 &0.03 \\
			&2 &0.08 &0.00 &0.02 &0.01 &0.01 &0.04 \\
			&3 &0.03 &0.00 &0.02 &0.04 &0.02 &0.03 \\
			&4 &0.04 &0.03 &0.01 &0.03 &0.03 &0.03 \\
			&5 &0.01 &0.05 &0.02 &0.02 &0.08 &0.05 \\
			\hline  
		\end{tabular}
	\end{minipage}%
	\begin{minipage}{.5\linewidth}
		\centering
					\begin{tabular}{|c|r|rrrrrr|}
			\hline 
			\multicolumn{2}{|c|}{CPC4}& \multicolumn{6}{c|}{MA order $q$} \\
			\cline{1-8}
			\multirow{8}*{\rotatebox{90}{AR order $p$}}		
			& & 0 & 1 & 2 & 3 & 4 & 5 \\ 
			\cline{3-8}
			&0 & 0.12 & 0.04 & 0.03 & 0.02 & 0.03 & 0.03 \\
			&1 & 0.06 & 0.02 & 0.01 & 0.02 & 0.02 & 0.04 \\ 
			&2 & 0.02 & 0.02 & 0.02 & 0.03 & 0.03 & 0.03 \\ 
			&3 & 0.03 & 0.01 & 0.02 & 0.01 & 0.03 & 0.03 \\ 
			&4 & 0.03 & 0.02 & 0.03 & 0.01 & 0.03 & 0.03 \\ 
			&5 & 0.03 & 0.03 & 0.01 & 0.03 & 0.03 & 0.02 \\ 
			\hline  
		\end{tabular}	
		\begin{tabular}{|c|r|rrrrrr|}
			\hline 
			\multicolumn{2}{|c|}{IPC4}& \multicolumn{6}{c|}{MA order $q$} \\
			\cline{1-8}
			\multirow{8}*{\rotatebox{90}{AR order $p$}}		
			& & 0 & 1 & 2 & 3 & 4 & 5 \\ 
			\cline{3-8}
			&0 &0.13 &0.05 &0.02 &0.02 &0.02 &0.04 \\
			&1 &0.05 &0.02 &0.01 &0.02 &0.02 &0.04 \\
			&2 &0.02 &0.02 &0.02 &0.02 &0.03 &0.03 \\
			&3 &0.02 &0.00 &0.01 &0.03 &0.02 &0.03 \\
			&4 &0.03 &0.01 &0.01 &0.02 &0.03 &0.03 \\
			&5 &0.04 &0.04 &0.03 &0.03 &0.04 &0.03 \\
			\hline  
		\end{tabular}
	\end{minipage} 
\caption{Distribution of $(p,q)$ pairs yielding the best ARIMA forecasts in the validation set. $p$ is the number of autoregressive terms and $q$ the number of moving average terms in the ARIMA model. $p=q=0$ is simply the geometric random walk with drift}
\label{t:arima}
\end{table}

\begin{figure}
	\caption{Prediction results based on \textbf{CPC 3-digit codes}.
		a) Average predictability gain based on $PG1$. Shaded areas indicate twice the standard error. b) Median predictability gain based on $PG1$. Shaded areas indicate the interquartile range.
		c) Average predictability gain based on $PG2$. Shaded areas indicate twice the standard error. d) Median predictability gain based on $PG2$. Shaded areas indicate the interquartile range.
	}
	\includegraphics[width=\textwidth]{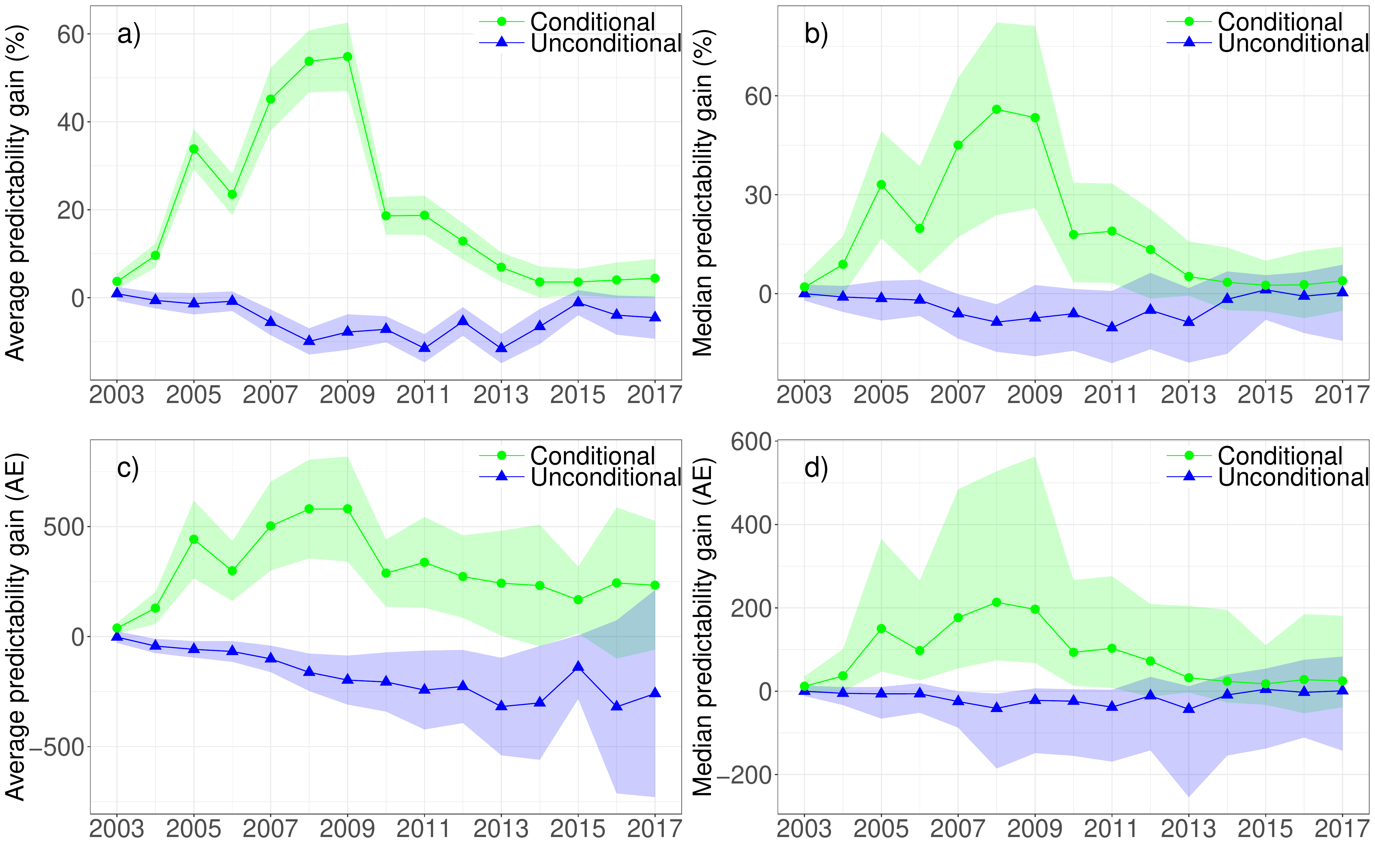}\label{fig:pg_cpc3}
		\caption{Prediction results based on \textbf{CPC 4-digit codes}.
		a) Average predictability gain based on $PG1$. Shaded areas indicate twice the standard error. b) Median predictability gain based on $PG1$. Shaded areas indicate the interquartile range.
		c) Average predictability gain based on $PG2$. Shaded areas indicate twice the standard error. d) Median predictability gain based on $PG2$. Shaded areas indicate the interquartile range.}
	\includegraphics[width=\textwidth]{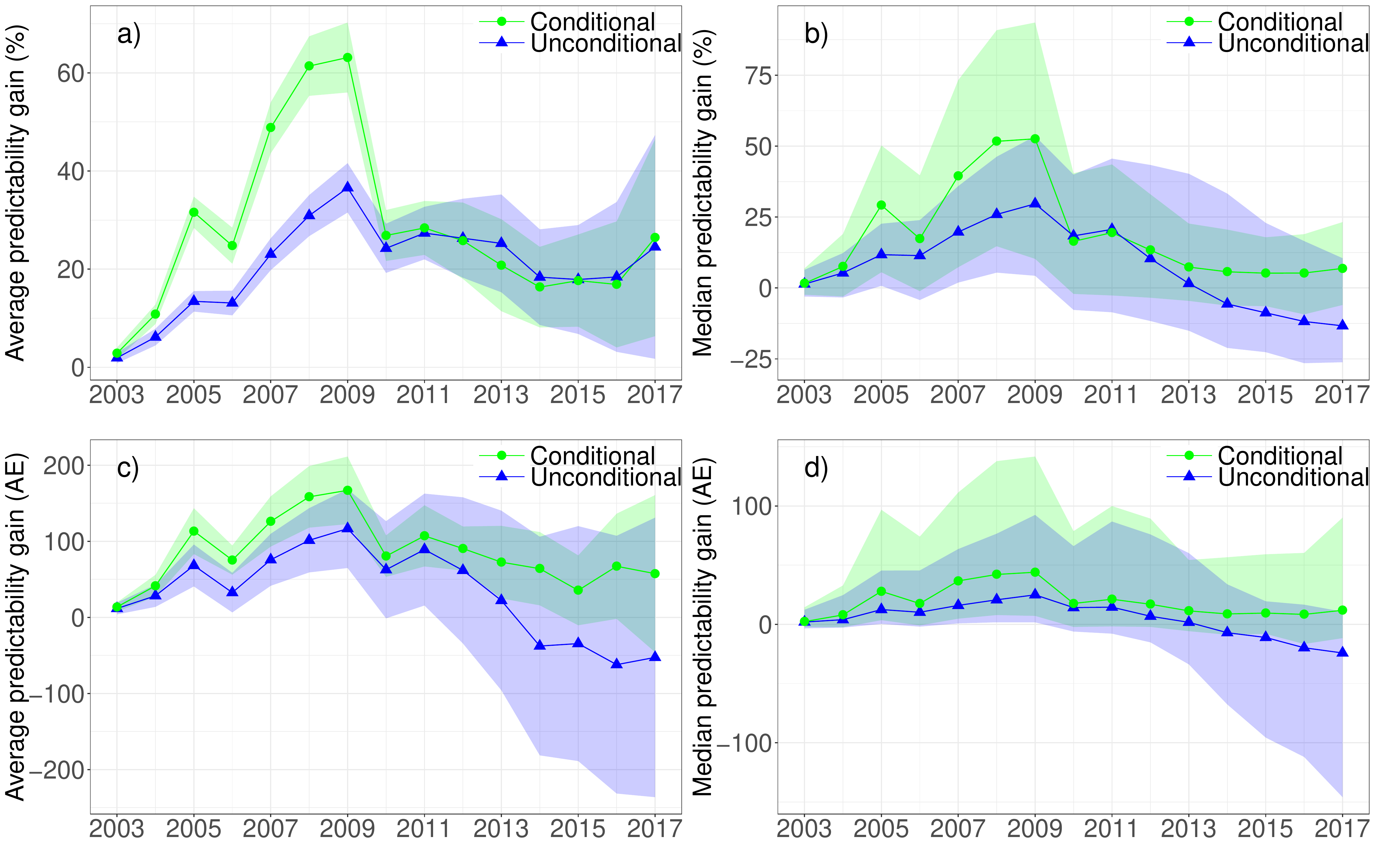}\label{fig:pg_cpc4}
\end{figure}

\begin{figure}
	\caption{Prediction results based on \textbf{IPC 3-digit codes}.
		a) Average predictability gain based on $PG1$. Shaded areas indicate twice the standard error. b) Median predictability gain based on $PG1$. Shaded areas indicate the interquartile range.
		c) Average predictability gain based on $PG2$. Shaded areas indicate twice the standard error. d) Median predictability gain based on $PG2$. Shaded areas indicate the interquartile range.
	}
	\includegraphics[width=\textwidth]{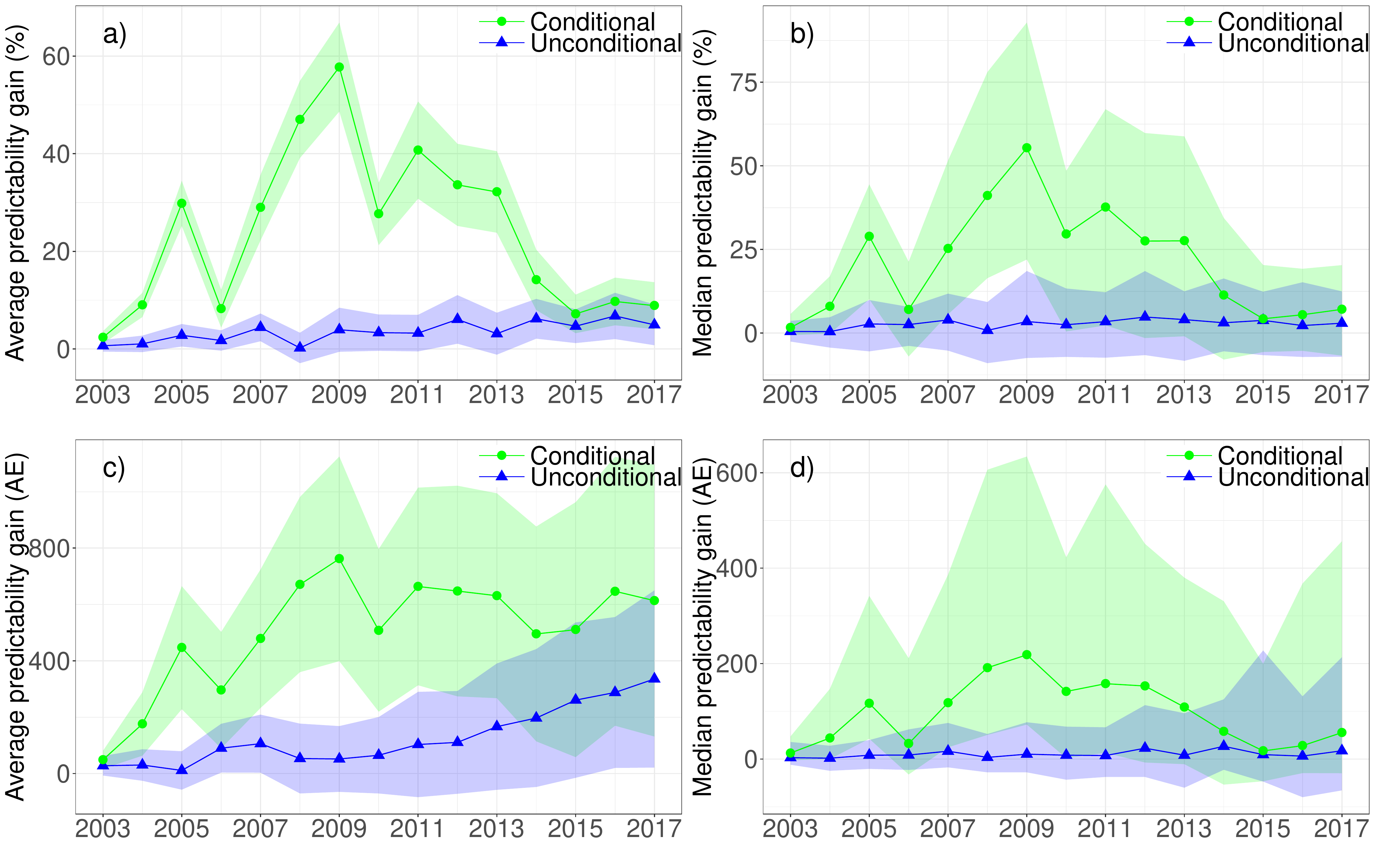}\label{fig:pg_ipc3}
	\caption{Prediction results based on \textbf{IPC 4-digit codes}.
		a) Average predictability gain based on $PG1$. Shaded areas indicate twice the standard error. b) Median predictability gain based on $PG1$. Shaded areas indicate the interquartile range.
		c) Average predictability gain based on $PG2$. Shaded areas indicate twice the standard error. d) Median predictability gain based on $PG2$. Shaded areas indicate the interquartile range.}
	\includegraphics[width=\textwidth]{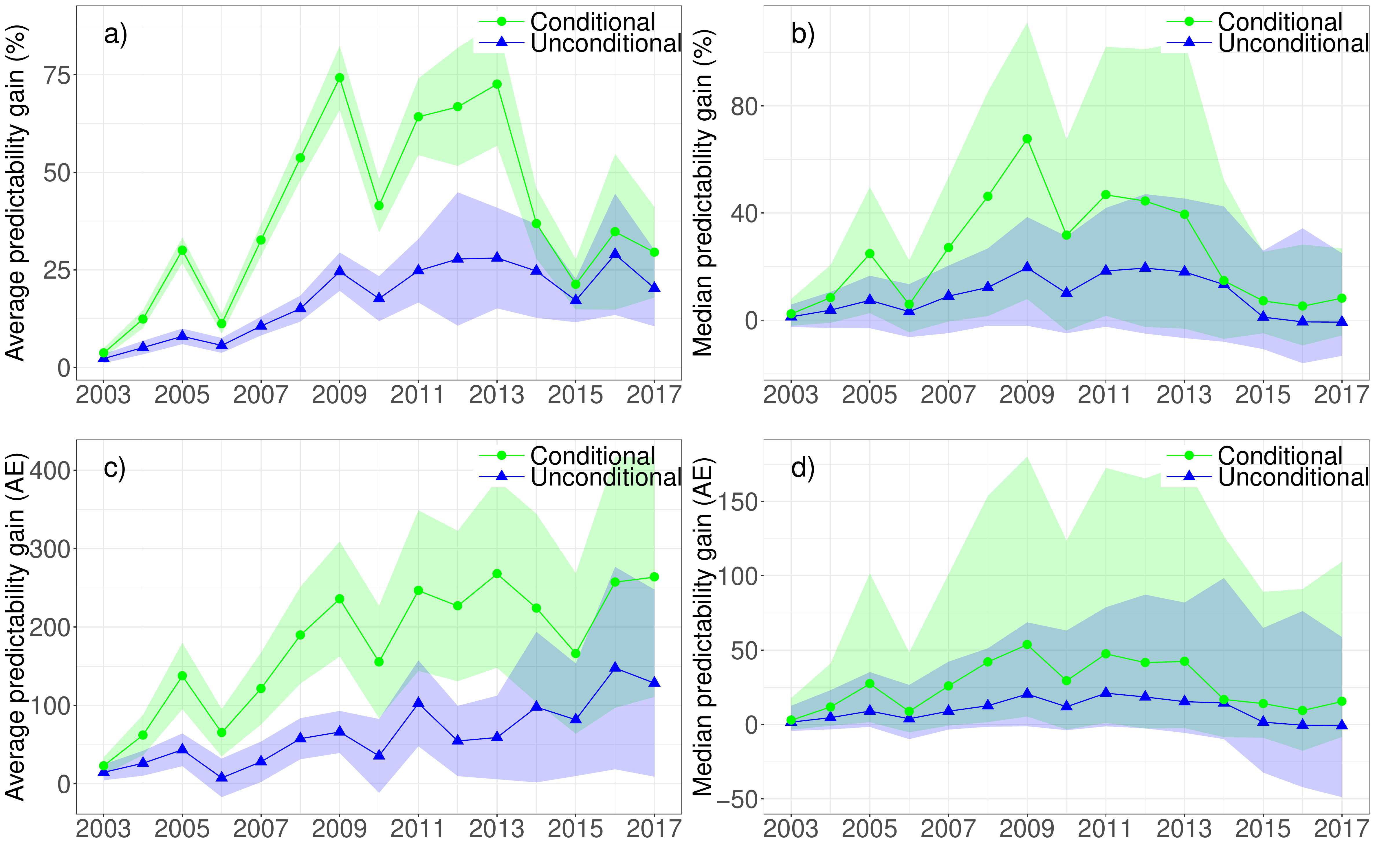}\label{fig:pg_ipc4}
\end{figure}

To obtain the unconditional forecasts, a similar model selection procedure is applied. Here, we first estimated the network model to the data in the training set and choose the parameter $k'$ such that the median absolute percentage error is minimized in the validation set. The model is then refitted to the data, including training and validation set (1947-2002), to predict patenting in the test set from 2003 to 2017 with given $k'$. When doing this for the four different technology aggregations, we obtain the following results: $k'=8$ for CPC 3-digits, $k'=2$ for CPC 4-digits, $k'=5$ for IPC 3-digits and $k'=1$ for IPC 4-digits. Thus, the applied model selection procedure always prefers $k'>0$, i.e. network effects are included in the predictions.

As in the main text, we then compare forecasts based on the predictability gain measure 
\begin{align}
PG1_{i,t} &=  \frac{ | {P}_{i,t} - \hat{P}_{i,t}^{ARIMA} | - | {P}_{i,t} - \hat{P}_{i,t}^{network} | }{ |{P}_{i,t} | }. \label{eq:pg}
\end{align} 
We also check the predictability gains based on the absolute errors instead of the absolute percentage errors by plotting
\begin{align}
PG2_{i,t} &=  | {P}_{i,t} - \hat{P}_{i,t}^{ARIMA} | - | {P}_{i,t} - \hat{P}_{i,t}^{network} |
\label{eq:pg_ae}
\end{align} 
over time. The results for both metrics and all four aggregation schemes are shown in Figures \ref{fig:pg_cpc3} - \ref{fig:pg_ipc4}. 
In every figure, panel a) plots the average predictability gains based on Equation \ref{eq:pg} as in the main text. Panel b) better illustrates the location of the predictability gain distribution by plotting the median and the interquartile range of Equation \ref{eq:pg}.
Averages and standard errors of Equation \ref{eq:pg_ae} are plotted in panel c) and the corresponding medians and interquartile ranges are shown in panel d).
We discuss the results for each aggregation scheme separately.\\[0.1cm]
 
\noindent
\textbf{CPC 3-digit codes (Figure \ref{fig:pg_cpc3}):} \\
Similarly as the results shown in the main text, predictions can be substantially improved in the conditional forecasts where the growth rates of a focal technology's neighborhood is known. This holds true for averages and medians, regardless of the considered performance metric. In contrast to the result in the main text, we find that the unconditional forecasts are not able to beat the ARIMA model systematically. The main reason why in this scenario the unconditional predictions underperform is that the optimal tuning parameter $k'$ is not stable between the validation and test set. While a large $k'=8$ is chosen based on the validation set forecasts, this choice is suboptimal when predicting patenting rates in the test set. Note that we also point out the possibility of this problem in the main text. \\[0.1cm]

\noindent
\textbf{CPC 4-digit codes (Figure \ref{fig:pg_cpc4}):} \\
Panel a) is the same plot as presented in the main text which plots average $PG1$ (solid lines) and twice the standard errors (shaded areas) over time. In panel b) we see that the largest part of the interquartile range of the predictability gain is positive, particularly when looking at the conditional forecasts. But also for the unconditional forecasts we find that most of the shaded area is above the zero line, although the median is getting closer to zero and slightly negative for the very long-term forecasts. Panels c) and d) show that the average predictability gain based on absolute errors exhibits somewhat more year to year fluctuations. On average, the network model outperforms the ARIMA forecasts in both prediction exercises, with the conditional forecasts yielding positive prediction gains in every year and the unconditional forecasts in 13 out of 15 years. Panel d) plots the median predictability gain and the corresponding interquartile range for $PG2$ which resembles the corresponding plot for $PG1$. As seen before, the median predictability gain is always positive in the conditional forecast scenario and positive in 12 out of 15 time points.\\[0.1cm]

\noindent
\textbf{IPC 3-digit codes (Figure \ref{fig:pg_ipc3}):} \\
The average predictability gains over the ARIMA models in the conditional forecast setting reach almost 60\% in 2009 when measured in percentages (panel a) or 750 patents when looking at the absolute errors (panel b). In the unconditional scenario, average predictability gains reach up to 10\% or 300 patents, respectively. Here, the qualitative patterns for the median predictability gains are similar. The median predictability gain based on the absolute percentage error in the conditional forecasts scenario lies between 1.6\% and 56\% as can be seen in panel c). In the unconditional scenario, the median is slightly negative in 2016 (-0.2\%) and else positive ranging from 0.8\% in the first year of the test set to roughly 9\% in 2012. Panel d) reveals a very similar picture as the one observed in panel c). 
\\[0.1cm]

\noindent
\textbf{IPC 4-digit codes (Figure \ref{fig:pg_ipc4}):}\\
For the IPC 4-digits codes we find average percentage predictability gains between 4\% in 2003 and 72\% in 2009 (panel a) when considering the conditional network forecasts. The predictability gains for the unconditional forecasts can get as high as 43\% in the year 2013. Panel b) shows that the median predictability gains are always positive for the conditional forecasts (ranging from 2\% to 66\%) and positive for the unconditional forecasts in 12 out of 15 forecasts. Only in the last three years, the median predictability gains are slightly negative.
When benchmarking the forecasts with Equation \ref{eq:pg_ae}, we find that on average predictability gains are always positive and tend to increase in time for both prediction scenarios (panel c). Qualitatively, panel d) resembles panel b) where median predictability gains are always positive, except for the last three years in the unconditional forecasts.

\clearpage

\section{Network model of \cite{acemoglu2016innovation}}
\label{section:acemogluetal2016}

In \cite{acemoglu2016innovation} a network model is presented to investigate whether patenting of ``downstream'' technologies can be predicted by patenting rates of ``upstream'' technologies. The authors use different technological classification schemes than we do in this study, namely USPTO classes (USPC) and subcategories (SCAT). A main difference to IPC or CPC codes is that here each patent is associated with a single USPC or SCAT code. On the SCAT level, 36 different technological domains can be distinguished. For the USPC codes, 353  different domains can be differentiated after excluding classes for which there is at least one year with less than five patents in that class. In total, they study around 1.8 million US utility patents which have been applied between 1975 and 2004 and slightly more than 1 million citations.

Predictions are based on a network model which can be written as
\begin{equation} \label{eq:aak}
	\hat{P}_{t} = \sum_{l=1}^{10} M_{l} P_{t-l},
\end{equation}
where $\hat{P}_{i,t}$ is the vector of predicted patenting rates at time $t$ and ${P}_{t-l}$ is the observed patenting at $t-l$. $M_{l}$ represents the innovation network which captures the knowledge spillover from one technological domain at a given time lag $l$ and is defined as
\begin{equation} \label{eq:aak_nw}
M_{ij,l} := \frac{normalized.cites_{ij,l}}{{P}_{j,75-84}},
\end{equation}
where ${P}_{j,75-84}$ is the total patenting in domain $j$ between the years 1975 to 1984.
The numerator $normalized.cites_{ij,l}$ is the total number of weighted citations between patents of domain $i$ and patents of domain $j$. Note that this is not the raw count of citations between two technological domains. Instead, a citation from patent $p$ to patent $q$ is weighted by one over the total number of citations made by patent $p$. We can simplify this by invoking the matrix algebra introduced in the main text (Appendix A). Let $H_{pq,l}$ be the weighted citation from patent $p$ to patent $q$ at time lag $l$ where the weight is obtained by taking the inverse of the total citations made by patent $p$.
Then the numerator of Equation \ref{eq:aak_nw} (ignoring lags) can be defined as
\begin{equation}
	normalized.cites_{ij,l} := B^\top H_l B,
\end{equation}
where $B$ is the bipartite patent-classification network (not row-normalized).

A drawback of this specification is that predictions are based on patenting levels which exhibit pronounced positive time trends. Since forecasts should be based on non-stationary time series, it would make sense to look at the differenced series instead. Additionally, the number of lags in the summation of Equation \ref{eq:aak} is arbitrary as well as the normalization constant in the denominator of Equation \ref{eq:aak_nw}. This renders it difficult to assess whether the predicted levels are due to network effects or due to normalization artifacts or number of lags included. Clearly, using an additional lag $l=11$ inflates the number of predicted patenting rates. Similarly, using a different time horizon for defining the denominator would impact the predicted patenting levels substantially. It should also be noted that forecasts are only one year ahead, since predicting ${P}_{i,t}$ includes data on patenting rates in domain $i$ at time $t-1$.

\begin{figure}[!t]
	\caption{Prediction results based on the model of \cite{acemoglu2016innovation}.
		a) Average predictability gain based on $PG1$. Shaded areas indicate twice the standard error. b) Median predictability gain based on $PG1$. Shaded areas indicate the interquartile range.
		c) Average predictability gain based on $PG2$. Shaded areas indicate twice the standard error. d) Median predictability gain based on $PG2$. Shaded areas indicate the interquartile range.
	}
	\includegraphics[width=\textwidth]{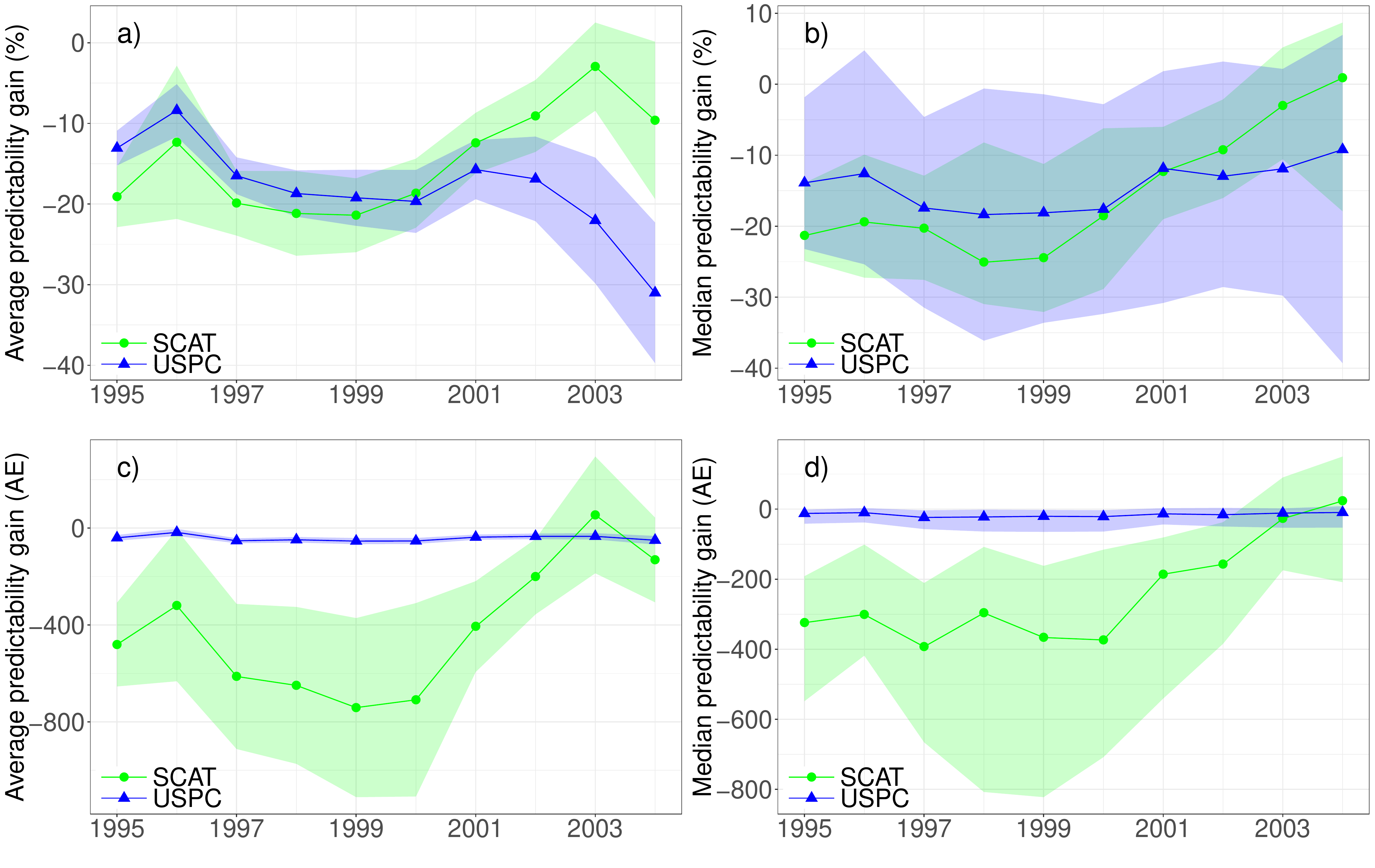}\label{fig:pg_aak}
\end{figure}

In Figure \ref{fig:pg_aak} we show the same performance metrics as discussed before. Here, we simply benchmark the predictions with the one-year forecasts of a geometric random walk model. Neither on average nor for medians we find any positive predictability gains of the network model predictions over the random walk predictions.

\FloatBarrier

\bibliography{../tech_ref} % refers to example.bib
\bibliographystyle{abbrvnat} % or try abbrvnat or unsrtnat